\documentclass[%
 reprint,
 amsmath,amssymb,
 aps,
floatfix,
]{revtex4-2}
\usepackage{xcolor}
\usepackage{graphicx}
\usepackage{dcolumn}
\usepackage{bm}
\usepackage{hyperref}
\usepackage{color}
\usepackage{bigints}

\begin{document}

\preprint{APS/123-QED}

\title{Temperature and its control in molecular dynamics simulations}

\author{M Sri Harish}
 \affiliation{Department of Civil Engineering, Indian Institute of Technology Kharagpur, West Bengal, India - 721302.}
\author{Puneet Kumar Patra}%
 \email{puneet.patra@civil.iitkgp.ac.in}
\affiliation{%
Department of Civil Engineering and Center for Theoretical Studies, Indian Institute of Technology Kharagpur, West Bengal, India - 721302}%

\date{\today}

\begin{abstract}
The earliest molecular dynamics simulations relied on solving the Newtonian or equivalently the Hamiltonian equations of motion for a system. While pedagogically very important as the total energy is preserved in these simulations, they lack any relationship with real-life experiments, as most of these tests are performed in a constant temperature environment that allows energy exchanges. So, within the framework of molecular dynamics, the Newtonian evolution equations need to be modified to enable energy exchange between the system and the surroundings. The prime motive behind allowing energy exchange is to control the temperature of the system. Depending on the temperature being controlled and the modifications made to the equations of motion, different evolution equations, or thermostat algorithms, can be obtained. This work reviews the recent developments in controlling temperature through deterministic algorithms. We highlight the physical basis behind the algorithms, their advantages, and disadvantages, along with the numerical methods to integrate the equations of motion. The review ends with a brief discussion on open-ended questions related to thermostatted dynamics.
\end{abstract}

\maketitle

\section{Introduction}
The computer revolution of the last century has fundamentally altered scientific research. With increasing computational power, researchers can build and test complex situations, which otherwise would have remained unexplored. The importance of computational modeling has been such that it has been adopted across disciplines as broad as physical sciences, chemical sciences, biological sciences, engineering, economics, etc. Now, computational techniques exist for solving problems at all known length scales -- smooth particle hydrodynamics at astronomical scales, finite element method at macroscales, Monte-Carlo, and molecular dynamics at the atomic scale, density functional theory at the quantum scale, etc. 

At atomistic scales, molecular dynamics (MD) has become one of the most popular techniques. MD simulations are compelling, only limited by the availability of computational resources. In MD, the temporal and spatial evolution of individual atoms and molecules (particles) are computed to obtain in-depth insights into the properties of a system and make several testable predictions about it. Consider an isolated system comprising $N$ particles, with a Hamiltonian given by:
\begin{equation}
H = \sum \limits_{i=1}^{3N} \frac{p_i^2}{2m} + \Phi \left( x_1,x_2,...x_{3N} \right),
\label{eq:two_one}
\end{equation}
where, $x_i$ and $p_i$ denote the position and the momentum of the $i^{\text{th}}$ particle, and $\Phi(\ldots)$ denotes the potential energy of the system. For simplicity, all particles are assumed to have the same mass $m$. The equations of motion of such a system are given by \cite{goldstein_01}:
\begin{equation}
\dfrac{dx_i}{dt} = \dfrac{\partial H}{\partial p_i}=\dfrac{p_i}{m}, \ \ \ \dfrac{dp_i}{dt} = -\dfrac{\partial H}{\partial x_i}= -\dfrac{\partial \Phi}{\partial x_i}.
\label{eq:two_two}
\end{equation}
Traditionally, in MD simulations these $6N$ equations are integrated in time to provide the system's trajectory. One of the simplest and most popular symplectic time integration algorithms is the Velocity-Verlet algorithm \cite{velocity_verlet_symplectic}, which propagates the positions and momenta in time through:
\begin{equation}
\begin{array}{ccl}
x_i(t + \Delta t) & = & x_i(t) + \dfrac{p_i(t)}{m} \Delta t+ \dfrac{1}{2} \dfrac{F_i(t)}{m} \Delta t^2, \\
p_i(t + \Delta t) & = & p_i(t) + \dfrac{1}{2} \left[ F_i(t) + F_i(t+\Delta t) \right] \Delta t.
\label{eq:two_three}
\end{array}
\end{equation}
Here $F_i = -\dfrac{\partial \Phi}{\partial x_i}$ is the force experienced by the $i^{\text{th}}$ particle. Since the system is isolated from the surroundings, these simulations preserve the Hamiltonian and the total energy, thereof. The trajectories obtained from these MD simulations are concomitant with the micro-canonical (NVE) ensemble encountered in statistical mechanics; the link between the two being provided by the ergodic hypothesis. 

The ability of MD simulations to accurately reflect the real-life properties of a system relies on the accurate modeling of -- (i) the interactions between the different particles that constitute the system, and (ii) the interaction of the system with its surroundings. While the accuracy of the interatomic interactions depends on the choice of potential ($\Phi(\ldots)$), modeling the interaction between the system and its surroundings depends on the equations of motion. Traditional MD, owing to it conserving the total energy, is well suited for modeling isolated systems. However, it falls short for modeling systems that exchange energy with the surroundings and have a constant temperature. Such scenarios occur very commonly in real-life. For example, most real-life experiments are performed in a constant temperature environment, where a continuous energy exchange occurs between the system and the heat reservoir. In order to model these scenarios through MD, the equations of motion (\ref{eq:two_two}) need to be supplemented in a manner that continuous energy exchange is permitted, a constant temperature is enforced and the dynamics is sampled from the Gibbs' distribution. Such alterations to the equations of motion, however, come with the cost of developing appropriate time-integration techniques.

Apart from accurately mimicking real-life experiments at constant temperatures, temperature control in MD is needed for -- (i) identifying the equilibrium properties of a system, (ii) calculating the transport characteristics of systems using Green-Kubo relations or by subjecting the system to dissipative fields, (iii) understanding temperature-dependent mechanical properties of materials, (iv) creating amorphous systems based on high-temperature quenching, and (v) extracting energy from a system on which external work is performed, etc. In fact, temperature control in MD is essential for studying non-equilibrium situations due to the lack of sound theoretical understanding of non-equilibrium statistical thermodynamics.  

Initial attempts at developing temperature control algorithms revolved around controlling the kinetic temperature of the system, and until as recently as the early part of this century, only kinetic temperature based thermostats were in existence. However, recent advances in statistical mechanics have rendered new ways of defining temperature that goes beyond the traditional kinetic temperature. The temperature in MD, which used to be invariably kinetic temperature, now can be defined solely in terms of the configurational variables. These advances have prompted the development of different classes of temperature control algorithms that go beyond controlling the kinetic temperature. It must be noted that each thermostat algorithm has its strength and weakness -- while some may be easier to implement and computationally less taxing, others may provide a closer approximation to reality -- but none is perfect. 

In this review, our objective is to provide a comprehensive guide for researchers trying to simulate a system whose (part or whole) temperature is controlled. We briefly describe the applicability of the different temperature control algorithms to different scenarios -- both equilibrium and non-equilibrium. Using the prototypical case of a single harmonic oscillator, under both equilibrium and non-equilibrium conditions, we demonstrate the similarities and the differences in the dynamical behavior of the different thermostats. The scope of this work is limited to deterministic thermostats. This review is organized as follows: in the next section, we discuss the properties of a good thermostat. We, subsequently, discuss the different ways of defining temperature -- thermodynamic, kinetic, generalized, configurational, Rugh's -- and algorithms for their control. The numerical integration of equations of motion for each temperature control algorithm (wherever accessible) is presented. The review ends with a set of open-ended questions, whose answers, once obtained, may further enrich thermostatted dynamics.

\section{Properties of a Good Deterministic Thermostat}
The role of a good deterministic thermostat goes beyond just controlling the temperature of the system. So, what makes a deterministic thermostat good? Some of the properties are listed below:
\begin{enumerate}
\item \textbf{Time-reversibility}: Consider a deterministic system which starts at the microstate $\mathbf{(x_0,p_0)}$, and evolves to the microstate $\mathbf{(x_\tau,p_\tau)}$ in time $\tau$. Hamiltonian mechanics suggests that if $\mathbf{p_\tau}$ is reversed instantaneously i.e., $\mathbf{p_\tau} \to -\mathbf{p_\tau}$ and the dynamics proceeds for a time duration of $\tau$, the initial microstate with reversed momenta i.e., $\mathbf{(x_0,-p_0)}$, is obtained. Alternatively, if the system starts from the microstate $\mathbf{(x_\tau,p_\tau)}$ and proceeds backward in time for a time duration of $\tau$, the initial microstate is reached. The equations of motion for a deterministic thermostat \textit{must} obey similar time-reversal symmetry.

\item \textbf{Ergodicity}: MD simulations are typically performed with one sample-path, which begins from a specific set of initial conditions. The properties measured from MD simulations are, as a result, time-averaged quantities obtained over the single sampled path. Our measurements in laboratory settings, on the other hand, are based on phase-space averages. So, there seems a disconnect -- on one side we have time-averaged quantities, and on the other, phase-averaged quantities. The disconnect can be removed by invoking the ``ergodic hypothesis'', which states that a phase-averaged variable is the same as its time-averaged counterpart. In essence, ergodicity of the dynamics is a prerequisite for obtaining statistical-mechanical properties from a single run of MD simulation. Thus, extracting any meaningful result from an MD simulation relies on the premise that the dynamics is ergodic. Throughout this work, we use Ehrenfests' ``\textit{quasiergodicity}'' equivalently with the traditional ergodicity. (Quasi)ergodicity says that a trajectory initiating from any microstate within the accessible phase-space must eventually come arbitrary close to all the microstates that lie within the accessible region \cite{sprott_hoover_14}. In the context of MD simulations, this implies that the phase-space trajectory must visit the accessible phase space in a frequency commensurate with the theoretical phase-space probability distribution. 

While no one doubts the necessity of ergodicity of dynamics in the context of MD simulations, it has been proved theoretically only for a handful of systems like hard billiard balls \cite{Domokos_ergodicity} and Lorentz gas \cite{sinai_79}. One usually employs numerical techniques to determine if the dynamics is ergodic. It has been customary to study the ergodic characteristics of thermostatted dynamics using a single harmonic oscillator \cite{legoll,bbk_pra,hoover_86,watanabe_07a} due to the oscillator's ``stiff'' nature and simplicity. To the best of our knowledge, no watertight proof of ergodicity exists for any thermostatted dynamics. In larger systems, comprising hundreds of thousands of degrees of freedom, the (non)ergodicity of dynamics takes a back seat owing to the large Poincar\'e recurrence time, which often is greater than the age of the universe \cite{zeh2001dieter}. Thus, although ideally, we \textit{need} algorithms that impart ergodicity to the dynamics, the effect of non-ergodicity becomes progressively less important as the system size increases. 

There are two ways of ascertaining ergodicity numerically. Both the approaches are briefly described, and the interested readers are referred to \cite{patra2014nonergodicity,patra2015deterministic} for a more comprehensive treatment. Note that in both approaches, we deal with a single harmonic oscillator.

\textit{Dynamical Systems Approach}: Non-ergodicity implies the partition of phase-space into two (or more) non-communicating regions. So, when the dynamics of a thermostatted single harmonic oscillator is limited to a (hyper-)torus for one or more initial conditions (except perhaps those which form a set of zero-measure), the thermostatted dynamics is non-ergodic. For a multidimensional phase-space flow, as is the case with thermostatted single harmonic oscillators, this can be assessed by studying the Lyapunov spectra of the dynamics \cite{posch_hoover}. A $d-$dimensional flow is associated with $d$ Lyapunov exponents, which are ordered: $L_1 > L_2 > \ldots > L_{d-1} > L_d$. The sum of these exponents describes the deformation of an infinitesimal hypercube in the $d-$dimensional space describing the motion. $L_1$ gives the time-averaged rate of separation of two neighboring trajectories. $L_1+L_2$ gives the rate of separation of the area defined by three neighboring trajectories. Similarly, $L_1+L_2+L_3$ describes the separation rate of volume, and so on. The pre-deformation infinitesimal hypervolume $\delta V(0)$ with the post-deformation hypervolume $\delta V(t)$ may be related as:
\begin{equation}
\delta V(t) = \delta V(0) e^{\left( L_1 + L_2 + \ldots + L_{d-1} + L_d \right)t}
\end{equation}
In equilibrium, $\sum \langle L_i \rangle = 0$, i.e. there is no change in phase-space volume in an averaged sense, and $\langle L_1 + L_d \rangle = \langle L_2 + L_{d-1} \rangle = \ldots = 0$, i.e. the Lyapunov exponents are conjugately paired. To ascertain the ergodic characteristics, Lyapunov spectra corresponding to millions of initial conditions need to be found. For non-ergodic dynamics, at least one initial condition yields a (hyper-)torus for which all the Lyapunov exponents are statistically insignificant from zero. 

\textit{Statistical Approach}: Consider the Maxwell-Boltzmann distribution given by equation (\ref{eq:two_nine}). For a single harmonic oscillator of unit mass and stiffness, the Maxwell-Boltzmann distribution reduces to a product of two independent normal distributions:
\begin{equation}
\begin{array}{ccl}
f(x,p) & = & \dfrac{1}{Z} \bigint \left[ \exp \left(-\dfrac{\beta_0}{2} x^2\right)\exp \left(-\dfrac{\beta_0}{2} p^2\right) \right] \\
\label{eq:three_one}
\end{array}
\end{equation}
In the presence of thermostat variables ($\eta_1, \eta_2, \ldots, \eta_n$), the distribution function gets augmented by additional terms. However, the conditional distribution of oscillator's position and momentum variables still remains jointly normal i.e.
\begin{equation}
\begin{array}{ccl}
f(x,p|\eta_1 = \eta_{1,0}, \eta_2 = \eta_{2,0}, \ldots, \eta_{n}=\eta_{n,0}) & = & \\
\dfrac{1}{Z'} \bigint \left[ \exp \left(-\dfrac{\beta_0}{2} x^2\right)\exp \left(-\dfrac{\beta_0}{2} p^2\right) \right] & & \\
\label{eq:statistical_method}
\end{array}
\end{equation}
The statistical approach of ascertaining ergodicity looks at this conditional joint distribution of position and momentum to assess if the dynamics samples the phase-space following the Maxwell-Boltzmann distribution. Any deviation from normality is an indicator of non-ergodicity. Numerically, this implies finding the joint probability distribution of position and momentum in a Poincar\'e section defined by $\eta_1 = \eta_{1,0}, \eta_2 = \eta_{2,0}, \ldots, \eta_{n}=\eta_{n,0}$, and performing a test for normality. 

\begin{figure*}
\centering \includegraphics[scale=0.375]{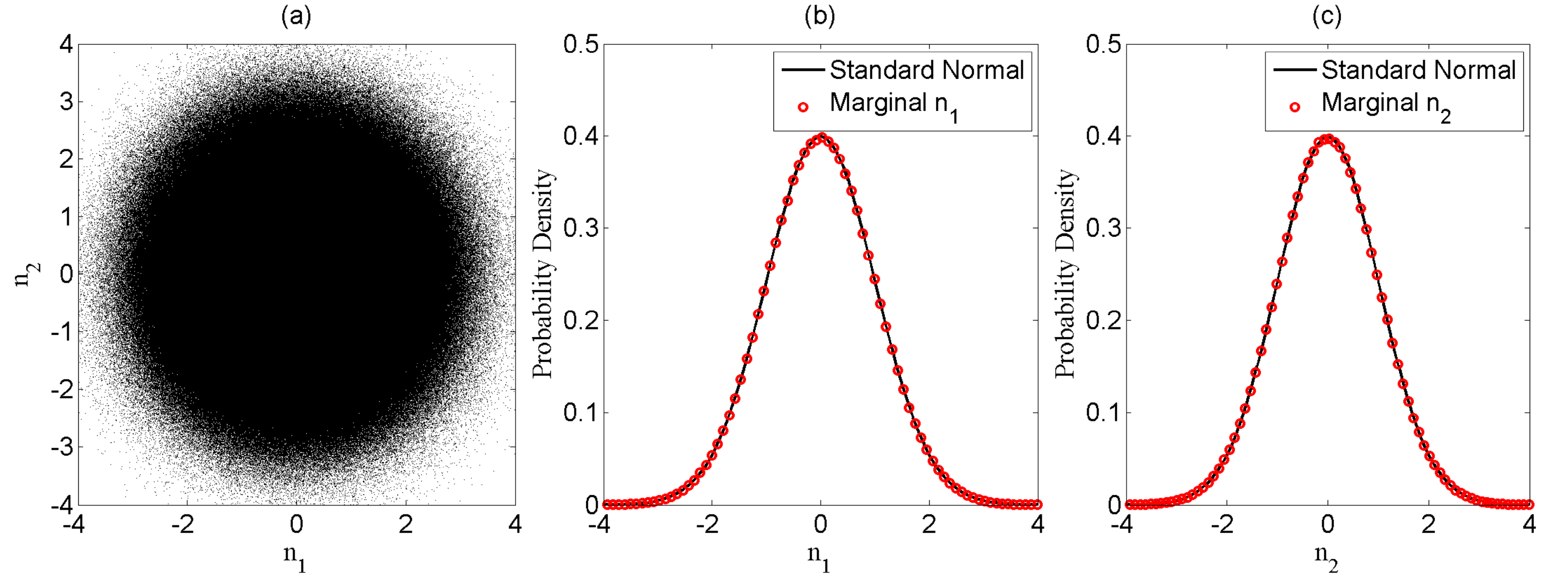}
\caption[Inability of projected variables to capture embedded holes]{\label{fig:3one} The inability of projected variables $n_1-n_2$ to capture the 4-dimensional hole of radius 0.25 forcefully embedded within a 4-dimensional joint standard normal. (a) Projected values of $n_1-n_2$ onto the $n_3 = 0, n_4 = 0$ plane. (b) Marginal distribution of $n_1$ and (c) marginal distribution of $n_2$. No difference in marginal distributions from that of standard normal can be observed.}
\end{figure*}

It is important to note that the marginal distribution of position and velocity are often incapable of capturing the deviation from normality. This is exemplified next. Consider a four-dimensional space filled with 10 million random samples drawn from the independent standard normal vectors ($n_1,n_2,n_3,n_4$). Let a small hole be embedded in this 4-D space by deleting the points that lie within the hyper-sphere $n_1^2 + n_2^2 +n_3^2 + n_4^2 \leq 0.0625$. Clearly, having lost all probability content around the origin of the 4-d space, the remaining points would no longer satisfy a joint normal density function. However, this system, when analyzed for holes by erroneously checking for departure from marginal normality, provides no clue at all about the hole. The phase-space plot projected on to $n_3 = 0, n_4 = 0$ plane is devoid any empty space (see figure \ref{fig:3one} (a)) and the marginal distributions of $n_1$ and $n_2$ agree marvellously with a standard normal distribution (see figures \ref{fig:3one} (b) and (c)). The first three even marginal and joint moments agree well with the standard normal distribution: $\langle n_1^2 \rangle = 1.000$, $\langle n_1^4 \rangle = 3.000$, $\langle n_1^6 \rangle = 15.007$, $\langle n_2^2 \rangle = 0.999$, $\langle n_2^4 \rangle = 2.999$, $\langle n_2^6 \rangle = 14.992$, $\langle n_1^2n_2^2 \rangle = 0.999$, $\langle n_1^4n_2^4 \rangle = 9.003$ and $\langle n_1^6n_2^6 \rangle = 227.496$. Thus, one needs to check ergodicity using conditional joint distributions rather than marginal distributions.

\item \textbf{Conformity with the Laws of Thermodynamics}: Being mathematical counterparts of the real-life thermal reservoirs, the thermostat algorithms \textit{must} satisfy the different laws of thermodynamics. Take the Zeroth Law of thermodynamics, for example, which says that if two bodies are in mutual thermal equilibrium with a third body, then the two bodies are in thermal equilibrium with each other. Now, consider a system in thermal equilibrium with known macroscopic properties. If this system is coupled with a thermostat (algorithm) at the same temperature, as per the Zeroth Law, the macroscopic properties of the system are time-invariant. Going one step further, if the system is coupled with more than one thermostat, the properties of not only the system but the individual thermostats are also time-invariant. This idea is central to testing if a thermostat algorithm satisfies the Zeroth Law: a single harmonic oscillator is simultaneously coupled with two thermostats kept at the same temperature. The resulting equations of motion are solved, and the phase-space of the oscillator is analyzed along with that of the thermostat. The Zeroth Law is not satisfied if: (i) the temperature of any of the thermostats differ from that of the system or with each other, (ii) there is a heat flow within the system, or (iii) the time-averaged phase-space compression, which is defined as,
\begin{equation}
    \langle \Lambda \rangle_t = \dfrac{1}{\tau}\dfrac{\partial \mathbf{\dot{\Gamma}}_{ex}}{\partial \mathbf{\Gamma}_{ex}},
    \label{eq:eq7}
\end{equation}
is not equal to zero. Here, $\tau$ represents the time, and $\mathbf{\Gamma}_{ex} \equiv (x,p,\eta_1,\eta_2,...,\eta_k)$ denotes the extended phase-space, with $\eta_i$s as the thermostat variables. 

The Second Law, on the other hand, demands heat to flow from a thermostat at a higher temperature to a thermostat at a lower temperature spontaneously. In essence, if one end of a system is thermostatted at a higher temperature and the other end at a lower temperature, then a spontaneous heat flow \textit{must} occur between the hotter and the colder thermostats. Thermostats unable to engineer such a heat flow fail to satisfy the Second Law, and hence, are not suitable for molecular dynamics simulations. Apart from the ability to ensure heat flow, the time-averaged phase-space compression, defined by (\ref{eq:eq7}), must be non-zero. In order to check if a thermostat satisfies the Second Law, we take a single harmonic oscillator and subject it to a position-dependent temperature field
\begin{equation}
    k_BT(x) = 1 + \epsilon \tanh(x),
    \label{eq:eq8}
\end{equation}
by coupling it with a thermostat. Here, $\epsilon$ denotes the strength of non-linearity. In this non-equilibrium problem, there is phase-space compression and the dynamics is such that a heat flow occurs. A good thermostat must be able to demonstrate these features. 

\item \textbf{Autonomous and Easy Implementation}: The equations of motion corresponding to the deterministic thermostats must be autonomous, i.e., they should not have an explicit dependence on time. An explicit dependence on time makes the equations lose their time-invariance characteristic, which is a prerequisite for equilibrium. Lastly, the equations of motion must be easy to implement, not be stiff, and simple to solve using existing numerical techniques. 

\item \textbf{Hamiltonian Basis}: Ideally, the equations of motion should have a Hamiltonian basis, and they may be obtained from Hamilton's equations. It is not necessary for the Hamiltonian to represent the ``true'' total energy of the system or evolve in ``Newtonian'' time. For example, the Hamiltonian time may be related to the Newtonian time through some transformation (for example, Sundman's transformation \cite{bond1999nose,sweet2004hamilton,leimkuhler2004canonical}).  

\end{enumerate}

\section{Thermodynamic Definition of Temperature}
Theoretically, the concept of temperature goes beyond mere perception of the degree of hotness or coldness of a body. Instead, it has a solid mathematical background. The idea of temperature begins with the Zeroth Law of thermodynamics. If two bodies are in mutual thermal equilibrium with a third body, then the two bodies are in thermal equilibrium with each other. The Zeroth Law enables us to define a class of equivalence relations that is symmetric, reflexive, and transitive. These equivalence relations are isotherms, each of which is associated with an empirical variable called temperature. However, the Zeroth Law in itself is not sufficient to identify the relative degree of hotness. For this purpose, we need both the First Law and the Second Law of thermodynamics. 

The First Law enables us to define heat energy in terms of the conservation of total energy. In a closed system, the sum of the change in internal energy and thermodynamic work done on the system equals heat energy supplied to the system:
\begin{equation}
\delta Q = dU + \delta W = dU + PdV,
\label{eq:two_four}
\end{equation}
where $\delta Q$ is the heat energy supplied to the system, $dU$ is the change in the internal energy of the system, and $dW$ is the work done on the system which can be expressed in terms of the pressure, $P$, and the change in volume, $dV$. 

The Second Law helps in identifying the relative hotness of two bodies in terms of the spontaneous flow of heat energy between them. For a reversible process (necessarily in equilibrium), one can replace the LHS of (\ref{eq:two_four}) and rewrite it as: 
\begin{equation}
TdS = dU + PdV,
\label{eq:two_five}
\end{equation}
where, $T$ = temperature of the system, and $dS$ is the change in entropy of the system. All three laws combined together give the thermodynamic definition of temperature,
\begin{equation}
\dfrac{1}{T}= \left(\dfrac{\partial S}{\partial U}\right)_V
\label{eq:two_six}
\end{equation}
One can understand temperature as the ratio of change in the internal energy of a system for a unit change in its entropy at constant volume \cite{haile_book}. Interestingly, the temperature may also be viewed as an integrating factor that converts the path differential variable $\delta Q$ to the total differential variable $\delta S$. Both these interpretations do not preclude the concept of negative temperature \cite{powles_63}. 

One can understand the concept of thermodynamic temperature in other ways as well. Consider Jaynes' framework of statistical mechanics \cite{jaynes_57,jaynes_57b}, where one maximizes Shannon's entropy functional \cite{shannon_48},
\begin{equation}
C = \int f(\mathbf{\Gamma}) \log(f(\mathbf{\Gamma}) \ d\mathbf{\Gamma},
\label{two_fourteen}
\end{equation}
subjected to the constraints imposed by the physics of the problem. For example, there are two constraints in a canonical ensemble -- energy constraint, $\langle H \rangle = E$, and normalization constraint, $\int f(\mathbf{\Gamma}) d \Gamma = 1$. The least biased distribution for the canonical ensemble can be found by maximizing:
\begin{multline}
C = \int f(\mathbf{\Gamma}) \log(f(\mathbf{\Gamma})) \ d\mathbf{\Gamma} - \lambda_0 \left( \int f(\mathbf{\Gamma})\ d\mathbf{\Gamma}  - 1\right) \\ - \lambda_1 \left( \int H(\mathbf{\Gamma}) f(\mathbf{\Gamma})  \ d\mathbf{\Gamma} - E \right),
\label{two_fifteen}
\end{multline}
where, $\lambda_0$ and $\lambda_1$ are the Lagrange multipliers associated with the normalization constraint and the energy constraint, respectively. It is straightforward to show that the least biased distribution is the canonical distribution, with $\lambda_1 = \beta = 1/k_BT$ \cite{casas_review_03}. Here, $k_B$ is the Boltzmann constant. Thus, the temperature of a system may be identified as the inverse of the Lagrange multiplier associated with the energy constraint. 

From an operational perspective, measurement and control of thermodynamic temperature are fraught with difficulties, both in equilibrium and non-equilibrium cases. This is because the computation of Gibbs' entropy in equilibrium comprises a $6N-$dimensional integral, performing which is a computational nightmare. The situation is more problematic in non-equilibrium, where the meaning of $S$ itself remains an open question. As a result, in MD we never use thermodynamic temperature. Instead, other definitions of temperature, some of which we describe next, are used. 

\section{Kinetic Temperature and Its Control}
In MD simulations, the temperature used to be invariably expressed in terms of the kinetic variables. Kinetic definition of temperature owes its origins to the kinetic theory of gases. Consider an isolated system comprising ideal gas particles that are confined to move within a container. The particles are assumed to be rigid and collide elastically with each other and with the walls. A quick comparison of the pressure exerted by the particles on the wall surface with the ideal gas equation of state reveals that the kinetic temperature \cite{huang_book}, $T_k$, is:
\begin{equation}
\dfrac{3}{2}k_BT_k = \left\langle \dfrac{p_i^2}{2m} \right\rangle.
\label{eq:two_seven}
\end{equation}
Here, $\langle \ldots \rangle$ denotes the phase-averaged quantity. Unlike thermodynamic temperature, the expression of $T_k$ is easy to compute and control in a computer simulation without expending too many computational resources. If the system is isotropic, as is the case with an ideal gas or a system without any directional dependence, one can employ equipartition theorem to show that the average of each component of velocity contributes equally towards temperature i.e.,
\begin{equation}
\dfrac{1}{2}k_BT_k = \left\langle \dfrac{p_{i,x}^2}{2m} \right\rangle = \left\langle \dfrac{p_{i,y}^2}{2m} \right\rangle = \left\langle \dfrac{p_{i,z}^2}{2m} \right\rangle
\label{eq:two_eight}
\end{equation}

Alternatively, $T_k$ may be derived from the canonical probability distribution function, wherein the phase space is sampled according to the Maxwell-Boltzmann distribution:
\begin{equation}
f(\mathbf{x},\mathbf{p}) = \dfrac{1}{Z}\exp \left(-\beta \Phi(\mathbf{x}) - \beta \left[ \sum_{i=1}^{3N} p_i^2 / 2m \right] \right).
\label{eq:two_nine}
\end{equation}
Here, $Z$ is the partition function and $\beta = (k_BT)^{-1}$. Equations (\ref{eq:two_seven}) and (\ref{eq:two_eight}) may be obtained by relating, $1/\beta$, the variance of the Maxwell-Boltzmann distribution with the momentum variables. 

From a statistical perspective, when a system is in thermal equilibrium with a reservoir, \textit{all} moments of the momentum distribution must agree with the kinetic temperature of the system. This statistical nature of the momentum variables may be exploited to obtain higher-order measures of $T_k$, as often, it has been observed that the control of the second-moment based $T_k$ alone is insufficient for effectively thermalizing small-scale systems such as single harmonic oscillators. With this in mind, higher-order moments of kinetic temperature -- $T_{k,2}$, the kinetic temperature calculated from the fourth moment of velocity distribution and $T_{k,3}$, the kinetic temperature calculated from the sixth moment of velocity distribution -- may be calculated as:
\begin{equation}
\begin{array}{cc}
k_BT_{k,2} = \langle \sqrt{\dfrac{p_i^4}{3m^2}} \rangle, & k_BT_{k,3} = \langle \sqrt[3]{\dfrac{p_i^6}{15m^3}} \rangle.
\end{array}
\label{eq:two_ten}
\end{equation}

An unconventional route to kinetic temperature is from the viewpoint of the kinetic energy of the system -- as particle velocities are normally distributed, the kinetic energy, $K$, follows a $\chi^2$ distribution. Based on this argument, $T_k$ shown in equation (\ref{eq:two_seven}) represents the mean of the $\chi^2$ distribution. Although $T_k$ calculated from the first even order moment of momentum distribution and the first moment of kinetic energy distribution agree with each other, the second-order kinetic temperature obtained from kinetic energy distribution, $T_{K,2}$, is different from that of $T_{k,2}$:
\begin{equation}
\begin{array}{rcl}
k_BT_{K,2} & = & \dfrac{\langle 4K^2 \rangle }{\langle 2K(N+2) \rangle}
\label{eq:2nd_order_kin_temp}
\end{array}
\end{equation}

It is important to note that all expressions of kinetic temperature written so far apply only to the translation motion of the individual particles. These expressions require modification if there is a center of mass translation or if rotational and internal degrees of freedom are present \cite{allen_tildesley}. 

\subsection{Velocity Rescaling}
Velocity rescaling utilizes the kinetic temperature, shown in equation (\ref{eq:two_seven}), to control the temperature. It is possibly the simplest of all temperature control algorithms. As the name suggests, the instantaneous velocities of the particles are rescaled in an ad-hoc manner such that the desired temperature, $T_0$, is obtained \cite{woodcock_71}. For a $d-$dimensional system, $T_0$ is related to the desired kinetic energy, $K_0$, as:
\begin{equation}
T_0 = \dfrac{2}{dNk_B}K_0.
\label{eq:two_sixteen}
\end{equation}
Let, at any time $t$, the instantaneous kinetic energy, $K_t$, be given by: $K_t = \sum p_i^2/2m$. The corresponding instantaneous kinetic temperature, $T_t$, is $T_t = 2K_t/(k_B dN)$. Evidently, if the instantaneous velocities of each particle are scaled by a factor $\alpha$, where,
\begin{equation}
\alpha = \sqrt{\dfrac{K_0}{K_t}}
\label{eq:two_eighteen}
\end{equation}
the instantaneous temperature of the system is forced at $T_0$. The rescaling may be performed after every few steps. 

\subsubsection{Solving the Equations:}
We now briefly describe the integration technique that may be adopted to solve the equations of motion for a large system:
\begin{enumerate}
    \item Initialize the positions and velocities of the particle as per the desired temperature
    \item Solve for $x_i(t + \Delta t)$ using equation (\ref{eq:two_three}).
    \item Compute the updated force on each particle.
    \item Solve for $p_i(t + \Delta t)$ using equation (\ref{eq:two_three}).
    \item Compute the instantaneous kinetic energy $K_t$ and scaling constant $\alpha$ after some iterations.
    \item Compute the rescaled velocities: $v_i \to \alpha v_i $.
\end{enumerate}
For a small system, the Runge-Kutta method can be used for solving the equations of motion.

\subsubsection{Phase-Space Characteristics using a Single Harmonic Oscillator}
\begin{figure}
\centering \includegraphics[scale=0.45]{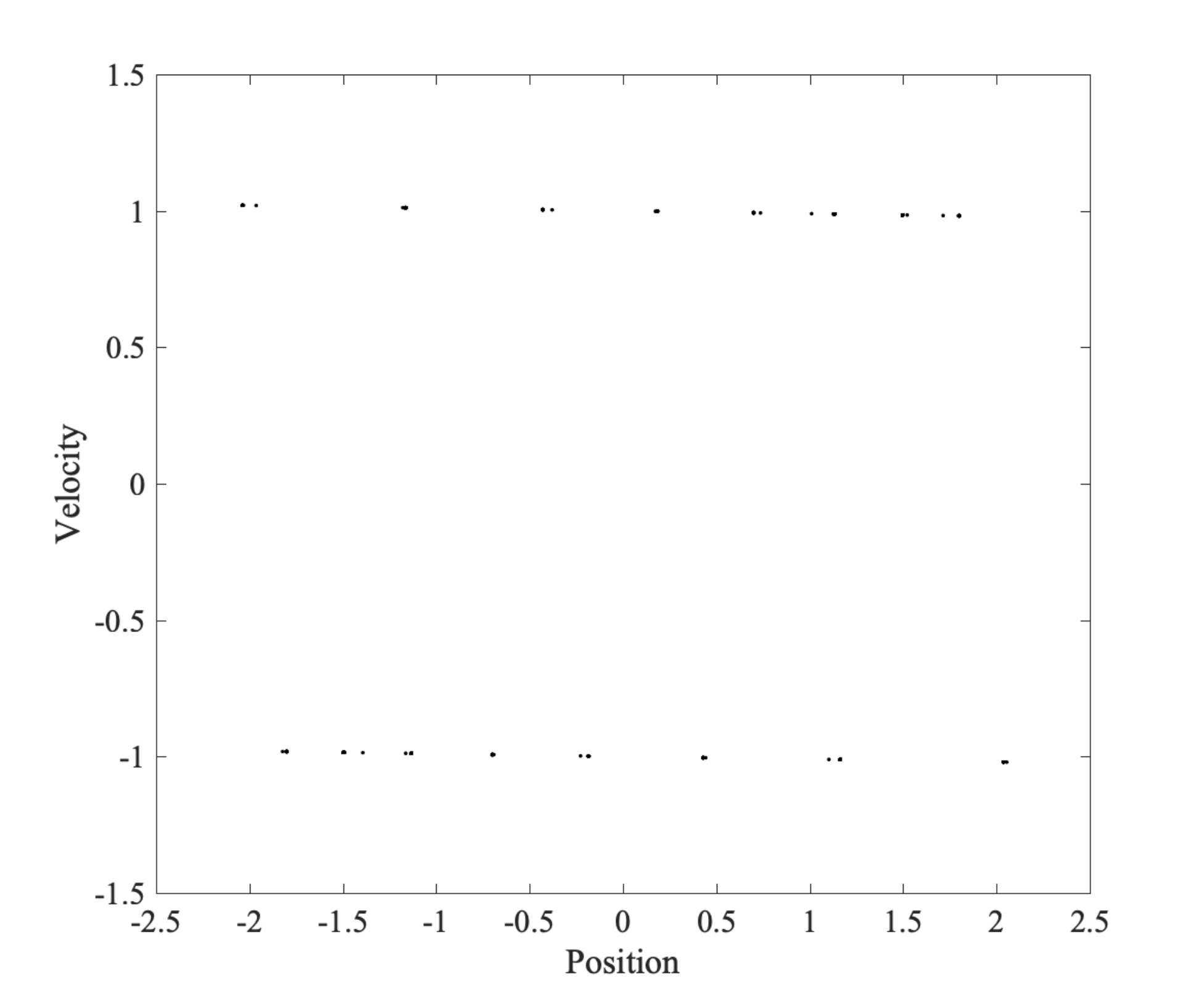}
\caption[Inability of projected variables to capture embedded holes]{\label{fig:velrescale} The phase trajectory for the velocity-rescaling algorithm. The velocities are rescaled every 1000 time-steps. Notice that the velocity jumps between +1 and -1 depending on if the instantaneous velocity is positive or negative. It is clear that the dynamics does not sample from a canonical ensemble.}
\end{figure}

Let us exemplify this algorithm for a single harmonic oscillator of unit mass and stiffness with $k_BT_0 = 1$. The equations of motion are:
\begin{equation}
    \dot{x} = p, \dot{p} = -x, p \to 1.
    \label{eq:vel_rescale_sho}
\end{equation}
Here we adopt the $4^{th}$ order Runge-Kutta method for solving the first two equations for 1 billion time-steps where each time step equals 0.001. Rescaling step performed every 1000 time steps. The resulting phase-space trajectory is shown in figure (\ref{fig:velrescale}). As is evident, the trajectory fails to sample from the correct canonical ensemble (see equation (\ref{eq:statistical_method})).  

The sampled ensemble is not canonical in momentum variables. However, in massive systems, this algorithm may provide satisfactory results due to the equivalence of different ensembles in the thermodynamic limit \cite{martin1979equivalence}. Apart from the inability to sample correctly, this algorithm is not physics preserving as the rescaling is done in an ad-hoc manner \cite{hunenberger} and does not follow any of the properties of a good thermostat highlighted before. The sampled space and the temperature fluctuations depend on the frequency of rescaling. For example, if the rescaling had been done every time step for the single harmonic oscillator (see equation (\ref{eq:vel_rescale_sho})), then the positions, $x$, reach unphysical values such as 100 units, and the fluctuations in temperature are zero. Because of these reasons, the velocity rescaling algorithm has fallen out of favor of researchers. However, a simple modification in the algorithm can remove some of the drawbacks just discussed. The idea is to treat $K_0$ not as a deterministic quantity but as a stochastic variable \cite{bussi_07}. Since all velocities in canonical ensemble are normally distributed, the desired kinetic energy follows a $\chi^2$ distribution. So, instead of keeping $K_0$ as a fixed quantity defined by equation (\ref{eq:two_eighteen}), one needs to sample $K_0$ from an appropriate $\chi^2$ distribution. 

\subsection{Gaussian Isokinetic Thermostat}
Gaussian isokinetic thermostat (GIK) is amongst the first physics-preserving deterministic thermostats. It was simultaneously developed by Hoover \cite{hoover_gik} and Evans \cite{evans_gik,evans_morriss_gik,evans_morriss_gik2}. As the name suggests, the GIK thermostat maintains constant kinetic energy, and hence, kinetic temperature. This is achieved through Gauss' principle of least constraint, which states that the true trajectory of a constrained system is the one that minimizes, in a least-squared manner, the difference between the true acceleration and the unconstrained acceleration of the constituent particles \cite{evans_book}: 
\begin{equation}
M=\dfrac{1}{2} \sum\limits_{i=1}^{3N} m\left( \ddot{x_i} - \dfrac{F_i}{m} \right),
\label{eq:two_nineteen}
\end{equation}
where $\ddot{x}_i$ is the true acceleration of a particle $i$ and $\frac{F_i}{m}$ is its unconstrained acceleration. This principle is applicable to any equality constraint, be it holonomic or non-holonomic \cite{evans2008statistical}. For GIK thermostatted dynamics, the constraint is non-holonomic \cite{morriss1998thermostats}, which can be written as:
\begin{equation}
g \left( x_i, \dot{x}_i, t \right) = \sum\limits_{i=1}^{3N} \dfrac{1}{2}m\dot{x}_i^2 - \dfrac{3}{2}Nk_BT_0 = 0.
\label{eq:two_twenty}
\end{equation}
Physically, equation (\ref{eq:two_twenty}) says that at every instant the kinetic energy of the system is constant and equals $3/2Nk_BT_0$. Differentiating the constraint once gives the differential form the constraint equation in the acceleration space: $G(\dot{x}_i,\ddot{x}_i, t) = \sum\limits_{i=1}^{3N}m\dot{x}_i.\ddot{x}_i=0$. Minimizing $M$ subjected to the constraint, $G$, through the Lagrangian multiplier, $\lambda$, gives:
\begin{equation}
\dfrac{\partial}{\partial \ddot{x}_i} \left( M + \lambda G \right) = 0,
\label{eq:two_twentyone}
\end{equation}
so that the equations of motion become:
\begin{equation}
m\ddot{x}_i = F_i - \lambda m \dot{x}_i.
\label{eq:two_twentytwo}
\end{equation}
The mathematical form of $\lambda$ can be obtained by substituting equation (\ref{eq:two_twentytwo}) in the differential constraint, $G$:
\begin{equation}
\lambda = \dfrac{\sum F_i \dot{x}_i}{\sum m \dot{x}_i^2} = \dfrac{\sum F_i p_i}{\sum p_i^2} .
\label{eq:two_twentythree}
\end{equation}
The system should be initialized carefully  -- the net initial kinetic energy \textit{must} equal $3Nk_B T_0$, failing which, the system gets thermalized at a different temperature since the evolution equations do not explicitly account for $T_0$. Equation (\ref{eq:two_twentytwo}) bears similarity to a damped system, where $\lambda$ acts like the damping coefficient. In such systems, $\lambda > 0$ and is a constant. As a result, energy is constantly extracted from the system. However, in the case of the GIK thermostat, $\lambda$ is neither a constant nor always positive. The magnitude and sign of $\lambda$ changes in time, depending on the instantaneous value of equation (\ref{eq:two_twentythree}): $\lambda > 0$ indicates that the reservoir extracts (heat) energy from the system, while $\lambda < 0$ suggests that the reservoir supplies (heat) energy to the system. The pseudo energy that is a constant of motion for the GIK thermostat is:
\begin{equation}
    E_{GIK} = \Phi(\mathbf{x}) + \sum \dfrac{p_i^2}{2m} + \int\limits_0^t \sum F_i p_i dt 
\end{equation}

Equations of motion of the GIK thermostat can also be derived using a Hamiltonian formulation \cite{dettmann_gik}. For simplicity, consider the case where $k_BT_0=1$ and $m=1$. Let the Hamiltonian, $H_{\text{GIK}}$, be given by: 
\begin{equation}
H_{\text{GIK}} (x_i,\pi_i,t') = \dfrac{1}{2} e^{\left[ \left( \gamma + 1 \right) \Phi \right]} \sum\limits_{i=1}^{3N}\pi_i^2 - \dfrac{1}{2}e^{\left[ \left( \gamma - 1 \right) \Phi\right]},
\label{eq:two_twentyfour}
\end{equation}
where $\Phi$ is the potential energy, $\pi_i$ is the momentum conjugate to $x_i$ and is different from the real momentum $p_i$, $t'$ is the Hamiltonian time, which is different from the real time $t$, and $\gamma$ is an arbitrary multiplier. Applying Hamilton's equations provide:
\begin{equation}
\begin{array}{lllll}
\dfrac{dx_i}{dt'} & = & \dfrac{\partial H_{\text{GIK}}}{\partial \pi_i} & = & e^{\left[ \left( \gamma + 1 \right)\Phi \right]} \pi_i \\
\dfrac{d\pi_i}{dt'} & = & \dfrac{-\partial H_{\text{GIK}}}{\partial x_i} & = & \dfrac{-1}{2}\dfrac{\partial \Phi}{\partial x_i} e^{\left[ \left( \gamma - 1 \right) \Phi \right]} \times \\
& & & & \left[ \left( \gamma + 1\right) e^{(2\Phi)} \sum\limits_{i=1}^{3N}\pi_i^2 - (\gamma - 1) \right]. \\
\label{eq:two_twentyfive}
\end{array}
\end{equation}
Equations of motion of the GIK thermostat -- equations (\ref{eq:two_twentytwo}) and (\ref{eq:two_twentythree}) -- can be obtained from equation (\ref{eq:two_twentyfive}), if the Hamiltonian variables and real-time variables are related as follows: 
\begin{equation}
\begin{array}{lll}
\dfrac{dt}{dt'} & = & \exp \left[ -\gamma \Phi \right] \\
p_i & = & \exp \left[ \Phi \right] \pi_i, \\
\label{eq:two_twentysix}
\end{array}
\end{equation}
This Hamiltonian formulation enables the study of GIK thermostatted dynamics within the framework of Hamiltonian mechanics, including conservation of phase-space volume and its symplectic structure. The GIK thermostatted dynamics is a part of the $\mu-$ thermostat family, under the special condition of $\mu = 1$ \cite{bright2005new}. The GIK thermostat is unique amongst the different $\mu-$ thermostat candidates, in the sense that --(i) it is the only one from the family for which the conjugate pairing rule holds, and (ii) it is the only one which generates an equilibrium state. 

For non-equilibrium problems, GIK thermostat has been extensively used for studying three-dimensional Couette flow \cite{evans1986shear,morriss1988lyapunov,evans1993equivalence}, as the peculiar kinetic energy can be made a constant of motion:
\begin{equation}
\begin{array}{rcl}
    \dot{\mathbf{r}}_i & = & \dfrac{\mathbf{p}_i}{m} + n_x \phi y_i \\
    \dot{\mathbf{p}}_i & = & \mathbf{F}_i - n_x \phi p_{y,i} - \lambda \mathbf{p}_i \\
    \lambda & = & \dfrac{\sum\limits_{i=1}^N{\mathbf{F}_i . \mathbf{p}_i - \phi p_{x,i}p_{y,i}} }{\sum\limits_{i=1}^N \mathbf{p}_i . \mathbf{p}_i}
\end{array}
\end{equation}
Here, $\mathbf{r} \equiv (x,y,z)$ and $\mathbf{p} \equiv (p_x,p_y,p_z)$ denote the position and momentum of the particles, respectively, $n_x$ is a unit vector in the $x-$direction and $\phi$ is the shear rate. Evidently, Couette flow is treated as a mechanical perturbation by writing the equations of motion in terms of peculiar momenta \cite{morriss1987application} within the framework of the GIK thermostat. However, the correctness of such enforcement has been questioned \cite{delhommelle2001configurational}. 

\subsubsection{Solving the Equations of Motion}
Considering the popularity and the broad applicability of the GIK thermostat, symplectic techniques have been proposed in the literature for solving the equations of motion. Note that methods such as Runge-Kutta and predictor-corrector are equally applicable, but both of them being non-symplectic are omitted from the discussion here. A symplectic algorithm using Lie-Trotter factorization is now presented for solving the Couette flow problem \cite{zhang1999kinetic}:
\begin{enumerate}
    \item Initialize the position $x_i$ and momentum $p_i$ of the particles as per the desired temperature $ T_0 $. Compute the target kinetic energy as $ K_0 $ using $T_0$, which remains constant throughout the simulation
    \item Run in a loop until the desired time, $t$, is reached with an incremental time step of $\Delta t$:
    \begin{enumerate}
	\item Propagate the position of $i^{th}$ particle by $\Delta t/2$: $ r_{j i} \to r_{j i} + (p_{j i} + \phi r_{y i}\delta_{j x})\frac{\Delta t}{2} +\phi p_{y i}\delta_{j x} (\frac{\Delta t}{2})^2$ where $\delta$ is the Kronecker-Delta function and $j=x,y,z$.
	\item Set the parameters $ \lambda^{old}$ and tolerance to appropriate values.
	\item Store the current momentum into a separate variable as $p_{j i}^{old} = p_{j i}; j=x,y,z$
    \item Calculate:
    \begin{equation}
        \text{num} =  \sum_{i} \left[ \dfrac{\sum\limits_{j=x,y,z} (F_{j i}(p_{j i}+p_{j i}^{old}))}{2}- \dfrac{\phi\prod\limits_{j=x,y} (p_{j i}+p_{j i}^{old})}{4} \right]
    \end{equation}  where the summation $i$ is performed over all particles.
    \item Calculate:
    \begin{equation}
	    \text{den} = \sum_{i} \left[ \dfrac{\sum\limits_{j=x,y,z} \phi(p_{j i}+p_{j i}^{old})^{2}}{4} \right]
    \end{equation}  where the summation $i$ is performed over all particles.
    \item Calculate $ \lambda = \frac{\text{num}}{\text{den}} $, $a_1 = 1.0/(1.0 + 0.5 \lambda \Delta t)$ and $ a_2 = 1.0 - 0.5 \lambda \Delta t$.
    \item Propagate the momentum of the $i^{th}$ particle by $\Delta t$: $ p_{j i} \to a_1 \left[ \Delta t(F_{j i}-0.5\phi(p_{y i}+p_{y i}^{old})\delta_{j x})  + a_2p_{j i}^{old} \right] $.
    \item Do a tolerance check: if $abs(\lambda - \lambda^{old})<$tolerance then proceed. Otherwise assign $\lambda^{old} = \lambda$ and repeat steps (e) to (p) 
	\item Propagate the position of $i^{th}$ particle by $\Delta t/2$: $ r_{j i} \to r_{j i} + (p_{j i} + \phi r_{y i}\delta_{j x})\frac{\Delta t}{2} +\phi p_{y i}\delta_{j x} (\frac{\Delta t}{2})^2$.
    \end{enumerate}
\end{enumerate}
Solution for equilibrium cases may be obtained by substituting $\phi = 0$ and keeping only the relevant steps.

\subsubsection{Phase-Space Characteristics}
Let us now look at the phase-space characteristics of the GIK thermostat using a single harmonic oscillator of unit mass and stiffness. If the desired temperature is such that $k_BT_0=1$, the GIK equations of motion become:
\begin{equation}
\begin{array}{rcl}
\dot{x} = p, & \dot{p} = -x - \lambda p, & \lambda = \dfrac{-xp}{p^2} \\
\label{eq:sho_gik}
\end{array}
\end{equation}
which can further be simplified to:
\begin{equation}
\begin{array}{rcl}
\dot{x} = p, & \dot{p} = 0, & \lambda = \dfrac{-x}{p} \\
\label{eq:sho_gik2}
\end{array}
\end{equation}
As is evident, for a single harmonic oscillator the GIK thermostat does not generate a canonical ensemble. In fact, since $\dot{v} = 0$, the position of the oscillator keeps increasing, resulting in an unphysical situation. An isokinetic distribution is generated for a system with \textit{at least} two particles, and for a larger system, a single GIK trajectory accurately samples from an isokinetic distribution \cite{evans_morriss_gik}. Note that the canonical distribution is different from an isokinetic distribution. Hence, the phase-space sampled by the GIK thermostat is not in accordance with a canonical ensemble. The situation changes in the thermodynamic limit, where it has been shown that all ensembles are equivalent. 

\begin{figure}
\centering \includegraphics[scale=0.45]{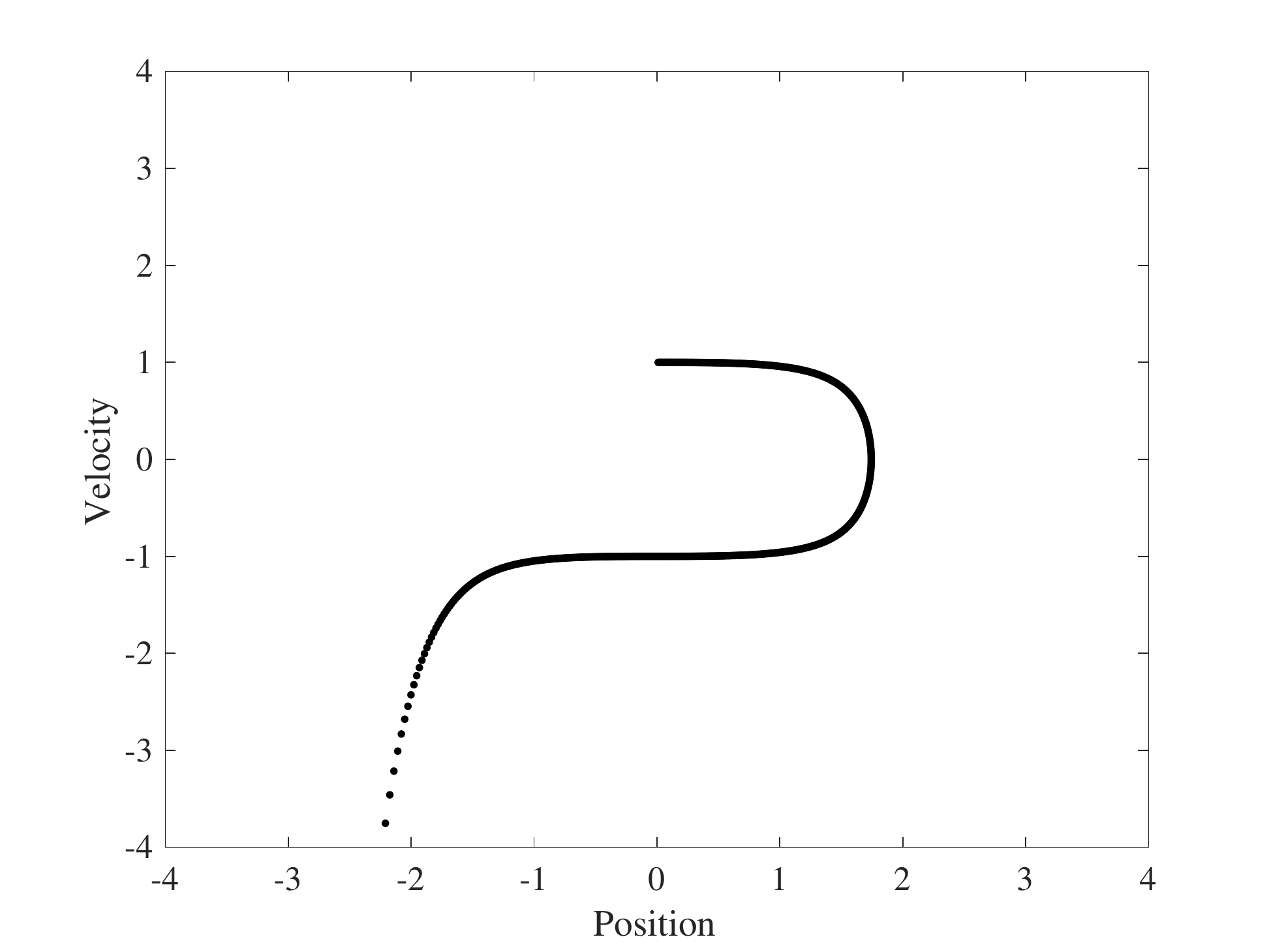}
\caption[GIK-NEQ]{\label{fig:gik_neq} Instability in dynamics of GIK thermostatted oscillator when subjected to a position-dependent temperature field. The initial conditions are: $(x,p) = (0,1)$. Because of the large feedback from the GIK multiplier $\lambda$, the dynamics becomes unstable.}
\end{figure}

We now bring the oscillator away from equilibrium by subjecting it to a position-dependent temperature field (see equation (\ref{eq:eq8}) with $\epsilon = 0.1$). The corresponding equations of motion become:
\begin{equation}
\begin{array}{rcl}
\dot{x} = p, & \dot{p} = -x - \lambda p, & \lambda = \dfrac{-xp}{1+0.1\tanh(x)} \\
\label{eq:sho_gik_neq}
\end{array}
\end{equation}
The equations of motion may be solved using the $4^{th}$ order Runge-Kutta with $\Delta t = 0.001$. If the oscillator is initialized at $(x,p) = (0,1)$, the phase-space plot, shown in figure (\ref{fig:gik_neq}), suggests that an instability occurs in the dynamics, primarily because of the large feedback from the Lagrangian multiplier $\lambda$. 
Fortunately, such instabilities do not occur for larger systems. 

Amongst the different properties of a good thermostat, GIK satisfies time reversibility, conforms with the different laws of thermodynamics, and is easy to implement as well. However, because of issues related to ergodicity, its inability to sample true canonical distribution in momentum variables, and problematic properties for small-sized systems, the use of GIK thermostat has become limited.

\subsection{Nos\'{e} Thermostat}
Nos\'e did the pioneering work on a Hamiltonian based thermostat \cite{nose,nose_thermostat}, which at the time was a significant breakthrough. Using his framework, one can now relate the rich Hamiltonian mechanics with the statistical-mechanical concepts of canonical distribution. The approach, which is now known as the extended-system method, has resulted in several different thermostats. The strength of Nos\'e's approach lies in treating the entire effect of a heat reservoir through a single variable \cite{nose_review}. The heat reservoir coupled with the system, therefore, constitutes a micro-canonical ensemble and has $6N+1$ degrees of freedom in total. 

Consider the extended system described by the variables $(x_i,p_i,s,p_s)$, where $x_i$ and $p_i$ have the usual meaning, $s$ is a time-scaling variable that models the reservoir variable, whose conjugate momentum is $p_s$. Similar to the GIK thermostat, let the Hamiltonian time, $t^\prime$, be different from the real-time, $t$; the relation between them given by $dt = s^{-1}dt'$. Likewise, let the Hamiltonian positions and the conjugate momenta, denoted by $x_i^\prime$ and $\pi_i$, respectively, be different from the real-time positions and momenta; their relation being:
\begin{equation}
\begin{array}{ccc}
x_i= x_i^\prime &, & p_i = \dfrac{\pi_i}{s}. \\
\label{eq:two_twentyseven}
\end{array}
\end{equation}
The real momenta, $p_i = mdx_i/dt$, may be written in terms of the Hamiltonian variables as:
\begin{equation}
\dfrac{dx_i}{dt}= \dfrac{dx_i^\prime}{dt} = s\dfrac{dx_i^\prime}{dt^\prime}.
\label{eq:two_twentyeight}
\end{equation}
Note that $\pi_i$ represents the momentum conjugate to $x_i^\prime$, i.e., $\pi_i = \dfrac{\partial L_\text{Nos\'e} }{ \partial \dot{x^\prime}_i}$, and is not necessarily equal to $m\dfrac{dx_i^\prime}{dt^\prime}$. Here, $L_\text{Nos\'e}$ denotes the Lagrangian of the extended system obtained by coupling the system to the Nos\'e thermostat:
\begin{equation}
L_{\text{Nos\'e}} = \sum\limits_{i=1}^{3N} \dfrac{m}{2}s^2 \dot{x^\prime}_i^2 - \Phi(\mathbf{x'}) + \dfrac{Q_s}{2} \dot{s}^2 - gk_BT_0 \log s.
\label{eq:two_twentynine}
\end{equation}
The overdot notation represents the derivatives with respect to the Hamiltonian time, $t^\prime$. The term $Q_s$ is a user-controlled parameter, which typically denotes the ``mass'' of the thermal reservoir, and $g$ is a constant that denotes the number of degrees of freedom within the system. From the Lagrangian, the momenta conjugate to the $x'$ and $s$ can be obtained as:
\begin{equation}
\begin{array}{ccc}
\pi_i = \dfrac{\partial L_{\text{Nos\'e}}}{\partial \dot{x'}_i} = ms^2\dot{x'}_i, & & p_s = \dfrac{\partial L_{\text{Nos\'e}}}{\partial \dot{s}} = Q_s \dot{s}.
\end{array}
\label{eq:two_thirty}
\end{equation}
Therefore, the Hamiltonian becomes:
\begin{equation}
H_{\text{Nos\'e}} = \sum\limits_{i=1}^{3N}\dfrac{{\pi}_i^2}{2ms^2} + \Phi(\mathbf{x}') + \dfrac{p_s^2}{2Q} + gk_BT_0\log(s),
\label{eq:two_thirtyone}
\end{equation}
from which the equations of motion can be derived:
\begin{equation}
\begin{array}{lllll}
\dfrac{dx'_i}{dt'} & = & \dfrac{\partial H_{\text{Nos\'e}}}{\partial \pi_i} & = & \dfrac{\pi_i}{ms^2} \\
\dfrac{d\pi_i}{dt'} & = & -\dfrac{\partial H_{\text{Nos\'e}}}{\partial x'_i} & = & -\dfrac{\partial \Phi}{\partial x'_i} \\
\dfrac{ds}{dt'} & = & \dfrac{\partial H_{\text{Nos\'e}}}{\partial p_s} & = & \dfrac{p_s}{Q_s} \\
\dfrac{dp_s}{dt'} & = & - \dfrac{\partial H_{\text{Nos\'e}}}{\partial s} & = & \dfrac{1}{s} \left[ \sum\limits_{i=1}^{3N} \dfrac{\pi_i^2}{ms^2} - gk_BT_0\right]
\end{array}
\label{eq:two_thirtytwo}
\end{equation}
This extended system constitutes a micro-canonical ensemble as it remains isolated from the environment, and all energy exchanges between the system and the reservoir are internal. Consequently, the Hamiltonian represented by equation (\ref{eq:two_thirtyone}) is a constant of motion. Under the assumption of ergodicity, the real phase-space is sampled as per the Maxwell-Boltzmann distribution, equation (\ref{eq:two_nine}), in both real as well as Hamiltonian time. Interested readers are referred to the review paper by H\"unenberger \cite{hunenberger} for a comprehensive treatment of the derivation. 

\subsubsection{Phase-Space Characteristics using a Single Harmonic Oscillators}
\begin{figure*}[htbp]
\includegraphics[scale = 0.3]{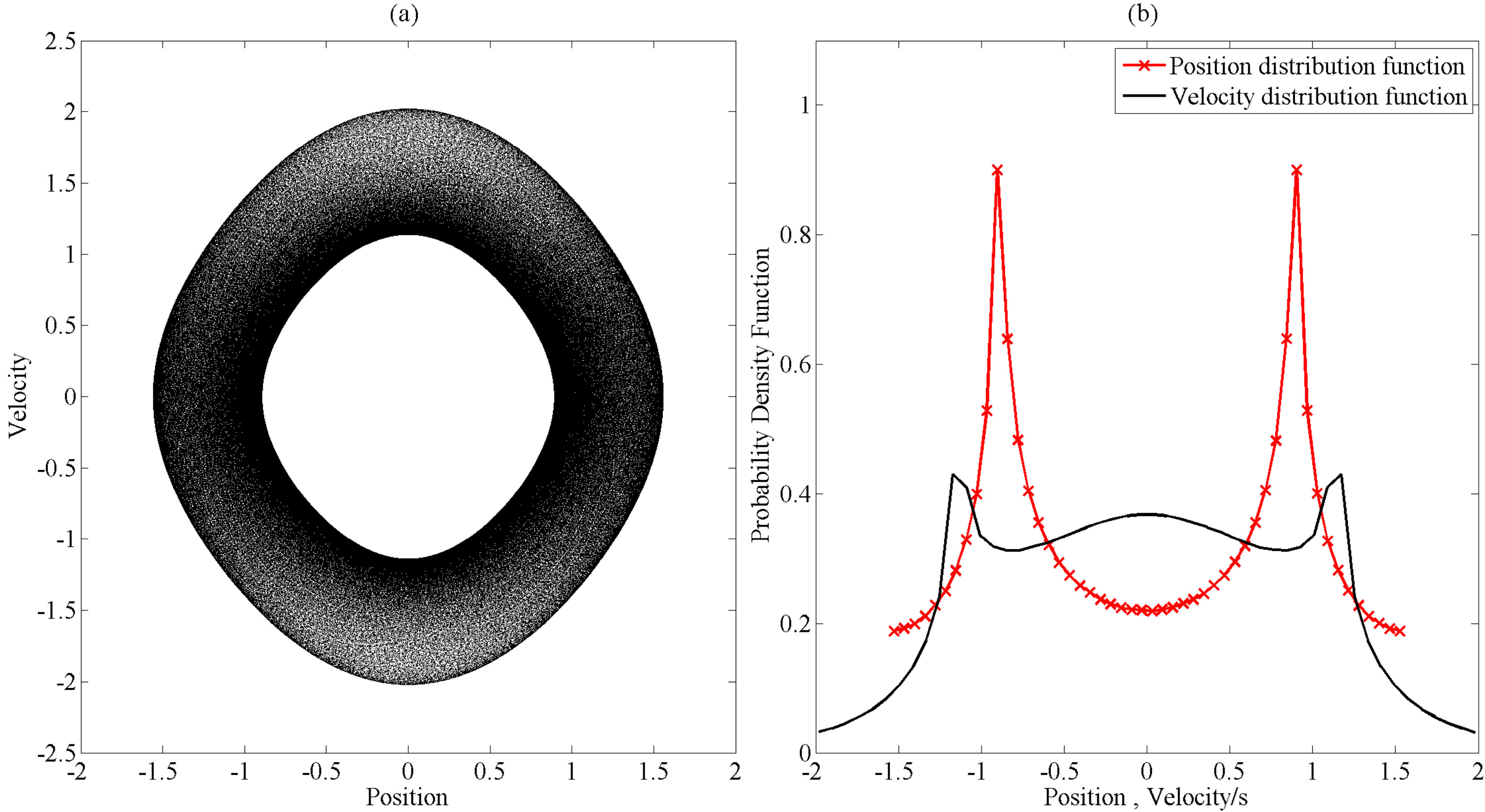}
\caption[non-ergodicity of Nos\'{e} dynamics]{\label{fig:nose_pos_vel_dist} Non-ergodicity of Nos\'{e} dynamics for the initial condition $(r',p',s,p_s) = (1,1,1,0)$ - (a) position-velocity plot of the oscillator and (b) probability distribution functions of the position and velocity/s. Notice the hole present in the dynamics in (a) and non-canonical nature of the distributions in (b).}
\end{figure*}

The Nos\'{e} thermostat, with $Q_s = 1$, when coupled to the single harmonic oscillator of unit mass, is governed by the equations:
\begin{equation}
\begin{array}{cccc}
\dot{x'} = \pi/s^2, \  & \dot{\pi} = -x', \  & \dot{s} = p_s, \  & \dot{p}_s = \dfrac{1}{s} \left(\dfrac{{\pi}^2}{s^2} - k_BT_0 \right).
\end{array}
\label{eq:two_thirtythree}
\end{equation}
Nos\'e dynamics represented by (\ref{eq:two_thirtythree}) is non-ergodic \cite{hoover_86}, and does not sample phase space according to the canonical distribution. For example, with $k_BT = 1$ and initial conditions as $(x',\pi,s,p_s) = (1,1,1,0)$, the phase space plot (see Figure \ref{fig:nose_pos_vel_dist} (a)) indicates that the dynamics is limited to a torus and the distribution is not Gaussian (Figure \ref{fig:nose_pos_vel_dist} (b)). 

The Nos\'e thermostat meets several properties of a good thermostat -- it is time-reversible, satisfies the different laws of thermodynamics, and has a Hamiltonian basis. However, it is neither ergodic nor easy to implement. As is evident from the equations of motion, in Nos\'e dynamics, phase-space sampling does not occur in real-time \cite{hunenberger}. Rather, it occurs in Hamiltonian time. Thus, when the thermostat is coupled to a non-Hamiltonian system (such that (\ref{eq:two_thirtyone}) does not exist), appropriate time-sampling becomes an issue. Further, due to sampling in Hamiltonian time, the trajectories in the real-time domain are available at non-uniform intervals. Any meaningful result obtained by time-averaging a variable, therefore, requires appropriately re-weighting the trajectories, which is not a trivial task. A simple remedy to these issues is to use Hoover's modification of Nos\'{e}'s dynamics - the Nos\'{e}-Hoover dynamics \cite{nose_hoover}. 

\subsection{Nos\'{e}-Hoover Thermostat}
Nos\'e's equations of motion can be written in the real-time domain using the transformations (\ref{eq:two_twentyseven}) and (\ref{eq:two_twentyeight}):
\begin{equation}
\begin{array}{cclcrrl}
\dfrac{dx_i}{dt} & = & \dfrac{p_i}{m}, & \dfrac{dp_i}{dt} & = & -\dfrac{\partial \Phi}{\partial x_i} - \left(\dfrac{1}{s} \dfrac{ds}{dt} \right)p_i, \\
\dfrac{ds}{dt} & = & \dfrac{sp_s}{Q_s}, & \dfrac{dp_s}{dt} & = & \left[ \sum\limits_{i=1}^{3N} \dfrac{{p}_i^2}{m} - 3Nk_BT_0 \right].
\end{array}
\label{eq:two_thirtyfour}
\end{equation}
Hoover introduced the variable $\eta = p_s/Q_s$, and rewrote the equations in terms of $\eta$ to obtain:
\begin{equation}
\begin{array}{cclclll}
\dfrac{dx_i}{dt} & = & \dfrac{p_i}{m}, & \dfrac{dp_i}{dt} & = & -\dfrac{\partial \Phi}{\partial x_i} - \eta p_i, \\
\dfrac{ds}{dt} & = & \eta s, & \dfrac{d\eta}{dt} & = & \dfrac{1}{Q_\eta}\left[ \sum\limits_{i=1}^{3N} \dfrac{{p}_i^2}{m} - 3Nk_BT_0 \right].
\end{array}
\label{eq:two_thirtyfive}
\end{equation}
For consistency, $Q_s$ has been replaced by $Q_\eta$. It is easy to see that the differential equation for $s$ is redundant, and may be omitted from further analysis. The remaining three differential equations that describe the extended-system are self-sufficient in defining the dynamics. These three equations are now known as the Nos\'{e}-Hoover equations of motion. 

The Nos\'{e}-Hoover equations of motion satisfy the steady-state Liouville's equation \cite{hoover_computational_stat_mech} in a comoving frame of reference:
\begin{equation}
\dfrac{\partial f_{\text{ex}}}{\partial t} + \dfrac{\partial f_{\text{ex}}}{\partial \mathbf{\Gamma}} . \mathbf{\dot{\Gamma}} = -f_{\text{ex}} \dfrac{\partial \dot{\mathbf{\Gamma}}}{\partial \mathbf{\Gamma}},
\label{eq:two_thirtysix}
\end{equation}
where, $\mathbf{\Gamma} = (\mathbf{x,p},\eta)$ and $f_{ex}$ denotes the extended phase-space distribution function given by:
\begin{equation}
f_\text{ex}(\mathbf{\Gamma}) = \dfrac{1}{Z^\prime} e^{\left[ -\beta \left( \sum\limits_{i=1}^{3N} \dfrac{{p}_i^2}{2m} + \Phi(\mathbf{x}) \right) \right]} \times e^{\left[ - \dfrac{\beta Q_\eta}{2}\eta^2 \right]}
\label{eq:two_thirtyseven}
\end{equation}

The extended system, on the virtue of being in thermodynamic equilibrium, follows the canonical distribution in $\eta$ as well. The term $ \dfrac{\partial \mathbf{\dot{\Gamma}}}{\partial \mathbf{\Gamma}} \equiv \left[ \dfrac{\partial \dot{\mathbf{x}}}{\partial \mathbf{x}} + \dfrac{\partial \dot{\mathbf{p}}}{\partial \mathbf{p}} + \dfrac{\partial \dot{\eta}}{\partial \eta} \right]$ of equation (\ref{eq:two_thirtysix}) denotes the rate of phase-space contraction and is usually denoted by $\Lambda$. The Nos\'e-Hoover equations of motion bear similarity to the GIK thermostat, the only difference being the dynamic evolution of $\eta$. 

An easier approach to derive the Nos\'e-Hoover equations, known as the \textit{guessing method}, makes use of the extended phase-space distribution and the Liouville's equation:
\begin{itemize}
    \item Begin with the extended phase-space distribution shown in equation (\ref{eq:two_thirtyseven}).
    \item Assume the evolution of $x_i$ and $p_i$ as: $\dfrac{dx_i}{dt} = \dfrac{p_i}{m}$ and $\dfrac{dp_i}{dt} = -\dfrac{\partial \Phi}{\partial x_i} - \eta p_i$, respectively. The evolution of $p_i$ is chosen in this manner to keep parity with the -- (i) Langevin equation which stochastically controls  $T_k$, and (ii) the GIK thermostat.
    \item Solve the steady-state Liouville's equation to obtain the temporal evolution of the variable $\eta$ under the assumption that $\eta$ and $\dot{\eta}$ are independent of each other.
\end{itemize}
The Nos\'e-Hoover equations of motion may directly be obtained from a Hamiltonian described by Dettmann and Morriss \cite{dettmann1997hamiltonian}:
\begin{equation}
    H_\text{D\&M} \equiv s H_\text{Nos\'e} \equiv 0
\label{eq:Hamiltonian_Dettmann}
\end{equation}
This Hamiltonian omits the problematic time-scaling used in Nos\'e's Hamiltonian. Bond and coworkers developed a more formal approach, the Nos\'e-Poincar\'e method, to come up with the same Hamiltonian \cite{nose_poincare}. Note that for some non-equilibrium problems of one-dimensional oscillators, the traditional Nos\'e-Hoover equations differ \cite{posch1997time} from those obtained using equation (\ref{eq:Hamiltonian_Dettmann}).

The pseudo energy, defined by:
\begin{equation}
    E_{NH} = \sum\limits_{i=1}^{3N} \dfrac{p_i^2}{2m} + \Phi(\mathbf{x}) + \dfrac{Q_\eta \eta^2}{2} + \int\limits_0^t \eta 3N k_B T_0 dt,
    \label{eq:NH_PsE}
\end{equation}
is a constant of motion for all $t$, and is different from the energy of the system. While $\dot{E}_{NH} = 0$, the rate at which the energy of the system changes is given by:
\begin{equation}
    \begin{array}{rcl}
         E & = & \sum \dfrac{p_i^2}{2m} + \Phi(\mathbf{x})  \\
    \implies    \dfrac{dE}{dt}     &  = & \sum\left[ \dfrac{p_i}{m} \dot{p}_i + \dfrac{\partial \Phi }{\partial x_i} \dot{x}_i  \right]\\
    & = & \sum\left[ \dfrac{p_i}{m} \left( -\dfrac{\partial \Phi}{\partial x_i} - \eta p_i \right)  + \dfrac{\partial \Phi }{\partial x_i} \dfrac{p_i}{m} \right] \\
    & = & -\sum \dfrac{\eta p_i^2}{m}.
    \end{array}
    \label{eq:NH_dE}
\end{equation}
 By definition, if no work is done on/by the system, it only exchanges heat energy with the Nos\'e-Hoover thermostat. Consequently, applying the First Law of thermodynamics provides the rate of heat flow ($\dot{Q}$): $\dot{E} = \dot{Q} \implies \dot{Q} = -\sum {\eta p_i^2}/{m}$. Under steady-state conditions: $\dfrac{d}{dt} \left\langle \dfrac{Q_\eta \eta^2}{2} \right\rangle = 0 \implies \left\langle \sum {\eta p_i^2}/{m} \right\rangle = 3N \langle \eta \rangle k_B T_0$, so that one can write:
\begin{equation}
    \begin{array}{rcl}
         \langle \dot{Q} \rangle & = & - \left\langle \sum \dfrac{\eta p_i^2}{m} \right\rangle = -3N\langle \eta \rangle k_BT_0
    \end{array}
    \label{eq:NH_Q}
\end{equation}
This simple equation has a lot of information embedded in it. In equilibrium, $\langle \dot{Q} \rangle = 0 \implies \langle \eta \rangle = 0$. However, instantaneously, $\dot{Q} \neq 0$ even in equilibrium (see fluctuation theorem \cite{searles2001fluctuation}), which suggests that $\eta(t) \neq 0 \forall t$. The heat-flow entropy rate, defined as: $\langle \dot{S} \rangle = -\langle \dot{Q} \rangle / T_0$ signifies that in non-equilibrium $\langle \eta \rangle > 0$. 

For Nos\'e-Hoover dynamics, the phase-space compression factor may be written as: $\Lambda = -3N \eta$. In equilibrium, since $\langle \eta \rangle = 0$, in an averaged sense, the phase-space neither contracts nor expands. However, in non-equilibrium, the phase-space volume continuously shrinks, and collapses to a dimension smaller than the embedding dimension. Note that $\langle \Lambda \rangle$ is closely related to the sum of Lyapunov exponents: $\langle \Lambda \rangle = \sum L_i$. Intuitively, this relation makes sense as the sum of Lyapunov exponents provides the rate at which the phase-space volume changes. Evidently, $\langle \dot{S} \rangle/k = -\langle \Lambda \rangle$, which makes it possible to relate the heat-flow entropy with $\sum L_i$. 

For non-equilibrium problems, this equivalence of the Gibbs' heat flow entropy with the phase-space compression factor, and $\sum L_i$, thereof, occurs under the special condition that the temperature field, $T_0$, is constant. If, however, one has a position-dependent temperature field such as the one shown in equation (\ref{eq:eq8}), this equivalence is violated. The Gibbs' heat flow entropy, in this case, is given by:
\begin{equation}
\begin{array}{rcl}
    \left\langle \dfrac{\dot{S}}{k_B} \right\rangle & = & \left\langle \dfrac{-\dot{Q}}{k_BT} \right\rangle =  \left\langle \dfrac{\sum \eta p_i^2}{mT} \right\rangle \\
    & \neq & \dfrac{\left\langle \sum \eta p_i^2/m \right\rangle}{\left\langle k_B T\right\rangle} = 3N \langle \eta \rangle = \langle -\Lambda \rangle
\end{array}
\label{eq:non-eq-entropy-lambda}
\end{equation}
However, a slight modification in the equations of motion (\ref{eq:two_thirtyfive}) can resolve this nuisance:
\begin{equation}
\begin{array}{cclclll}
\dfrac{dx_i}{dt} & = & \dfrac{p_i}{m}, & \dfrac{dp_i}{dt} = -\dfrac{\partial \Phi}{\partial x_i} - \eta p_i, \dfrac{d\eta}{dt} & = & \dfrac{1}{Q_\eta}\left[ \dfrac{\sum {p}_i^2/m}{3N k_BT} - 1 \right].
\end{array}
\label{eq:two_thirtyfiveone}
\end{equation}
Imposing steady-state conditions: $d/dt \langle (Q_\eta \eta^2/2) \rangle= 0 \implies \left\langle \dfrac{\sum \eta p_i^2 / m}{k_B T} \right\rangle = \langle 3N \eta \rangle$ on the time-averaged rate of heat-flow entropy we get: $\left\langle \dfrac{\dot{S}}{k_B} \right\rangle = \left\langle \dfrac{\sum \eta p_i^2 / m }{k_BT}\right\rangle = \langle 3N \eta \rangle = \langle -\Lambda \rangle$. The remarkable ability of the Nos\'e-Hoover thermostat to link the dynamical variables with their thermodynamic counterparts has made it very popular amongst researchers. 

Nos\'e-Hoover thermostat has spurred the development of several thermostat algorithms, each with its own merit. Watanabe and Kobayashi \cite{watanabe2007ergodicity} relaxed the assumption of $\partial \dot{\eta} / \partial \eta = 0$, and generalized the Nos\'e-Hoover equations. Working with Jellinek and Berry's generalization of Nos\'e Hamiltonian \cite{jellinek_berry}, which results in a more efficient mixing of phase-space trajectories, Bra\'nka and Wojciechowskie generalized the Nos\'e-Hoover dynamics \cite{branka2000generalization} to obtain improved thermalizing characteristics in a single harmonic oscillator. Using Hoover's guessing method, Bravetti and Tapias \cite{bravetti2016thermostat} developed equations of motion that generate any target density distribution. 

\subsubsection{Solving the Equations of Motion}
For a large system, one can use Gear's predictor-corrector method \cite{sadus2002molecular} for solving the Nos\'e-Hoover equations of motion (\ref{eq:two_thirtyfive}). In the predictor-corrector algorithm, the variables are predicted based on Taylor's series expansion of each variable. These variables are then corrected with respect to a higher-order derivative of acceleration. Since such a correction is, to some extent, ad hoc, the time-reversibility of the equations of motion is lost along with the symplectic property of the dynamics. While for a small system, the $4^{th}$ order Runge-Kutta method may be used, the method is time-consuming as well as not symplectic, and there is a long term energy drift. Tuckerman and coworkers have developed a symplectic algorithm for Nos\'e-Hoover dynamics based on Trotter's factorization and Liouville's operators. This algorithm is not only symplectic, but stable as well. A possible implementation of the algorithm with $m  = 1$ is described below:
\begin{enumerate}
    \item Initialize the position and the momentum of each particle, and set $\eta = 0$. Set $k_B = 1$.
    \item Run in a loop until the desired time, $t$, is reached with an incremental time step of $\Delta t$:
    \begin{enumerate}
        \item Compute $K$, the kinetic energy of the system.
        \item Compute $G = (2K - 3N T_0)/Q_\eta$.
        \item Propagate $\eta$ by $\Delta t /4$: $\eta \to \eta + G \Delta t/4$.
        \item Define $\zeta$, the scale parameter: $\zeta=\exp(-\eta \Delta t/2.0)$.
        \item Scale all momentum: $p_i \to p_i \times \zeta$.
        \item Compute scaled $K$: $K \to K \zeta^2$.
        \item Compute $G = (2K - 3N T_0)/Q_\eta$.
        \item Propagate $\eta$ by $\Delta t /4$: $\eta \to \eta + G \Delta t/4$.
        \item Propagate all momenta by $\Delta t/2$: $p_i \to p_i + F_i \Delta t / 2$.
        \item Propagate all positions by $\Delta t$: $x_i \to x_i + p_i \Delta t$.
        \item Propagate all momenta by $\Delta t/2$: $p_i \to p_i + F_i \Delta t / 2$.
        \item Compute $K$.
        \item Compute $G = (2K - 3N T_0)/Q_\eta$.
        \item Propagate $\eta$ by $\Delta t /4$: $\eta \to \eta + G \Delta t/4$.
        \item Define $\zeta$, the scale parameter: $\zeta=\exp(-\eta \Delta t/2.0)$.
        \item Scale all momentum: $p_i \to p_i \times \zeta$.
        \item Compute scaled $K$: $K \to K \zeta^2$.
        \item Compute $G = (2K - 3N T_0)/Q_\eta$.
        \item Propagate $\eta$ by $\Delta t /4$: $\eta \to \eta + G \Delta t/4$.
    \end{enumerate}
\end{enumerate}
If the equations are implemented correctly the pseudo energy defined by equation (\ref{eq:NH_PsE}) remains constant in time.

\subsubsection{Phase-space characteristics using a single harmonic oscillator}
The Nos\'e-Hoover thermostat satisfies almost all qualities of a ``good'' thermostat -- (i) it is time-reversible: if we reverse each momentum term such that $p_i \to -p_i$ and $\eta \to -\eta$, the path is traced back, (ii) it conforms with the laws of thermodynamics and allows heat to flow from a hotter thermostat to a colder thermostat, (iii) a Hamiltonian function is associated with it, and  (iii) it is easy to implement. 

\begin{figure}
\includegraphics[scale=0.45]{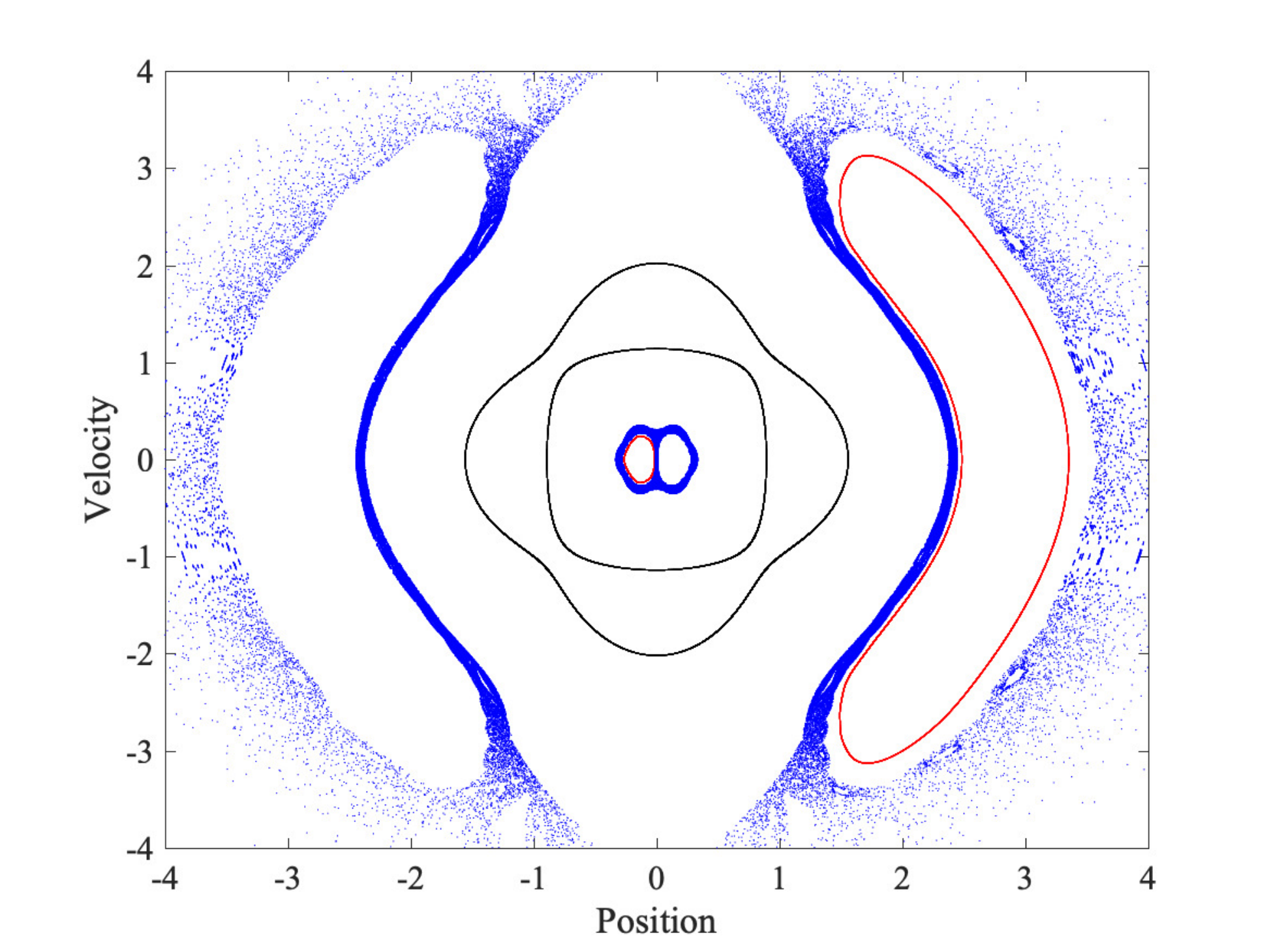}
\caption[Phase space plots for Nos\'{e}-Hoover dynamics]{\label{fig:NH_phase_plot} Poincar\'e section plots for the Nos\'e-Hoover dynamics at $\eta = 0$ cross-section for three different initial conditions: (a) black = $(x,p,\eta) = (1,1,0)$, (b) red = $(x,p,\eta) = (2,2,1)$, and (c) blue = $(x,p,\eta) = (3,3,3)$. The different initial conditions result in different nature of trajectories, with none being phase-space filling. The lack of ergodicity, and consequently the inability of NH thermostat to thermalize the single harmonic oscillator is self-evident. The equations of motion are solved using the classic $4^{\text{th}}$ order Runge-Kutta algorithm for 1 billion time steps, each of 0.001.}
\end{figure}

But, the dynamics is not ergodic \cite{hoover_86}. For this purpose, let us look at a single harmonic oscillator of unit mass and stiffness constant which is coupled to a Nos\'e-Hoover thermostat kept at $k_BT_0 = 1$ and has $Q_\eta = 1$. The equations of motion are:
\begin{equation}
\begin{array}{ccc}
\dot{x} = p,\  & \dot{p} = -x -\eta p,\  &  
\dot{\eta} = \left( p^2 - 1 \right). 
\end{array}
\label{eq:NH_SHO}
\end{equation}
The Poincar\'e section plot at $\eta = 0$ cross-section, for three different initial conditions, are shown in figure (\ref{fig:NH_phase_plot}): case (a) with $(x,p,\eta) = (1,1,0)$, case (b) with $(x,p,\eta) = (2,2,1)$, and case (c) $(x,p,\eta) = (3,3,3)$. For cases (a) and (b), two kneaded tori are obtained, while in case (c) the dynamics is chaotic. The plots clearly indicate that the phase-space can be easily partitioned into multiple non-communicating regions, and hence, the dynamics is not ergodic. Although there is chaoticity in the dynamics for different initial conditions, the chaotic space makes up for only 6\% of the entire phase-space \cite{patra2016equivalence}. In fact, none of the initial conditions sample the phase-space as per equation (\ref{eq:three_one}). The poor ergodic characteristics of the Nos\'e-Hoover thermostat may be explained by the periodic dynamics of the variable $\eta$ \cite{watanabe_07b}, and the presence of conserved quantities that cause the energy of the system to be bounded \cite{watanabe_07a}. 

\begin{figure*}[htp]
\includegraphics[width = 0.80\textwidth]{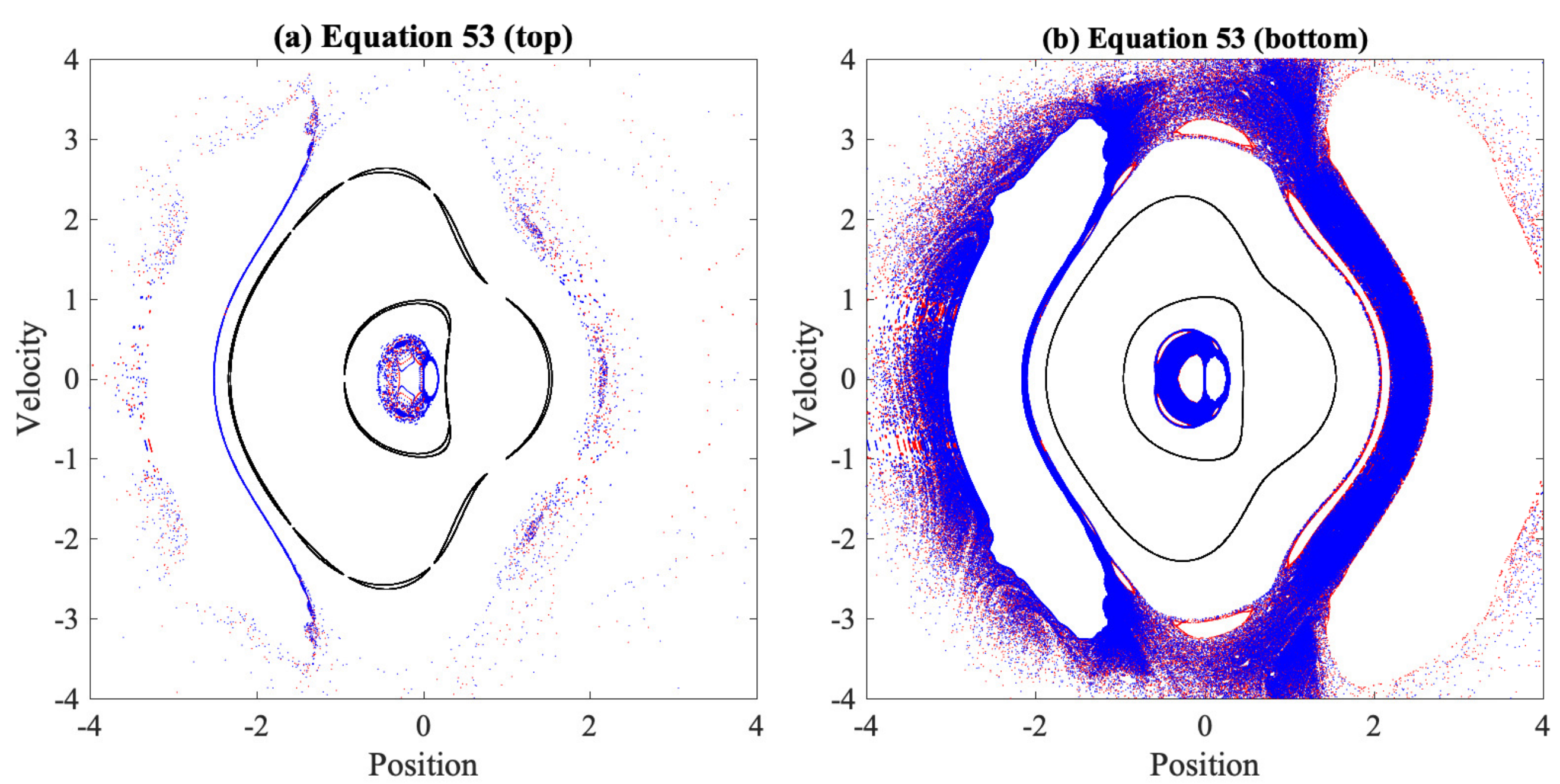}
\caption[Poincare cross-section NH Neq]{\label{fig:NH_Neq} Poincar\'e plot of Nos\'e-Hoover dynamics subjected to a position dependent temperature field: $T(x) = 1 + 0.30 \tanh(x)$ at the cross-section $\eta \in [-0.001,0.001]$ for three initial conditions: black = $(x,p,\eta) = (1,1,0)$, red = $(x,p,\eta) = (2,2,1)$, and blue = $(x,p,\eta) = (3,3,3)$. Figure (a) corresponds to the evolution where $\dot{\eta} = \left( p^2 -  T(x)\right)$, while figure (b) is for the evolution where $\dot{\eta} = \left( {p^2}/{T(x)} -  1\right)$.}
\end{figure*}

We now subject the Nos\'e-Hoover thermostatted oscillator to a position-dependent temperature field, as defined in equation (\ref{eq:eq8}).  The resulting equations of motion according to equations (\ref{eq:two_thirtyfive}) and (\ref{eq:two_thirtyfiveone}) are, respectively:
\begin{equation}
\begin{array}{ccc}
\dot{x} = p,\  & \dot{p} = -x -\eta p,\  &  
\dot{\eta} = \dfrac{1}{Q_\eta} \left( p^2 -  T(x)\right); \\
\dot{x} = p,\  & \dot{p} = -x -\eta p,\  &  
\dot{\eta} = \dfrac{1}{Q_\eta} \left( \dfrac{p^2}{T(x)} -  1\right). 
\end{array}
\label{eq:NH_SHO_NEQ}
\end{equation}
where, $T(x) = 1.0 + 0.30 \tanh(x)$
The three initial conditions used in the equilibrium case, have been studied here. When solved using the $4^{th}$ order Runge-Kutta method for 10 billion time steps using $\Delta t = 0.001$, the phase-space portrait at the Poincar\'e section defined by $\eta = 0$, is shown in figure (\ref{fig:NH_Neq}). The plots differ from that of equilibrium as well as with each other. 

\begin{table}[]
    \centering
    \begin{tabular}{c|c|c|c|c|c|c} \hline
         & $\langle \dot{Q}_T \rangle$ & $\langle \dot{Q}_B \rangle$ & $\langle \dot{S}_T \rangle$ & $\langle \dot{S}_B \rangle$ & $\langle \Lambda_T \rangle$ & $\langle \Lambda_B \rangle$  \\ \hline \hline
Set (a)        & 0.0000 & 0.0000 & 0.0000 & 0.0000 & 0.0000 & 0.0000 \\ \hline
Set (b)        & 0.0000 & 0.0000 & -0.0061 & -0.0002 & -0.0114 & -0.0002 \\ \hline
Set (c)        & 0.0000 & 0.0000 & -0.0061 & -0.0002 & -0.0113 & -0.0002 \\ \hline \hline
    \end{tabular}
    \caption{Different thermodynamic quantities for the three sets of initial conditions: black = $(x,p,\eta) = (1,1,0)$, red = $(x,p,\eta) = (2,2,1)$, and blue = $(x,p,\eta) = (3,3,3)$. $\langle \dot{Q} \rangle$ denotes the average rate of heat flow, $\langle \dot{S} \rangle$ is the average rate of heat-flow entropy and $\langle \Lambda \rangle$ is the average phase-space compression. Each quantity has a subscript: $_T$ denotes the equations of motion with $\dot{\eta} = \left( p^2 -  T(x)\right)$, while $_B$ is for the evolution with $\dot{\eta} = \left( {p^2}/{T(x)} -  1\right)$. Notice that while $\langle \dot{S}_T \rangle \neq \langle \Lambda_T \rangle$, $\langle \dot{S}_B \rangle = \langle \Lambda_B \rangle$, as has been discussed in the text.}
    \label{tab:my_label1}
\end{table}
The different thermodynamic quantities, for the two types of evolution and the three initial conditions, are summarized in the table \ref{tab:my_label1}. In stationary states, when work is absent, the total time-averaged heat flow must vanish, which implies $\langle \dot{Q} \rangle = 0$. The results, shown in table \ref{tab:my_label1}, indicate that $\langle \dot{Q} \rangle = 0$ for all cases. For Nos\'e-Hoover oscillator, some of the trajectories are conservative, wherein $\langle \dot{S} \rangle = \langle \Lambda \rangle = 0$, while the others are dissipative, for which $\langle \dot{S} \rangle <0$ and $\langle \Lambda \rangle < 0$. In table \ref{tab:my_label1} we can see both types of trajectories -- a conservative trajectory with a three-dimensional torus occurs for initial conditions $(x,p,\eta) = (1,1,0)$ while dissipative trajectories occur for the remaining two initial conditions. The co-existence of conservative and dissipative features \cite{sprott2014heat} is a unique feature of the Nos\'e-Hoover dynamics which occurs due to the non-ergodicity of the dynamics. That being said, Nos\'e-Hoover thermostat performs well for large equilibrium as well as non-equilibrium problems, having been verified experimentally as well \cite{yu2016simulation}. This is because the effect of non-ergodicity is superseded by (prohibitively) large Poincar\'e recurrence time for even a moderately large system. 

In general, it is thought that the issue of non-ergodicity of thermostatted dynamics of very small-scale systems can be tackled by using multi-variable thermostats \cite{sprott_hoover_14, hoover_kum, watanabe_07a, mkt,hh}. We now describe these multi-variable thermostats.

\subsection{Martyna-Klein-Tuckerman Thermostat}
One of the major breakthroughs in improving the ergodic characteristics of the Nos\'e-Hoover dynamics was the development of the Nos\'e-Hoover chain (MKT) method \cite{mkt,tobias_93}. In this approach, one controls the kinetic temperature of the system, along with those of the reservoir by means of additional variables. In other words, while the thermostat variable $\eta_1$ controls the kinetic temperature of the system similar to that in the Nos\'e-Hoover thermostat, the fluctuations of the variable $\eta_1$ are controlled by the second thermostat variable $\eta_2$. Likewise, the fluctuations of $\eta_2$ are controlled by the third thermostat variable, $\eta_3$, and so on. Thus, a chain of thermostats ($\eta_1, \eta_2, \ldots, \eta_k$) is formed. The MKT equations of motion for a system are: 
\begin{equation}
\begin{array}{rcl}
\dot{x_i} & = & \dfrac{p_i}{m}, \\
\dot{p_i} & = & \dfrac{\partial \Phi}{\partial x_i}-p_i\dfrac{\eta_1}{Q_{\eta_1}}, \\
\dot{\eta_1} & = & \left[\sum\limits_{i=1}^{3N}\dfrac{p_i^2}{m_i}-3NNk_BT_0\right]-\eta_1\dfrac{\eta_2}{Q_{\eta_2}}, \\
. & & \\
. & & \\
. & & \\
\dot{\eta_j} & = & \left[ \dfrac{\eta_{j-1}^2}{Q_{\eta_{j-1}}} - k_BT_0 \right] - \eta_j\dfrac{\eta_{j+1}}{Q_{\eta_{j+1}}} \\
. & & \\
. & & \\
. & & \\
\dot{\eta_k} & = & \left[ \dfrac{\eta_{k-1}^2}{Q_{\eta_{k-1}}} - k_BT_0 \right]. 
\end{array}
\label{eq:MKT_Full}
\end{equation}
The variables $Q_{\eta_i}$ may be thought of as the mass associated with the thermostat variable $\eta_i$. An empirical rule of selecting the thermostat masses is:  $Q_{\eta_1}=3Nk_BT_0/\omega^2$ and $Q_{\eta_{j\neq 1}}=k_BT_0/\omega^2$ \cite{mkt,tobias_93}. The frequency, $\omega$,  describes the frequency with which the kinetic energy oscillates between the system and the reservoirs. A lot of approximations have gone in developing this relationship, and a suitable choice is usually problem-dependent. We will see later how the dynamics changes substantially depending on the choice of $Q_{\eta_i}$, especially for small-scale systems. 

The dynamics due to the MKT thermostat satisfies the following extended phase-space distribution:
\begin{equation}
f_\text{ex}(\mathbf{\Gamma}) \propto \exp\left[-\beta_0 \left(H + \sum\limits_{i=1}^k \dfrac{\eta_i^2}{2Q_{\eta_i}} \right)\right],
\label{eq:measure_NHC}
\end{equation}
where, $H = \sum p_i^2/2m + \Phi(\mathbf{x})$ and $\mathbf{\Gamma} = (\mathbf{x,p},\eta_1,\eta_2,\ldots,\eta_k) $. Corresponding to this extended phase-space, the pseudo energy, which is a constant of motion, is given by: 
\begin{equation} 
E_{\text{MKT}} = \left(H + \sum\limits_{i=1}^k \dfrac{\eta_i^2}{2Q_{\eta_i}} \right) + k_BT_0 \int\limits_0^t \left[ 3N\dfrac{\eta_1}{Q_{\eta_1}} +\sum\limits_2^k\dfrac{\eta_i}{Q_{\eta_i}} \right] dt.
\label{eq:MKT_PsE}
\end{equation}
Like the Nos\'e-Hoover thermostat, the dynamical variables of an MKT thermostatted system may be linked with the thermodynamic quantities:
\begin{equation}
    \begin{array}{rcl}
        \dot{E} & = & \dot{Q} = \sum\limits_{i=1}^{3N} \left(\dfrac{\partial \Phi}{\partial x_i}\dot{x}_i + \dfrac{p_i}{m}\dot{p}_i\right) = \sum\limits_{i=1}^{3N} \dfrac{-\eta_1 p_i^2}{mQ_{\eta_1}}, \\
        \dot{S} & = & -\dfrac{\dot{Q}}{T_0} = \dfrac{-\eta_1}{mQ_{\eta_1}}\sum\limits_{i=1}^{3N} \dfrac{p_i^2}{T_0}, \\
        \Lambda & = & -3N\dfrac{\eta_1}{Q_{\eta_1}} - \sum\limits_{i=2}^{k}\dfrac{\eta_i}{Q_{\eta_i}}
    \end{array}
\label{eq:MKT_Thermo_Vars}
\end{equation}
Recalling that in steady-state:
\begin{equation}
    \left\langle \dfrac{d}{dt} \left( \sum\limits_{i=1^k} \dfrac{\eta_i^2}{2Q_{\eta_1}}\right) \right\rangle = 0,
\end{equation}
$\langle \dot{Q} \rangle$ and $\langle \dot{S} \rangle$ may be written as:
\begin{equation}
    \begin{array}{rcl}
        \langle \dot{Q} \rangle & = & -k_BT_0\left[ 3N\dfrac{\langle \eta_1 \rangle}{Q_{\eta_1}} +\sum\limits_2^k\dfrac{\langle \eta_i \rangle}{Q_{\eta_i}} \right], \\
        \left\langle \dfrac{\dot{S}}{k_B} \right\rangle & = & -\left\langle\dfrac{ \dot{Q}}{k_BT_0}\right\rangle = \left[ 3N\dfrac{\langle \eta_1 \rangle}{Q_{\eta_1}} +\sum\limits_2^k\dfrac{\langle \eta_i \rangle}{Q_{\eta_i}} \right] = -\langle \Lambda \rangle
    \end{array}
\label{eq:MKT_Thermo_Vars2}
\end{equation}
Note that like the Nos\'e-Hoover thermostat, the formulation of the MKT thermostat shown in equation (\ref{eq:MKT_Full}) satisfies equations (\ref{eq:MKT_Thermo_Vars2}) in equilibrium and non-equilibrium states where the desired temperature $T_0$ does not change. In cases where $T_0$ is position-dependent, the MKT equations require a modification similar to that shown in equation (\ref{eq:two_thirtyfiveone}). Again, like the Nos\'e-Hoover thermostat, $\langle \Lambda \rangle$ is zero in equilibrium and non-zero in non-equilibrium. In both the situations, $ \langle \Lambda \rangle$ can be related to $\sum L_i$. However, unlike the Nos\'e-Hoover thermostat, the MKT thermostat does not have any known Hamiltonian from which the equations of motion can be derived.

\subsubsection{Solving the Equations of Motion}
The most popular variant of the MKT thermostat is the two-chain variant, for which the equations of motion are:
\begin{equation}
\begin{array}{rcl}
\dot{x_i} & = & \dfrac{p_i}{m}, \\
\dot{p_i} & = & \dfrac{\partial \Phi}{\partial q_i}-p_i\dfrac{\eta}{Q_{\eta}}, \\
\dot{\eta} & = & \left[\sum\limits_{i=1}^{3N}\dfrac{p_i^2}{m_i}-3Nk_BT_0\right]-\dfrac{\eta\xi}{Q_{\xi}}, \\
\dot{\xi} & = & \left[ \dfrac{\eta^2}{Q_{\eta}} - k_BT_0 \right]. 
\end{array}
\label{eq:MKT_2_Variable}
\end{equation}
These equations of motion may be solved using Runge-Kutta technique for small-scale systems like a harmonic oscillator. However, for a large system, using Runge-Kutta is time-consuming. In view of this, Martyna and coworkers \cite{martyna1996explicit} developed a symplectic technique based on Lie-Trotter factorization and Liouville's operators. The algorithm is summarized below:
\begin{enumerate}
    \item Initialize the system with suitable initial conditions of positions and momenta, $\eta = 0$, and $ \xi = 0$.
    \item Run in a loop until the desired time, $t$, is reached with an incremental time step of $\Delta t$:
    \begin{enumerate}
        \item Compute $K$, the kinetic energy of the system.
        \item Compute $G_2 = (Q_{\eta} \eta^2 - T_0)/Q_{\xi}$.
        \item Propagate $\xi$ by $\Delta t /4$: $\xi \to \xi + G_2 \Delta t/4$.
        \item Propagate $\eta$ by $\Delta t /8$: $\eta \to \eta \times \exp(-\xi \Delta t/8 ) $.
        \item Compute $G_1 = (2K - 3N T_0)/Q_{\eta}$.
        \item Propagate $\eta$ by $\Delta t /4$: $\eta \to \eta + G_1 \Delta t/4$.
        \item Propagate $\eta$ by $\Delta t /8$: $\eta \to \eta \times \exp(-\xi \Delta t/8 ) $.
        \item Define $\zeta$, the scale parameter: $\zeta=\exp(-\eta \Delta t/2.0)$.
        \item Scale all momenta: $p_i \to p_i \times \zeta$.
        \item Compute scaled $K$: $K \to K \zeta^2$.
        \item Propagate $\eta$ by $\Delta t /8$: $\eta \to \eta \times \exp(-\xi \Delta t/8 ) $.
        \item Compute $G_1 = (2K - 3N T_0)/Q_{\eta}$.
        \item Propagate $\eta$ by $\Delta t /4$: $\eta \to \eta + G_1 \Delta t/4$.
        \item Propagate $\eta$ by $\Delta t /8$: $\eta \to \eta \times \exp(-\xi \Delta t/8 ) $.
        \item Compute $G_2 = (Q_{\eta} \eta^2 - T_0)/Q_{\xi}$.
        \item Propagate $\xi$ by $\Delta t /4$: $\xi \to \xi + G_2 \Delta t/4$.
        \item Propagate all momenta by $\Delta t/2$: $p_i \to p_i + F_i \Delta t / 2$.
        \item Propagate all positions by $\Delta t$: $x_i \to x_i + p_i \Delta t$.
        \item Propagate all momenta by $\Delta t/2$: $p_i \to p_i + F_i \Delta t / 2$.
        \item Compute $K$.
        \item Propagate $\xi$ by $\Delta t /4$: $\xi \to \xi + G_2 \Delta t/4$.
        \item Propagate $\eta$ by $\Delta t /8$: $\eta \to \eta \times \exp(-\xi \Delta t/8 ) $.
        \item Compute $G_1 = (2K - 3N T_0)/Q_{\eta}$.
        \item Propagate $\eta$ by $\Delta t /4$: $\eta \to \eta + G_1 \Delta t/4$.
        \item Propagate $\eta$ by $\Delta t /8$: $\eta \to \eta \times \exp(-\xi \Delta t/8 ) $.
        \item Define $\zeta$, the scale parameter: $\zeta=\exp(-\eta \Delta t/2.0)$.
        \item Scale all momenta: $p_i \to p_i \times \zeta$.
        \item Compute scaled $K$: $K \to K \zeta^2$.
        \item Propagate $\eta$ by $\Delta t /8$: $\eta \to \eta \times \exp(-\xi \Delta t/8  $.
        \item Compute $G_1 = (2K - 3N T_0)/Q_{\eta}$.
        \item Propagate $\eta$ by $\Delta t /4$: $\eta \to \eta + G_1 \Delta t/4$.
        \item Propagate $\eta$ by $\Delta t /8$: $\eta \to \eta \times \exp(-\xi \Delta t/8 ) $.
        \item Compute $G_2 = (Q_{\eta} \eta^2 - T_0)/Q_{\xi}$.
        \item Propagate $\xi$ by $\Delta t /4$: $\xi \to \xi + G_2 \Delta t/4$
    \end{enumerate}
\end{enumerate}
If the equations are implemented correctly the pseudo energy defined by equation (\ref{eq:MKT_PsE}) remains constant in time. For a chain with more than two thermostat variables, a similar algorithm can be developed. 

\subsubsection{Phase-space characteristics from single harmonic oscillator}
\begin{figure}[htp]
\includegraphics[width=0.5\textwidth]{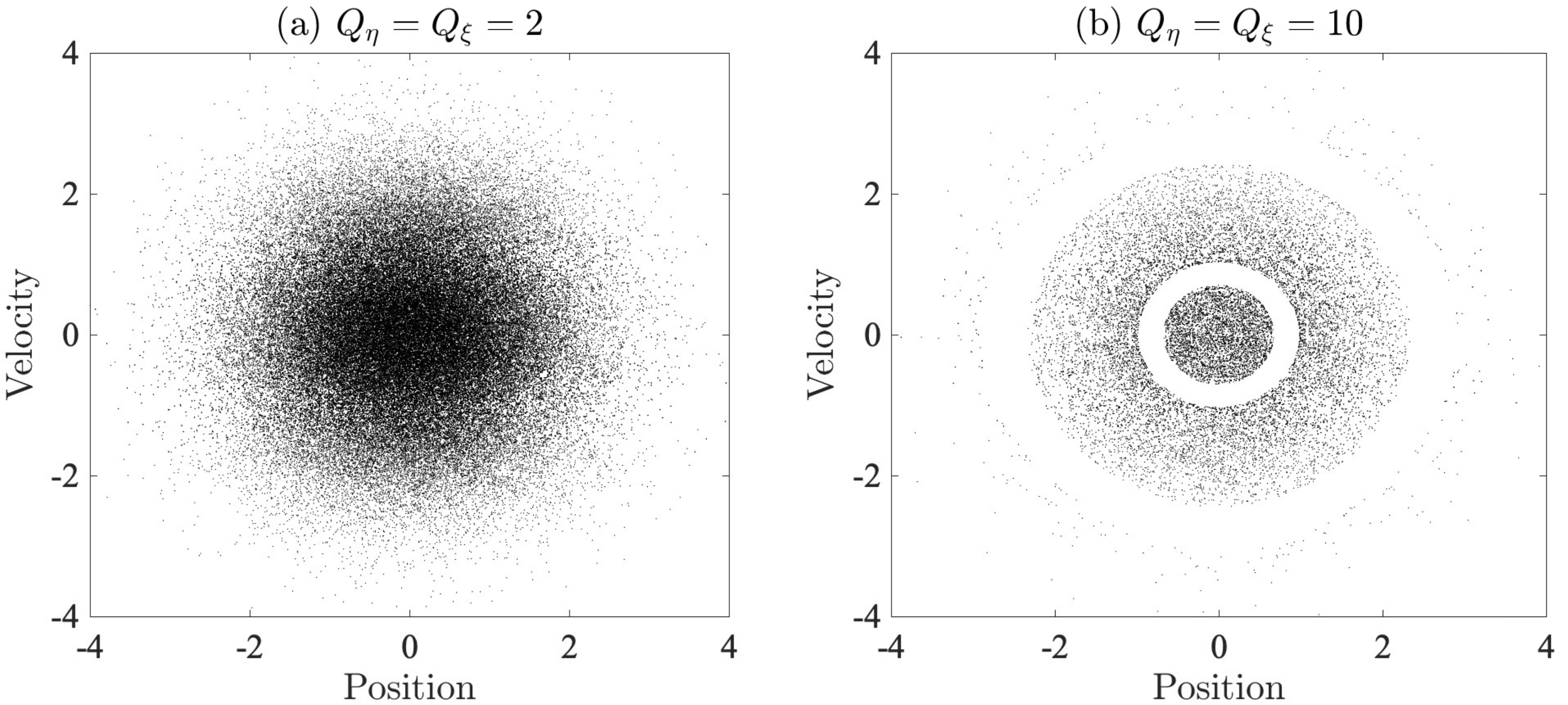}
\caption[Poincare cross-section MKT-EQ]{\label{fig:MKT_Poincare} Poincar\'e section at $|\eta| = |\xi| < 0.001$ cross-section for: (a) $Q_\eta = Q_\xi = 2$ and (b) $Q_\eta = Q_\xi = 10$. As is evident, for case (b) the dynamics is non-ergodic while for case (a) the dynamics is ergodic. This highlights the importance of choosing the correct value of $Q_\eta$ and $Q_\xi$. The exact value of $Q_\eta$ and $Q_\xi$ where the dynamics transitions from being ergodic to non-ergodic is yet to be discovered. } 
\end{figure}

\begin{figure}[htp]
\includegraphics[width=0.5\textwidth]{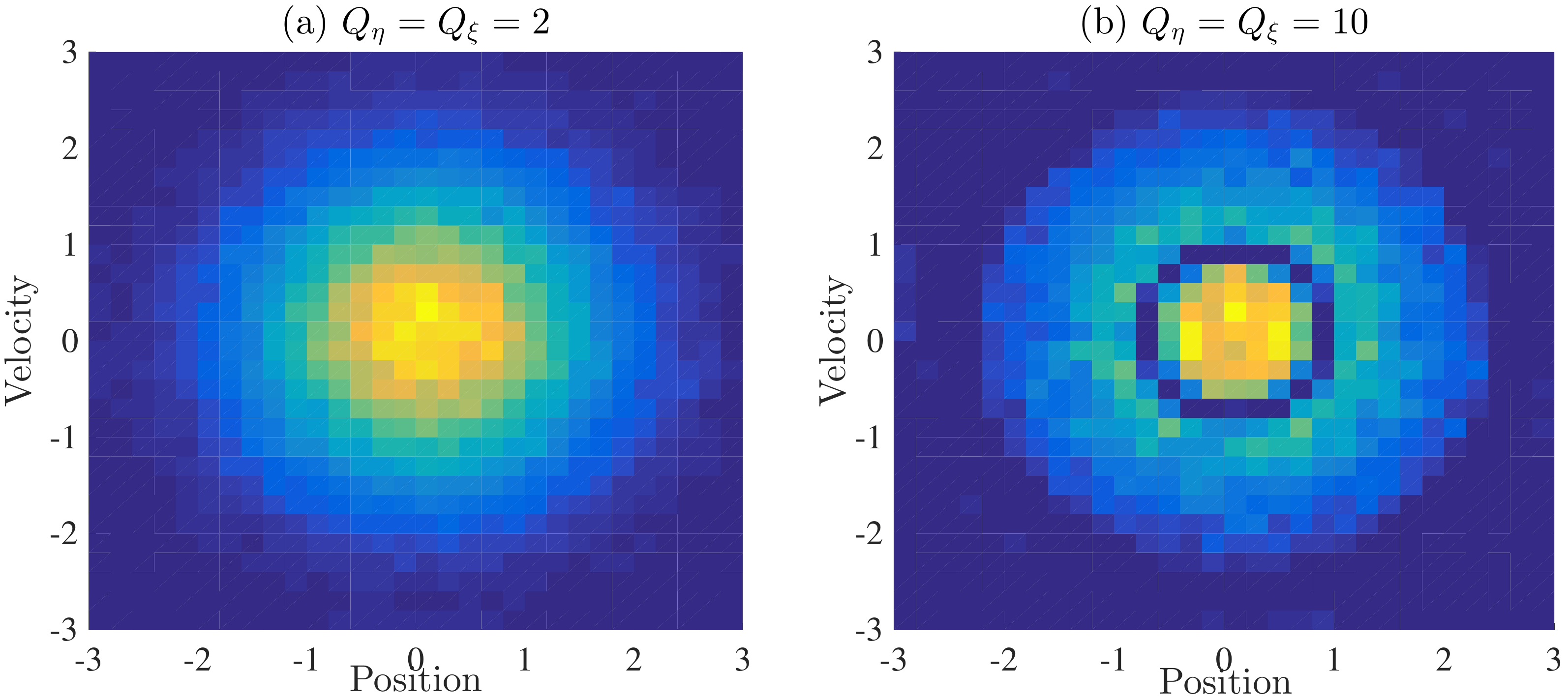}
\caption[Poincare cross-section NH Neq]{\label{fig:MKT_JPDF} JPDF of MKT thermostatted oscillator at the Poincar\'e section given by $|\eta| = |\xi| < 0.001$ with: (a) $Q_\eta = Q_\xi = 2$ and (b) $Q_\eta = Q_\xi = 10$. The deviation from normality is obvious in case (b). } 
\end{figure}

Let us now study the phase-space characteristics of a two-variable MKT thermostatted single harmonic oscillator having unit mass and spring constant. The resulting dynamics may be written as:
\begin{equation}
\begin{array}{cc}
\dot{x} = p, &
\dot{p} = -x - \dfrac{\eta p}{Q_{\eta}},\\
\dot{\eta} = p^2 - k_{B}T_0 - \dfrac{\eta\xi}{Q_{\xi}}, &
\dot{\xi} = \dfrac{\eta^2}{Q_{\eta}} - k_{B}T_0.
\end{array}
\label{eq:mkt_two}
\end{equation}
In a bid to ascertain the ergodic properties of the MKT thermostat, Patra and Bhattacharya \cite{patra2014nonergodicity} performed a series of computational runs with different values of $Q_\eta$, $Q_\xi$ and initial conditions. Working with the argument that if the dynamics due to the MKT thermostat is canonical, then the following conditional joint probability distributions hold true:
\begin{equation}
\begin{array}{ccl}
f(x,p|\eta=\eta_0, \xi=\xi_0) & \propto & \exp \left(-\dfrac{\beta_0}{2} x^2\right)\exp \left(-\dfrac{\beta_0}{2} p^2\right) \\
f(\eta,\xi|x=x_0,p=p_0) & \propto & \exp \left(-\dfrac{\beta_0}{2Q_\eta} \eta^2\right)\exp \left(-\dfrac{\beta_0}{2Q_\xi} \xi^2\right), \\
\label{eq:three_one_half}
\end{array}
\end{equation}
Patra and Bhattacharya showed that there is an appreciable difference between the joint PDFs calculated from the simulations with those obtained from theory. For example, if $\beta_0$ is chosen as unity, theory suggests that the joint PDF of $(x,p)$ is bivariate standard normal while that of ($\eta,\xi$) is bivariate normal with a variance of $Q_\eta$ and $Q_\xi$, respectively. Figure (\ref{fig:MKT_Poincare}) shows the Poincar\'e section plot at $|\eta| = |\xi| < 0.001$ cross-section for: (a) $Q_\eta = Q_\xi = 2$ and (b) $Q_\eta = Q_\xi = 10$. It is evident that in case (b) the dynamics is non-ergodic while in case (a) the dynamics is ergodic. Figure (\ref{fig:MKT_JPDF}) shows the joint PDFs for these cases, where the difference from a joint normal distribution is clearly visible for $Q_\eta=Q_\xi = 10$. 

\begin{figure}
\includegraphics[width=0.5\textwidth]{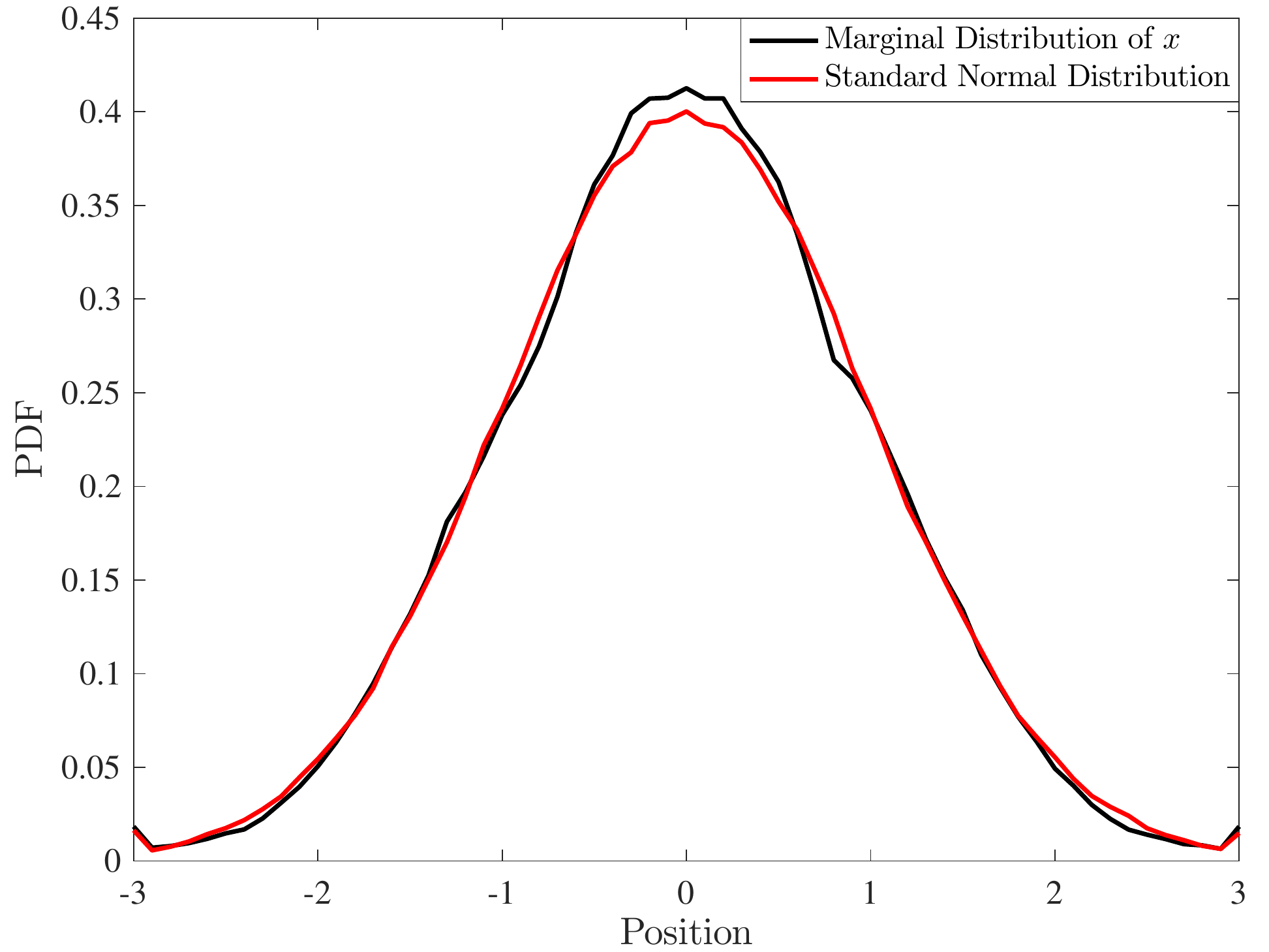}
\caption[Poincare cross-section NH Neq]{\label{fig:MKT_marginal} Marginal distribution of $x$ obtained by projecting all trajectory points and its comparison with a standard normal distribution. Notice the good agreement between the two distributions. A similar conclusion can be reached from the marginal distribution of $p$ as well.} 
\end{figure}
Recently, the non-ergodicity of MKT thermostat has been demonstrated by Legoll \cite{legoll2007non,legoll2009non} for $Q_\eta = Q_\xi \neq 1$. There is no denying the fact that as $Q_\eta \to 1$ amd $Q_\xi \to 1$, the ergodic characteristics of the MKT thermostat become superior to the Nos\'e-Hoover thermostat. However, if one looks at the marginal distribution functions, as had been done previously \cite{mkt,tuckerman_00}, one erroneously reaches the conclusion that the MKT thermostat is ergodic for all values of $Q_\eta$ and $Q_\xi$. This is illustrated in figure (\ref{fig:MKT_marginal}), where the marginal distribution agrees well with that of the standard normal probability distribution function when $Q_\eta = Q_\xi = 10$. 

A similar conclusion was reached from the dynamical perspective by Hoover and coworkers \cite{patra2015deterministic}. They investigated the MKT dynamics where $Q_\eta = Q_\xi = 1$, with millions of different initial conditions, to search for an initial condition that results in a conservative hyper-dimensional torus. Such a search is conducted as follows:
\begin{enumerate}
    \item Divide the ($x,p,\eta,\xi$) space into $100 \times 100 \times 100 \times 100$, where each variable lies between [-2.5,2.5]. The coordinate of each corner of the resulting hypercube serves as an initial condition for a simulation run. 
    \item Apart from the true trajectory, four additional sets of satellite trajectories are solved in tangent space. 
    \item Gram-Schmidt ortho-normalization follows every simulation time-step.
    \item The spectrum of Lyapunov exponents is then computed. 
\end{enumerate}
The resulting Lyapunov spectrum is: $\langle L_1 \rangle = +0.066_5, \langle L_2 \rangle = +0.000_0, \langle L_3 \rangle = -0.000_0, \langle L_4 \rangle = -0.066_5$ \cite{patra2015deterministic}. None of the initial conditions resulted in statistically significant deviation from the mentioned Lyapunov spectrum. 

It is evident from this discussion that the ergodicity, and hence, the dynamical properties, of a small-scale system change, depending on the thermostat masses $Q_\eta$ and $Q_\xi$. Some questions still remain open, though, -- for what values of $Q_\eta$ and $Q_\xi$, the dynamics of an MKT thermostatted oscillator is ergodic, and the exact value of $Q_\eta$ and $Q_\xi$ at which the stable periodic orbit disappears. 

\begin{figure}[htp]
\includegraphics[width=0.5\textwidth]{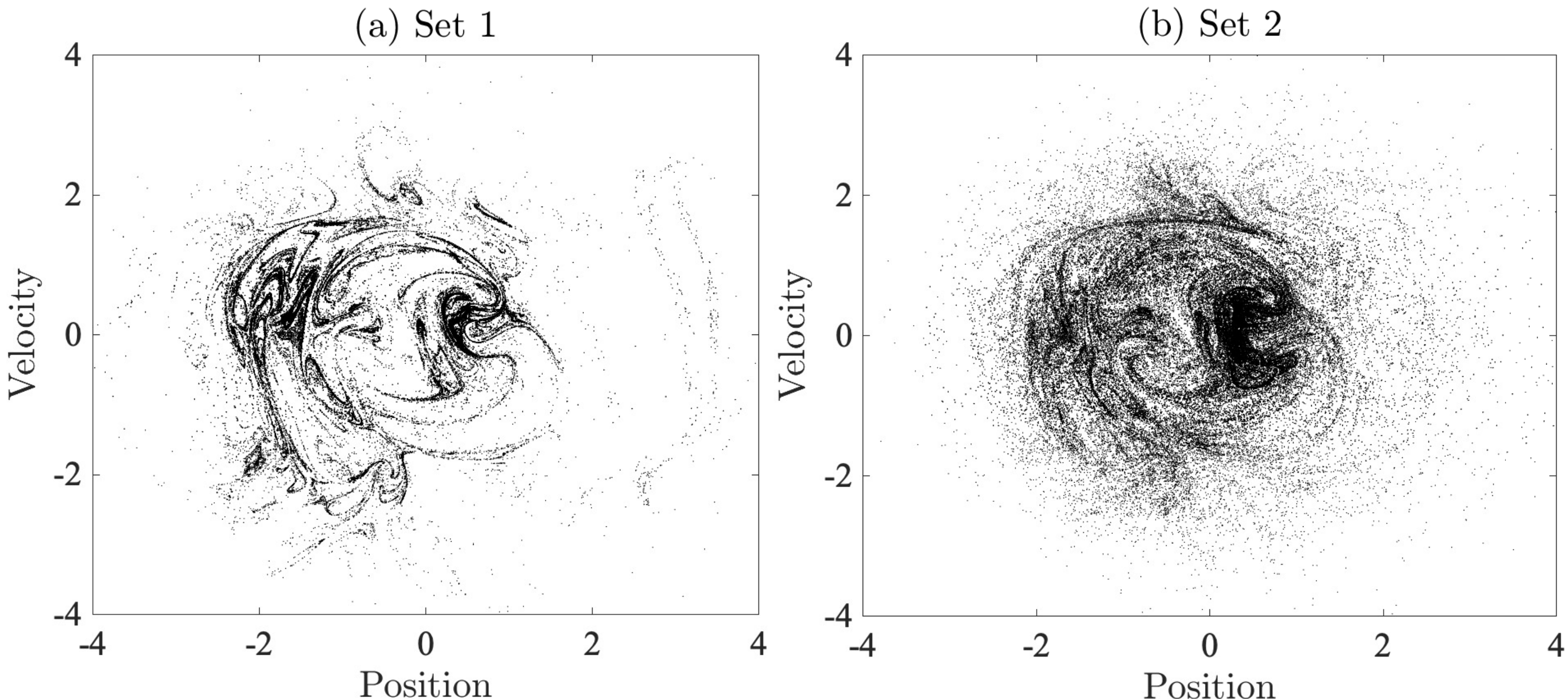}
\caption[Poincare cross-section MKT Neq]{\label{fig:MKT_Neq} Poincar\'e section plot of the Martyna-Klein-Tuckerman thermostatted oscillator subjected to the position dependent temperature field: $T(x) = 1 + 0.2 \tanh(x)$ at the cross-section $|\eta| = |\xi| < 0.001$. The initial conditions are $(x,p,\eta,\xi) = (1,0,0,0)$. Notice the multifractral nature of the dynamics. The multifractal nature indicates that phase space has ``compressed''. }
\end{figure}

We now subject the single harmonic oscillator with $Q_\eta = Q_\xi = 1$ to the position-dependent temperature field: $T(x) = 1.0 + 0.2 \tanh(x)$. Like the Nos\'e-Hoover thermostat, two sets of equations of motion can be written for the MKT thermostatted oscillator -- (i) the usual equations of motion, where $\langle \dot{S}/k_B \rangle \neq -\langle \Lambda \rangle$ (Set 1):
\begin{equation}
\begin{array}{cc}
\dot{x} = p, &
\dot{p} = -x - \eta p,\\
\dot{\eta} = p^2 - T(x) - \eta\xi, &
\dot{\xi} = \eta^2 - T(x),
\end{array}
\label{eq:mkt_set1}
\end{equation}
and (ii) the modified equations of motion,  where $\langle \dot{S}/k_B \rangle = -\langle \Lambda \rangle$ (Set 2):
\begin{equation}
\begin{array}{cc}
\dot{x} = p, &
\dot{p} = -x - \eta p,\\
\dot{\eta} = p^2/T(x) - 1 - \eta\xi, &
\dot{\xi} = \eta^2 - 1,
\end{array}
\label{eq:mkt_set2}
\end{equation}
The resulting equations of motion are solved using the $4^{th}$ order Runge-Kutta method for 100 billion time steps. The Poincar\'e section plots at the cross-section $|\eta| = |\xi| < 0.001$ are shown in figure (\ref{fig:MKT_Neq}). Notice that the phase-space filling nature of the dynamics has given way to a complicated multi-fractal for both the equation sets. The presence of such a multi-fractal is a signature of satisfying the second law of thermodynamics \cite{hoover1994second,hoover1998multifractals,hoover1998liouville,hoover2007second}. For the two sets of equations, the different thermodynamic variables are summarized in table \ref{tab:my_label2}.
\begin{table}[]
    \centering
    \begin{tabular}{|c|c|c|c|c|c|c} \hline
$\langle \dot{Q}_1 \rangle$ & $\langle \dot{Q}_2 \rangle$ & $\langle \dot{S}_1 \rangle$ & $\langle \dot{S}_2 \rangle$ & $\langle \Lambda_1 \rangle$ & $\langle \Lambda_2 \rangle$  \\ \hline \hline
0.0000 & 0.0000 & -0.0164 & -0.0038 & -0.0322 & -0.0038 \\ \hline
    \end{tabular}
    \caption{Different thermodynamic quantities for the two equation sets with initial conditions $(x,p,\eta,\xi) = (1,0,0,0)$. Each quantity has a subscript: $_1$ denotes the equations of motion (\ref{eq:mkt_set1}), while $_2$ is for the denotes the equations of motion (\ref{eq:mkt_set1}). Notice that while $\langle \dot{S}_1 \rangle \neq \langle \Lambda_1 \rangle$, $\langle \dot{S}_2 \rangle = \langle \Lambda_2 \rangle$.}
    \label{tab:my_label2}
\end{table}
The time-averaged Lyapunov exponents are: $\langle L_1 \rangle = 0.069_2, \langle L_2 \rangle = 0, \langle L_3 \rangle = -0.015_9, \langle L_4 \rangle = -0.085_6$ such that $\sum \langle L_i \rangle < 0$. The negative sum indicates that the phase-space has compressed and the information dimension (=3.624) is not equal to the embedding dimension (=4.0).

To summarize, MKT thermostat has improved ergodic characteristics along with the ability to show phase-space compression and heat flow in non-equilibrium. Thus, it conforms to the Second Law of thermodynamics. Further, it satisfies the Zeroth Law and is time-reversible. Although the equations of motion have, so far, not been derived from a Hamiltonian approach, its ease of implementation has made the MKT thermostat very popular for investigating a variety of equilibrium and non-equilibrium situations. In fact, MKT thermostat with a chain length of two is a standard library function in several well-known molecular dynamics software such as LAMMPS \cite{plimpton1993fast}, Gromacs \cite{berendsen1995gromacs}, etc.

\subsection{Hoover-Holian Thermostat}
The Hoover-Holian thermostat takes a different approach than the MKT thermostat to improve the ergodic characteristics of the Nos\'e-Hoover dynamics. This thermostat is based on the kinetic-moments method \cite{hh}, which controls simultaneously the first two moments of the kinetic energy (see equation (\ref{eq:2nd_order_kin_temp}) ). Consequently, errors associated with the first two moments of the kinetic energy are removed. Each moment is controlled by a different reservoir variable so that there are two thermostat variables. It must be noted that for a system to follow the ``true'' Maxwell-Boltzmann distribution, errors associated with \textit{all} moments must be removed. If the dynamics is ergodic, it is expected that given sufficient time the dynamics is sampled in a manner that errors associated with the higher-order moments get removed as well. 

The Hoover-Holian thermostat may be developed using the guessing method (see Nos\'e-Hoover section). The salient steps are indicated below. Let us assume that the coupling between the thermostat and the system is as follows:
\begin{equation}
\begin{array}{rcl}
    \dot{x}_i & = & p_i \\
    \dot{p}_i & = & F_i - \eta p_i - \xi (K/K_0)p_i \\
    \dot{\eta} & = & ? \\
    \dot{\xi} & = & ?
    \label{eq:HH_1}
\end{array}
\end{equation}
Here, $K_0 = (3N/2) k_B T_0$, is the desired kinetic energy, and $K$, the instantaneous kinetic energy. At this stage, the time evolution of $\dot{\eta}$ and $\dot{\xi}$ are not known. The coupling shown in equation (\ref{eq:HH_1}) may be viewed as two independent thermal reservoirs simultaneously acting on the system -- one for controlling the first moment of kinetic energy ($\eta$) and the other for controlling the second moment ($\xi$). Being in thermal equilibrium, the PDFs of the reservoir variables, $\eta$ and $\xi$, must also satisfy Maxwell-Boltzmann distribution, so that the extended phase-space distribution function becomes:
\begin{equation}
f_{\text{ex}} \left(\mathbf{\Gamma}\right) \propto \exp \left({-\beta \left[ H - K_0Q_\eta \eta^2 + K_0 Q_\xi \xi^2 \right]} \right),
\label{eq:hh_canonical_distribution}
\end{equation}
where, $H = \sum p_i^2/2m + \Phi(\mathbf{x})$ and $\mathbf{\Gamma} = (\mathbf{x,p},\eta,\xi)$. The steady state Liouville's equation (\ref{eq:two_thirtysix}) is solved to obtain the evolution of $\dot{\eta}$ and $\dot{\xi}$. The final equations of motion are:
\begin{equation}
\begin{array}{rcl}
    \dot{x}_i & = & p_i \\
    \dot{p}_i & = & F_i - \eta p_i - \xi \left( \dfrac{K}{K_0} \right)p_i \\
    \dot{\eta} & = & \dfrac{1}{Q_\eta} \left[\dfrac{K}{K_0} - 1\right] \\
    \dot{\xi} & = &  \dfrac{1}{Q_\xi} \dfrac{K}{K_0} \left[ \dfrac{K}{K_0} - 1 - \dfrac{2}{3N}\right]
    \label{eq:HH_2}
\end{array}
\end{equation}
These equations are applicable to equilibrium and nonequilibrium many-body simulations. They are time-reversible, ergodic, conform with the different laws of thermodynamics and easy to implement. However, a Hamiltonian basis for these equations of motion is yet to be discovered, even though a constant of motion exists:
\begin{equation}
\begin{array}{rcl}
    E_\text{HH} & = & \Phi(\mathbf{x}) + \sum\limits_{i=1}^{3N} \dfrac{p_i^2}{2m} + K_0 \left( Q_\eta \eta^2 + Q_\xi \xi^2 \right) \\
    & & + \int\limits_{0}^t \left[ 2K_0 \eta + K \xi \left(1+\dfrac{2}{3N} \right)  \right] dt\\
    \label{eq:HH_COM}
\end{array}
\end{equation}
In comparison to the MKT thermostat, the equations of motion for the Hoover-Holian thermostat are stiffer since they have terms involving $K^2$. In fact, these equations can be readily modified to incorporate higher order moments of kinetic energy, but the resulting differential equations become very stiff (since they have $K^3$ or higher order terms), necessitating very small integration time-steps and increasing the computational cost. Just like in the MKT thermostat, where additional thermostat variables are introduced to control the fluctuations of thermostat variables, the Hoover-Holian thermostat can be generalized to include additional thermostat variables to control the fluctuations of $\eta$ and $\xi$ \cite{liu2000generalized} . 

\subsubsection{Solving the Equations of Motion}
To solve the equations we use trotter's factorization. 
\begin{enumerate}
    \item Initialize the system with suitable initial conditions of positions and momenta, $\eta = 0$, and $ \xi = 0$.
    \item Run in a loop until the desired time, $t$, is reached with an incremental time step of $\Delta t$:
    \begin{enumerate}
        \item Compute $K$, the kinetic energy of the system and $ K_0$ as the desired kinetic energy.
        \item Compute $G_1 = \dfrac{1}{Q_\eta} \left[\dfrac{K}{K_0} - 1\right]$
        \item Propagate $\eta$ by $\Delta t /4$: $\eta \to \eta + G_1 \Delta t/4$
        \item Compute $G_2 = \dfrac{1}{Q_\xi} \dfrac{K}{K_0} \left[ \dfrac{K}{K_0} - 1 - \dfrac{2}{3N}\right]$
        \item Propagate $\xi$ by $\Delta t /4$: $\xi \to \xi + G_2 \Delta t/4$
        \item Define $\zeta$, the scale parameter: $\zeta=\exp(-\lambda \frac{\Delta t}{8.0})$ where $ \lambda = \frac{\eta}{Q_{\eta}} $.
        \item Scale all momenta: $p_i \to p_i \times \zeta$.
        \item Compute scaled $K$: $K \to K \zeta^2$.
        \item Define $\zeta$, the scale parameter: $\zeta=(1+\frac{2\xi K}{K_0}\frac{\Delta t}{4.0})^{-1/2}$.
        \item Scale all momenta: $p_i \to p_i \times \zeta$.
        \item Compute scaled $K$: $K \to K \zeta^2$.
        \item Define $\zeta$, the scale parameter: $\zeta=\exp(-\lambda \frac{\Delta t}{8.0})$ where $ \lambda = \frac{\eta}{Q_{\eta}} $.
        \item Scale all momenta: $p_i \to p_i \times \zeta$. 
        \item Compute scaled $K$: $K \to K \zeta^2$.
        \item Define $\zeta$, the scale parameter: $\zeta=\exp(-\lambda \frac{\Delta t}{8.0})$ where $ \lambda = \frac{\eta}{Q_{\eta}} $.
        \item Scale all momenta: $p_i \to p_i \times \zeta$.
        \item Compute scaled $K$: $K \to K \zeta^2$.
        \item Define $\zeta$, the scale parameter: $\zeta=(1+\frac{2\xi K}{K_0}\frac{\Delta t}{4.0})^{-1/2}$.
        \item Scale all momenta: $p_i \to p_i \times \zeta$.
        \item Compute scaled $K$: $K \to K \zeta^2$.
        \item Define $\zeta$, the scale parameter: $\zeta=\exp(-\lambda \frac{\Delta t}{8.0})$ where $ \lambda = \frac{\eta}{Q_{\eta}} $.
        \item Scale all momenta: $p_i \to p_i \times \zeta$. 
        \item Compute scaled $K$: $K \to K \zeta^2$.
        \item Compute $G_1 = \dfrac{1}{Q_\eta} \left[\dfrac{K}{K_0} - 1\right]$
        \item Propagate $\eta$ by $\Delta t /4$: $\eta \to \eta + G_1 \Delta t/4$
        \item Compute $G_2 = \dfrac{1}{Q_\xi} \dfrac{K}{K_0} \left[ \dfrac{K}{K_0} - 1 - \dfrac{2}{3N}\right]$
        \item Propagate $\xi$ by $\Delta t /4$: $\xi \to \xi + G_2 \Delta t/4$
        \item Propagate all momenta by $\Delta t/2$: $p_i \to p_i + F_i \Delta t/2$.
        \item Propagate all positions by $\Delta t$: $x_i \to x_i + p_i \Delta t$.
        \item Propagate all momenta by $\Delta t/2$: $p_i \to p_i + F_i \Delta t/2$.
        \item Redo steps (a) to (aa)
    \end{enumerate}
\end{enumerate}
Here propagation of $ \eta$ and $\xi $ can be done in any order since they are independent of each other.

\subsubsection{Phase space characteristics from single harmonic oscillator}
Now let us look at the phase-space characteristics of the Hoover-Holian thermostat by coupling it with a single harmonic oscillator of unit mass and stiffness. A slightly modified form of the Hoover-Holian equations of motion, for this case, are:
\begin{equation}
\begin{array}{rcl}
\dot{x} & = & p\\
\dot{p} & = & -x -\eta p -\xi p^3 \\
\dot{\eta} & = & \dfrac{1}{Q_\eta} \left( p^2 - k_BT_0 \right) \\
\dot{\xi} & = & \dfrac{1}{Q_\xi} \left( p^4 - 3 k_BT_0 p^2 \right),
\end{array}
\label{eq:HH_SHO}
\end{equation}
Note that this modified form has the same problem as that of the Nos\'e-Hoover and the MKT thermostat -- $\langle \dot{S}/k_B \rangle \neq -\langle \Lambda \rangle$. The situation may be resolved by writing the equations of motion as per the equation (\ref{eq:HH_2}).

\begin{figure}[htp]
\includegraphics[width=0.5\textwidth]{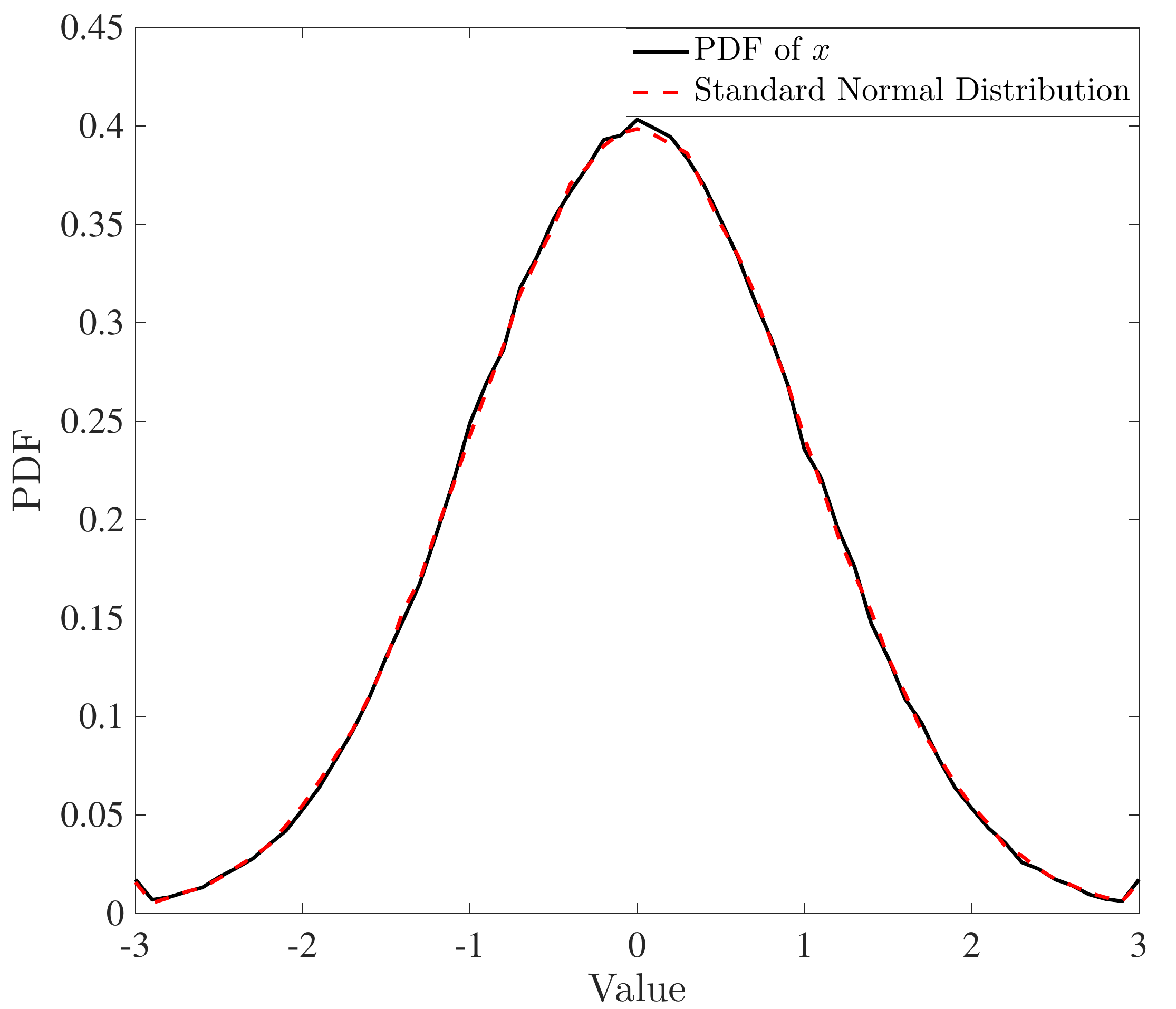}
\caption[Poincare cross-section HH Eq]{\label{fig:HH_Eq} PDF of $x$ of the Hoover-Holian thermostatted oscillator at the Poincar\'e section $|\eta| = |\xi| < 0.001$ when it is subjected to $k_BT_0 = 1$. The initial conditions are $(x,p,\eta,\xi) = (1,0,0,0)$. Notice the good agreement of the PDF of $x$ with the standard normal distribution.}
\end{figure}

Choosing $k_BT_0 = Q_\eta = Q_\xi = 1.0$ results in a scenario where the oscillator comes to thermal equilibrium at a temperature of unity, and the extended phase-space distribution becomes a product of four independent standard normal variables. The Hoover-Holian thermostatted single harmonic oscillator correctly samples the phase-space from a canonical distribution both in the projected space as well as at Poincar\'e sections. The marginal distribution of $x$, at the Poincar\'e section of $|\eta| = |\xi| < 0.001$, has been plotted in figure (\ref{fig:HH_Eq}). The results correspond to the initial conditions $(x,p,\eta,\xi) = (1,0,0,0)$, and have been obtained by solving equation (\ref{eq:HH_SHO}) using the $4^{th}$ order Runge-Kutta method for 100 billion time steps, with $\Delta t = 0.001$. The equivalence of the PDF with that of a standard normal PDF suggests that that the Hoover-Holian thermostat is ergodic from statistical perspective. 

A similar conclusion may also be reached from the dynamical perspective. Working with millions of different initial conditions, Hoover and coworkers \cite{patra2015deterministic} made a futile search for an initial condition which results in a conservative hyper-dimensional torus. The spectrum of Lyapunov exponents obtained by them is: $\langle L_1 \rangle = +0.068_0, \langle L_2 \rangle = +0.000_0, \langle L_3 \rangle = -0.000_0, \langle L_4 \rangle = -0.068_0$, with no initial condition for which the largest Lyapunov exponent is zero. The failure of this extensive brute-force search to find such an initial condition suggests that the dynamics is ergodic. But, it must be noted that this method is not \textit{full proof}, and there may still exist an initial condition for which the largest Lyapunov exponent is zero. 

A closely related approach to the Hoover-Holian thermostat is the algorithm that controls the first two even moments of velocity, for which the temperatures being controlled are as per equations (\ref{eq:two_seven}) and (\ref{eq:2nd_order_kin_temp}). When this thermostat is coupled with a single harmonic oscillator, the resulting equations of motion are identical to that shown in equations (\ref{eq:HH_SHO}), and consequently all thermodynamic properties are identical. However, for a multi-particle system, the equations of motion for the two thermostats are different.

\begin{figure}[htp]
\includegraphics[width=0.5\textwidth]{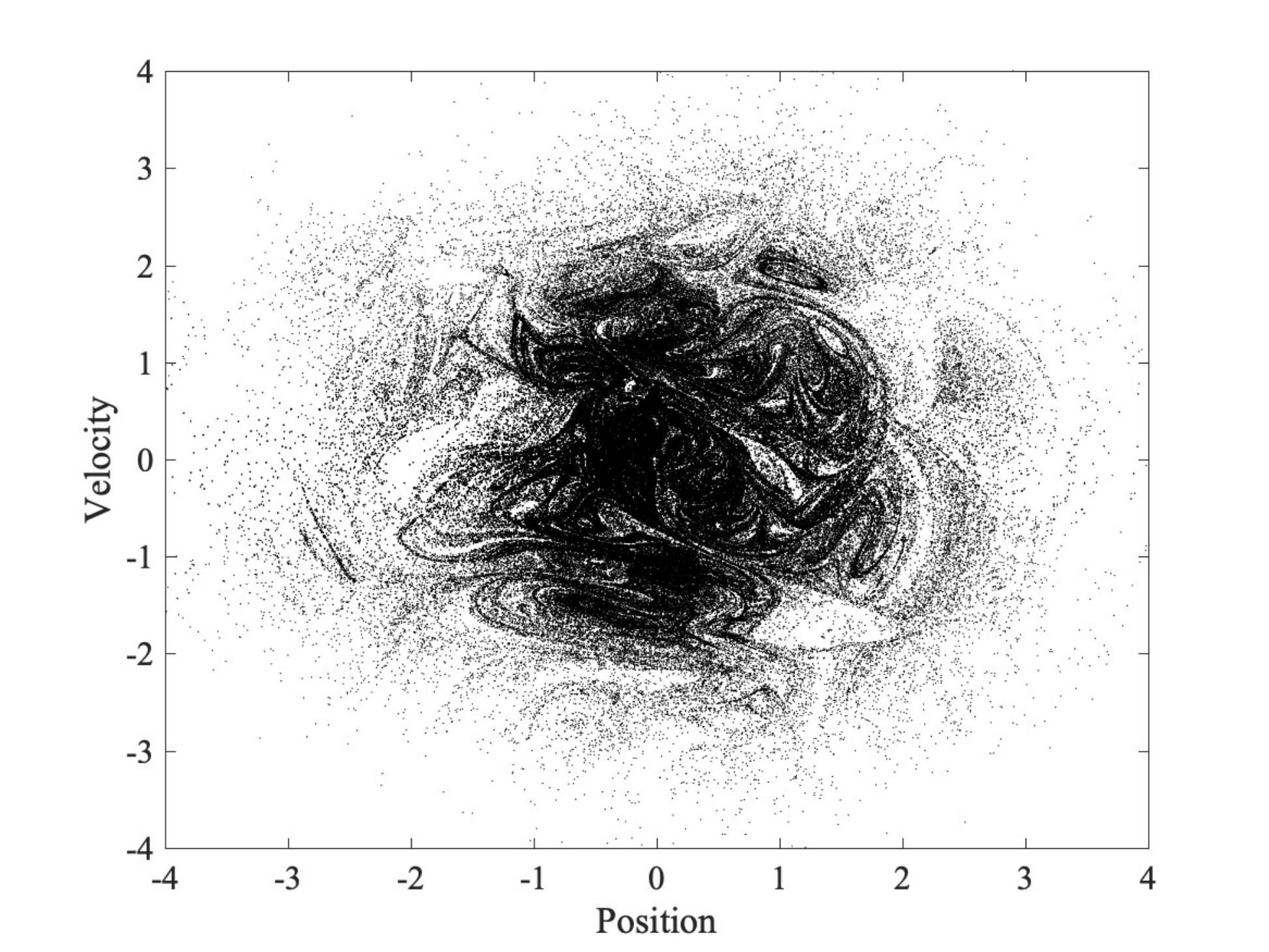}
\caption[Poincare cross-section HH Neq]{\label{fig:HH_NEq} Poincar\'e section plot of the Hoover-Holian thermostatted oscillator subjected to the position dependent temperature field: $T(x) = 1 + 0.2 \tanh(x)$ at the cross-section $|\eta| = |\xi| < 0.001$. The initial conditions are $(x,p,\eta,\xi) = (1,0,0,0)$. The dynamics is multifractral like the MKT thermostatted oscillator, but the nature of the multifractal is different. Additionally, $\langle \Lambda \rangle$ and the information dimension of the multifractal are different from those of the MKT thermostatted oscillator. }
\end{figure}
Let us now subject the Hoover-Holian thermostatted oscillator to the non-equilibrium position dependent temperature field. The equations of motion in this case get modified as:
\begin{equation}
\begin{array}{rcl}
\dot{x} & = & p\\
\dot{p} & = & -x -\eta p -\xi p^3 \\
\dot{\eta} & = & p^2 - \left[1.0 + 0.2 \tanh(x) \right] \\
\dot{\xi} & = & p^4 - 3 p^2 \left[1.0 + 0.2 \tanh(x) \right].
\end{array}
\label{eq:HH_SHO_NEQ}
\end{equation}
As is expected from a good thermostat, figure (\ref{fig:HH_NEq}) shows that under the imposed temperature field, the dynamics represented by the equation (\ref{eq:HH_SHO_NEQ}) is multi-fractal. The presence of a multi-fractal is a signature of the information dimension being smaller than the embedding dimension (in this case 4) and conformity with the Second Law of thermodynamics. However, the nature of the multi-fractal is significantly different from that of the MKT thermostatted oscillator under the same imposed temperature field. The spectrum of Lyapunov exponents obtained here: $\langle L_1 \rangle = +0.065_0, \langle L_2 \rangle = +0.000_3, \langle L_3 \rangle = -0.001_2, \langle L_4 \rangle = -0.070_2$, is also different from that of MKT thermostatted oscillator. The Kaplan-Yorke dimension, calculated by linearly interpolating between the last positive sum of Lyapunov exponents and the first negative sum, is 3.913, again differs from that of MKT thermostatted dynamics. These differences make us ask the question -- which of the two thermostats result in a closer approximation of reality. The question remains open for answering. 

Overall, despite the good ergodic characteristics, Hoover-Holian thermostat has not seen widespread use in MD community, except in pedagogical cases involving single harmonic oscillators.

\subsection{Campisi-Zhan-Talkner-Hangii Thermostat}
Most of the thermostats described so far are further developments of the original thermostat proposed by Nos\'e. The thermostat proposed by Campisi et al. \cite{CZTH_thermostat,CZTH_thermostat2} takes a different route. This Hamiltonian thermostat possesses the property of infinite heat capacity while simultaneously avoiding the troublesome time-scaling factor of the Nos\'e thermostat. When coupled weakly to a microscopic system, the equations of motion can be obtained directly from the Hamilton's equations. This thermostat, also known as the log thermostat, comprises an oscillator of mass $m_s$ that is governed by the Hamiltonian:
\begin{equation}
H_{\text{CZTH}} = \dfrac{p_s^2}{2m_s} + \dfrac{k_BT_0}{2} \log \left( s^2 + \delta \right),
\label{eq:CZTH_Hamiltonian}
\end{equation}
where, $s$ and $p_s$ denote the position and momentum of the oscillator, respectively. In order to prevent singularity of the potential energy at the origin, a small constant $\delta$ is usually added. Equation (\ref{eq:CZTH_Hamiltonian}) has an interesting property -- no matter the energy of the oscillator, the kinetic temperature of the oscillator always equals $k_BT_0$. One can show this by employing the virial theorem under the assumption that $\delta \ll 1$:
\begin{equation}
\begin{array}{cccccc}
\left\langle p_s \dfrac{\partial H_{\text{CZTH}}}{\partial p_s} \right\rangle & = & \left\langle s \dfrac{\partial H_{\text{CZTH}}}{\partial s} \right\rangle &
\implies \left\langle \dfrac{p_s^2}{m} \right\rangle & = & k_BT_0 .
\label{eq:CZTH_Virial}
\end{array}
\end{equation}
Further, it can be shown that the momentum-space sampled by the dynamics is as per the Maxwell-Boltzmann distribution. Being Hamiltonian, the dynamics is time-reversible as well. So, the log thermostat satisfies several properties of a good thermostat -- it's time reversible, possesses a Hamiltonian and samples the phase-space according to the Maxwell-Boltzmann distribution.

Consider a log thermostat coupled to another microscopic system comprising $N$ particles, and having a Hamiltonian $H = \sum p_i^2/2m + \Phi(\mathbf{x})$. Let the coupling be defined by $h(\mathbf{x},s)$, such that this extended system is governed by the Hamiltonian:
\begin{equation}
    H_{\text{ex}} = H_{\text{CZTH}} + H(\mathbf{x,p}) + h(\mathbf{x},s).
\end{equation}
The equations of motion can then be obtained as:
\begin{equation}
\begin{array}{rc}
\dot{x}_i = \dfrac{p_i}{m}, & \dot{p}_i = - \dfrac{\partial \Phi}{\partial x_i} - \dfrac{\partial h(\mathbf{x},s)}{\partial x_i}, \\
\dot{s} = p_s, & \dot{p}_s = \dfrac{k_BT_0s}{s^2+\delta} - \dfrac{\partial h(x,s)}{\partial x}.
\end{array}
\label{eq:CZTH_EOM}
\end{equation}
The equations of motion can be integrated using the Velocity-Verlet algorithm (see equation (\ref{eq:two_three}) as they have been obtained by applying Hamilton's equations. 

The presence of $\delta$, which was added to $H_\text{CZTH}$ to prevent any singularity as $s \to 0$, causes deviation of the momentum distribution from the Maxwell-Boltzmann distribution \cite{CZTH_thermostat2}, with the effect being more pronounced as $N$ increases. This limits the usefulness of log thermostat to systems comprising very few particles. The coupling function, $h(\mathbf{x},s)$, ensures the interaction between the system and the log thermostat. In its absence, the system and the log thermostat undergo microcanonical dynamics independently. Ideally, $h(\mathbf{x},s)$ should be chosen such that ergodicity is imparted within the system. However, the selection of appropriate coupling functions remains a problem open to research. 

The theoretical advantages of the log thermostat disappear when one tries to employ it in standard molecular dynamics simulations. Some of the salient problems are listed below:
\begin{enumerate}
    \item The log thermostat does not conform with the Zeroth Law of thermodynamics \cite{patra2018zeroth}. When a single harmonic oscillator, in thermal equilibrium at $T_0$, is coupled using harmonic springs with two log thermostats, also kept at $T_0$, the temperature of the log thermostats no longer remains at $T_0$, which is a clear violation of the Zeroth Law of thermodynamics. Such a problem is not observed in other thermostats. 
    \item The log thermostat fails to promote heat flow [\cite{hoover_czth}]. When a one-dimensional $\Phi^4$ chain is coupled to two log thermostats kept at different temperatures, no heat flow occurs, which is a clear violation of the Second Law of thermodynamics.
    \item The configurational temperature associated with the log thermostat is negative in one dimension and zero in two dimension [\cite{sponseller_14}], which is an unphysical situation. 
    \item The log thermostat violates the equipartition theorem along with the virial theorem for strong coupling to the system. \cite{chen2017violation}.
\end{enumerate}

\subsubsection{Phase-space characteristics from single harmonic oscillator }
\begin{figure}[htp]
\includegraphics[scale = 0.30]{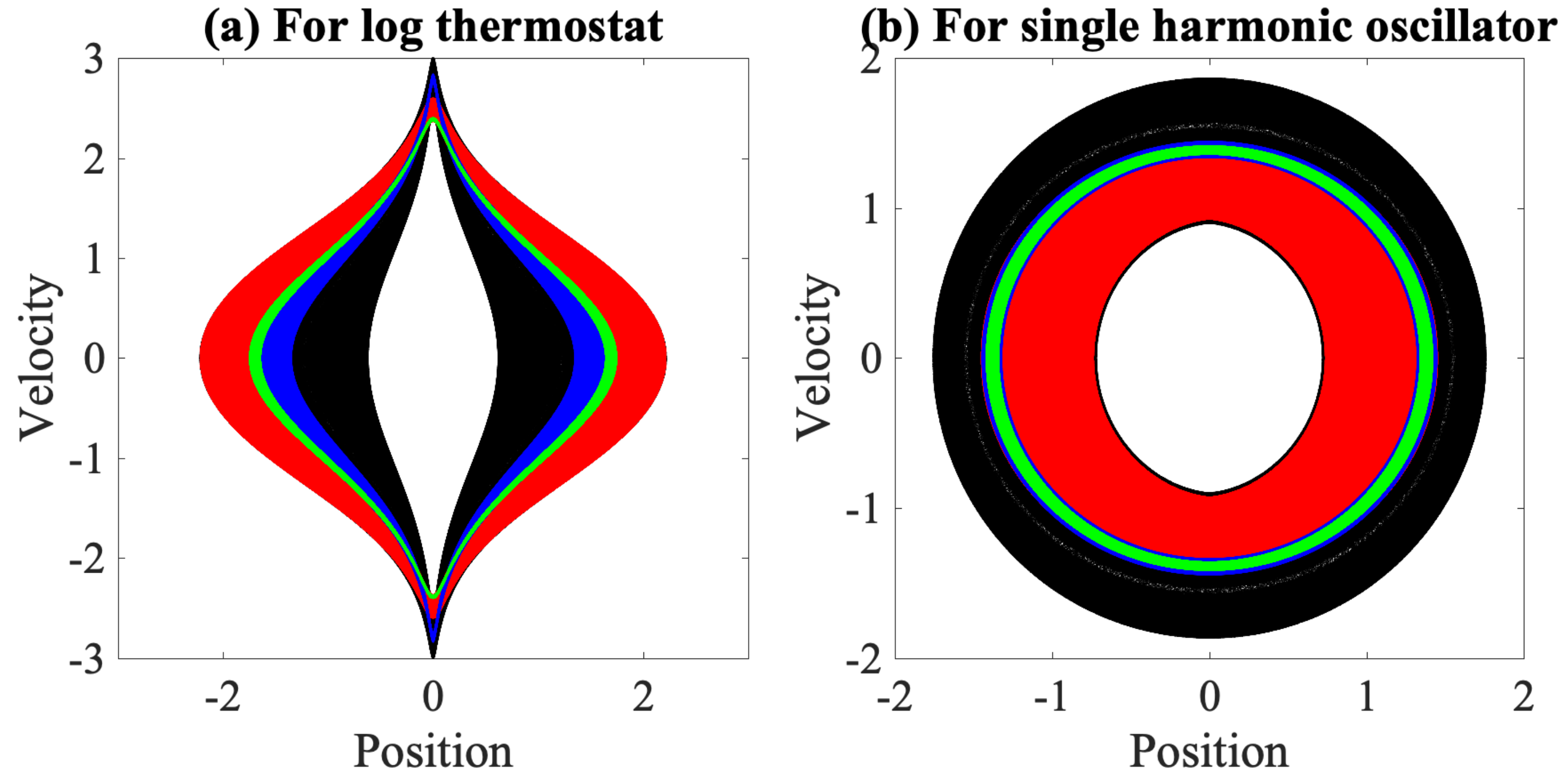}
\caption[CZTH plots]{\label{fig:CZTH_POS_VEL} Plot of position and velocity for: (a) log thermostat and (b) single harmonic oscillator. The different colors correspond to: black -- $k=0.1,\delta = 0.001$, red -- $k=0.1,\delta = 0.01$, blue -- $k=0.01,\delta = 0.001$, green -- $k=0.01,\delta = 0.01$. Notice that for none of the cases, the trajectory of the single harmonic oscillator is such that the entire phase-space is filled. Consequently, the dynamics is not ergodic.}
\end{figure}

When a single harmonic oscillator is coupled to a log thermostat, kept at $k_BT_0 = 1$, through a Hookean spring $h(x,s) = 0.5k(x-s)^2$, the equations of motion are:
\begin{equation}
\begin{array}{rl}
\dot{x} = p, & \dot{p} = -x - k(x-s), \\
\dot{s} = p_s, & \dot{p}_s = \dfrac{k_BT_0s}{s^2+\delta} + k(x-s).
\end{array}
\label{eq:CZTH_EOM2}
\end{equation}
We now solve these equations using the $4^{th}$ order Runge-Kutta method, as $k$ and $\delta$ change, for understanding the phase-space characteristics. The results are shown in figure (\ref{fig:CZTH_POS_VEL}). The different colors correspond to: black -- $k=0.1,\delta = 0.001$, red -- $k=0.1,\delta = 0.01$, blue -- $k=0.01,\delta = 0.001$, and green -- $k=0.01,\delta = 0.01$. 

Notice that none of the cases resulted in a situation where the entire phase-space is filled. Consequently, the dynamics is not ergodic.

\section{Rugh's Temperature and Its Control}
Let us take a relook at the Maxwell-Boltzmann distribution, shown in equation (\ref{eq:two_nine}) -- $\beta$ is associated with both momentum and configurational variables. So, far we have discussed about the kinetic temperature, which relates $\beta$ with \textit{only} the momentum variables. Rugh's temperature, on the other hand, does not treat momentum variables as the preferred candidate for expressing temperature, and includes both configurational and momentum variables for defining temperature. In Rugh's approach, the temperature is determined from the global geometric structure of the energy surface \cite{Rugh,rugh_98}. It was later observed that the Rugh's temperature is a particular case of a more generalized situation: consider a continuous differentiable phase functional, $B$. The temperature of a system in equilibrium is related to the phase functional $B$ through the expression \cite{jepps_00}:
\begin{equation}
\dfrac{1}{k_BT} = \left\langle \dfrac{\nabla . \nabla B}{\nabla B . \nabla H} \right\rangle \equiv \dfrac{ \langle \nabla . \nabla B \rangle}{\langle \nabla B . \nabla H \rangle}.
\label{eq:two_eleven}
\end{equation}
Here, $\nabla$ is the gradient with respect to both configurational and momentum variables. This expression of temperature is independent of the choice of ensemble, and is equally applicable to microcanonical, canonical and molecular dynamics ensembles \cite{jepps_00}. Under very general conditions, it has been shown that, in equilibrium, the choice of $B$ does not influence the numerical value of the temperature. It is, therefore, possible to obtain a family of temperature definitions by choosing an appropriate functional form of $B$. In fact, the different measures of kinetic temperature can be obtained from equation (\ref{eq:two_eleven}): by choosing $B = \sum p_i^2/2$, one can recover the expression for the usual kinetic temperature shown in equation (\ref{eq:two_seven}), while choosing $B = \sum p_i^4/4$ leads to $T_{k,2}$. Likewise, one can obtain $T_{K,2}$ by choosing $B = K^2$.

Rugh's temperature, $T_R$, is obtained by choosing $B = H = K + \Phi$ in (\ref{eq:two_eleven}):
\begin{equation}
\dfrac{1}{k_BT_R} = \dfrac{ \langle \nabla^2 K + \nabla^2 \Phi \rangle}{\langle (\nabla K)^2 + (\nabla \Phi)^2 \rangle},
\label{eq:two_twelve}
\end{equation}
As can be seen from equation (\ref{eq:two_twelve}), $T_R$, apart from taking into account the momentum variables through the terms $\nabla K$ and $\nabla^2 K$, separately considers the configurational variables through the terms $\nabla \Phi$ and $\nabla^2 \Phi$. A closer look at equation (\ref{eq:two_twelve}) reveals a dimensional inconsistency that can be corrected by multiplying a unit constant of appropriate dimensions to both numerator and denominator \cite{morriss_99}. Extension of equilibrium Rugh's temperature to nonequilibrium cases also exists \cite{morriss_99}. 

As of now, only two approaches are available for controlling Rugh's temperature -- the Patra-Bhattacharya thermostat and the Bauer-Bulgac-Kusnezov thermostat. Let us briefly describe these algorithms.  

\subsection{Patra-Bhattacharya Thermostat}
The need to control the Rugh's temperature arose for correctly simulating the near-equilibrium problems. In the near-equilibrium regime, where local thermodynamic equilibrium holds, all postulates of equilibrium thermodynamics can be applied \textit{locally}. Important consequences of local thermodynamic equilibrium are: (i) velocity distribution must follow Maxwell-Boltzmann distribution locally, and (ii) different ways of defining temperature must yield the same numerical value locally. It is observed that, while using the kinetic temperature based thermostats, although the velocity distribution agrees locally with the Maxwell-Boltzmann distribution, there is a significant difference between the kinetic and the Rugh's temperature \cite{hoover2008nonequilibrium,hoover2007hamiltonian,hoover_czth}, which is against the postulates of the local thermodynamic equilibrium hypothesis. Molecular dynamics simulations of isothermal Couette flow reveal that a heat flow occurs even in the absence of any temperature gradient when kinetic thermostats are used \cite{baranyai1992isothermal}. To accurately account for the heat flow, one needs to bring in Rugh's temperature \cite{ayton1999validity}. Further, the dynamical properties of a microscopic system subjected to near-equilibrium conditions depend on the definition of temperature being controlled \cite{daivis2012effect,basconi2013effects}. These issues prompted the development of thermostats that controlled Rugh's temperature. One such thermostat is the Patra-Bhattacharya thermostat. 

The Patra-Bhattacharya thermostat uses two thermostat variables -- $\eta$ and $\xi$ -- for separately controlling both momentum and configurational variables in such a manner that the Rugh's temperature, $T_R$, gets controlled. The thermostat algorithm may be developed using the guessing method of Hoover. We highlight the important steps now. Let us begin with the assumption that each thermostat independently couples to the system such that the configurational and momentum variables evolve as follows:
\begin{equation}
\begin{array}{cclcclcclccl}
\dot{x}_i & = & \dfrac{p_i}{m} - \xi \dfrac{\partial \Phi}{\partial x_i}, &
\dot{p}_i & = & -\dfrac{\partial \Phi}{\partial x_i} - \eta p_i, &
\dot{\xi} & = &? &
\dot{\eta} & = &? 
\end{array}
\label{eq:four_one}
\end{equation}
At this stage, the temporal evolution of $\dot{\xi}$ and $\dot{\eta}$ are not known. Since, $\eta$ and $\xi$ are assumed to be two independent reservoirs, they must follow Maxwell-Boltzmann distribution, so that the extended phase-space distribution is given by:
\begin{equation}
f_{\text{ex}} \left(\mathbf{\Gamma} \right) \propto \exp \left({-\beta \left[H + \dfrac{Q_\eta \eta^2}{2} + \dfrac{Q_\xi \xi^2}{2} \right] } \right),
\label{eq:PB_extended_distribution}
\end{equation}
where, $H = \sum p_i^2/2m + \Phi(\mathbf{q})$. The steady state Liouville's equation (\ref{eq:two_thirtysix}) is solved, where $\mathbf{\Gamma} = (\mathbf{x,p},\eta,\xi)$, to obtain the evolution of $\dot{\eta}$ and $\dot{\xi}$. The final equations of motion become:
\begin{equation}
\begin{array}{ccl}
\dot{x}_i & = & \dfrac{p_i}{m} - \xi \dfrac{\partial \Phi}{\partial x_i}, \\
\dot{p}_i & = & -\dfrac{\partial \Phi}{\partial x_i} - \eta p_i, \\
\dot{\xi} & = & \dfrac{1}{Q_\xi} \sum\limits_{i=1}^{3N} \left[ \left(\dfrac{\partial \Phi}{\partial x_i} \right)^2 - k_BT_0 \dfrac{\partial^2 \Phi}{\partial x_i^2} \right], \\
\dot{\eta} & = & \dfrac{1}{Q_\eta} \sum\limits_{i=1}^{3N} \left[ \dfrac{p_i^2}{m} -k_BT_0 \right].
\end{array}
\label{eq:four_twelve}
\end{equation}
Just like most other thermostats mentioned in the previous section, the rate at which a system approaches a canonical ensemble is dependent on $Q_\eta$ and $Q_\xi$ \cite{cho_joannopoulos,tobias_93}. The fluctuation of $T_R$ is significantly influenced by these parameters. It is important to note that for a large-system, $Q_\eta \neq Q_\xi$, since the changes in the configurational variables occur over a longer time-scale than the momentum variables. So depending on the problem being simulated, $Q_\xi$ can be one (or more) order of magnitude smaller than $Q_\eta$. 

The Patra-Bhattacharya thermostat satisfies several properties of a good thermostat -- it is time-reversible, conforms to the different laws of thermodynamics, including the spontaneous flow of heat between a hotter and a colder thermostat, and is simple to implement for systems with pair-wise interacting particles. Apart from giving more consistent results in non-equilibrium thermal conduction, the Patra-Bhattacharya thermostat can be utilized for creating a thermal gradient between the configurational and the kinetic degrees of freedom. Such unique thermostatting capability provides an ability to engineer thermal rectification \cite{patra2016heat}. 

However, for systems with multi-body interactions, or where the analytical form of interaction potential is not readily available, this thermostat is computationally more expensive than the kinetic temperature based thermostats. This is because of the necessity to numerically compute the diagonal elements of the Hessian matrix, $\nabla^2 \Phi$, which requires $O(N^3)$ operations. Like the Hoover-Holian and the MKT thermostat, the Patra-Bhattacharya thermostat does not possess a Hamiltonian, but a pseudo-energy, which is a constant of motion, exists:
\begin{equation}
\begin{array}{rcl}
E_\text{PB} & = & \Phi(\mathbf{x}) + \sum\limits_{i=1}^{3N} \dfrac{p_i^2}{2m} + \dfrac{Q_\eta \eta^2}{2} + \dfrac{Q_\xi \xi^2}{2} \\
& & + k_BT_0 \int\limits_0^t \left( 3N \eta + \sum\limits_{i=1}^{3N} \dfrac{\partial^2 \Phi}{\partial x_i^2} \right) dt
\end{array}
\end{equation}
While the equations of motion (\ref{eq:four_twelve}) do not result in: $\langle \dot{S} \rangle /k_B= \langle \Lambda \rangle$ for non-equlibrium cases with position-dependent $T$, a small modification in the equations of motion can improve the situation.

\subsubsection{Solving the Equations of Motion}
As of now, no symplectic technique has been developed to solve these equations of motion. The equations may be integrated using Gear's predictor-corrector algorithm, which is summarized below:
\begin{enumerate}
    \item Use Taylor's series for predicting the positions, velocities and accelerations of the particles:
    \begin{equation}
        \begin{array}{rcl}
             x_p(t+\Delta t) & = & x(t) + v(t) \Delta t + \dfrac{1}{2} a(t) \Delta t^2 + \dfrac{1}{6} b(t) \Delta t^3 \\
             v_p(t+\Delta t) & = & v(t) + a(t) \Delta t + \dfrac{1}{2} b(t) \Delta t^2 \\
             a_p(t+\Delta t) & = & a(t) + b(t) \Delta t 
        \end{array}
        \nonumber
    \end{equation}
    \item Compute the updated accelerations, $a_n$, based on the predicted positions, $x_p$. Incorporate the effect of $\eta$ on the acceleration: $a_n \to a_n - \eta v_p$.
    \item Compute the updated velocities, $v_n$, through: $v_n = v_p - \xi \dfrac{\partial \Phi}{\partial x_p}$.
    \item The error in acceleration is: $\Delta a = a_n - a_p$. Based on this error, the following corrections are made:
    \begin{equation}
        \begin{array}{rcl}
             x(t+\Delta t) & = & x_p(t+ \Delta t) + c_0 \Delta a \\
             v(t+\Delta t) & = & v_n(t+ \Delta t) + c_1 \Delta a \\
             a(t+\Delta t) & = & a_n(t + \Delta t) \\
             b(t+\Delta t) & = & b(t) + c_2 \Delta a
        \end{array}
    \nonumber
    \end{equation}
    The constants $(c_0,c_1,c_2) = (1/6,5/6,1/3)$.
    \item Update $\eta$ and $\xi$ using the new positions, velocities and accelerations.
\end{enumerate}

\begin{figure}[h]
\includegraphics[scale = 0.45]{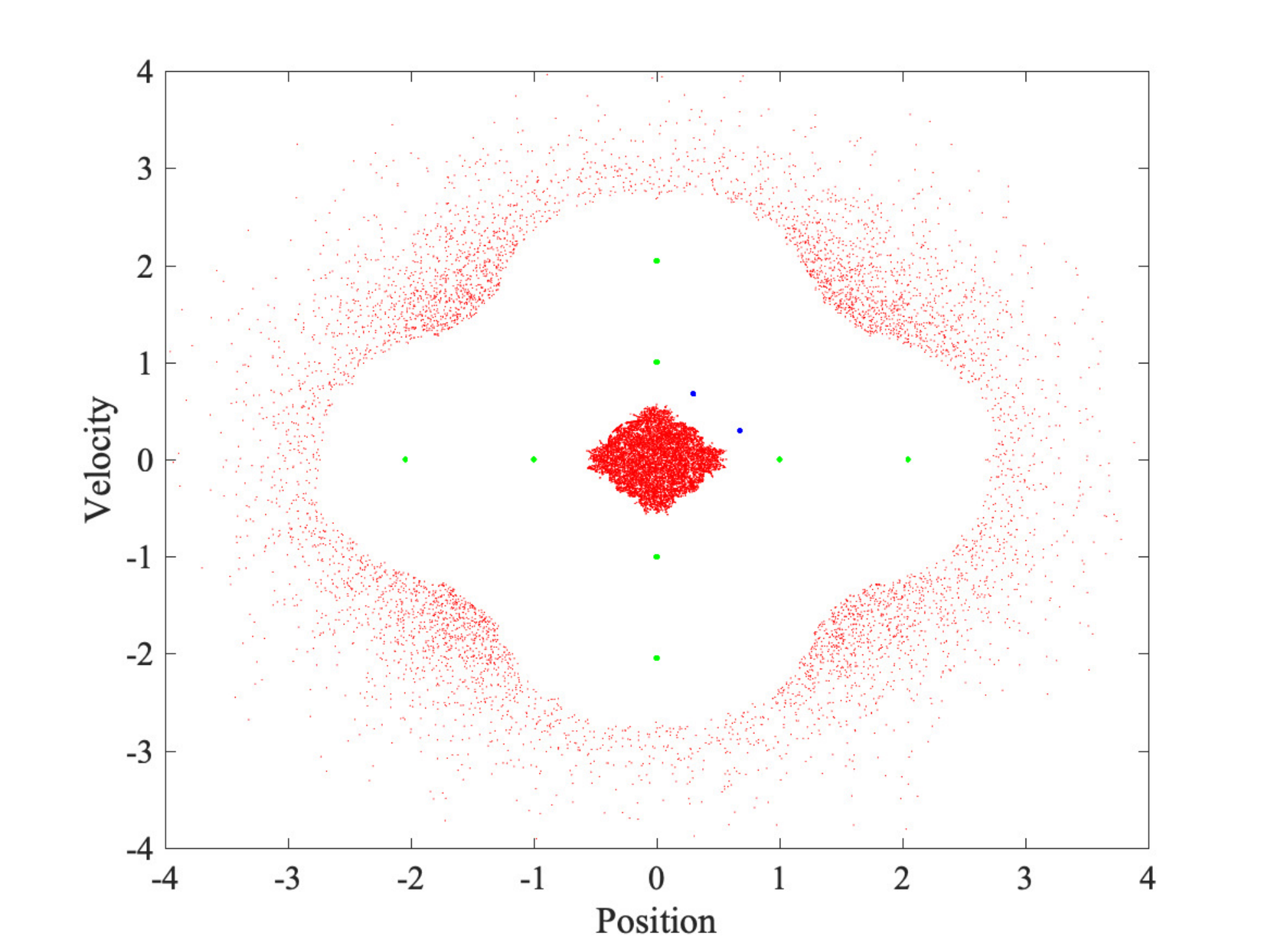}
\caption[CZTH plots]{\label{fig:PB_EQ} Plot of position and velocity of the Patra-Bhattacharya thermostatted single harmonic oscillator at the double Poincar\'e section $|\eta|=|\xi|<0.001$ for three different initial conditions: green -- $(x,p,\eta,\xi) = (1,0,0,0)$, red -- $(x,p,\eta,\xi) = (4,1,0,0)$ and blue -- $(x,p,\eta,\xi) = (0,1,0,1)$. The trajectories corresponding to green and blue colors are such that hyper-dimensional tori are formed, while that corresponding to red is chaotic. However, the entire phase-space is not filled by a single trajectory, and consequently, the dynamics is non-ergodic.}
\end{figure}
\subsubsection{Phase-space characteristics from single harmonic oscillator}
The Patra-Bhattacharya thermostat is non-ergodic. Prior to its development, it was thought that any two-variable thermostat is ergodic. The Patra-Bhattacharya thermostat is the first two-variable thermostat which was shown to be non-ergodic. Let us see the phase-space characteristics by coupling the thermostat with a single harmonic oscillator of unit mass and stiffness constant. If the temperature and thermostat masses are such that $k_B T_0 = 1$ and $Q_\eta = Q_\xi = 1$, the equations of motion are given by:
\begin{equation}
    \dot{x} = p -\xi x, \dot{p} = -x - \eta p, \dot{\eta} = p^2 - 1.0, \dot{\xi} = x^2 - 1.0
\end{equation}
Solving these equations using the $4^{th}$ order Runge-Kutta method shows the presence of hyper-dimensional tori distributed within a chaotic sea. The double Poincar\'e section plot at $|\eta|=|\xi|<0.001$ for three different initial conditions are shown in figure (\ref{fig:PB_EQ}): green -- $(x,p,\eta,\xi) = (1,0,0,0)$, red -- $(x,p,\eta,\xi) = (4,1,0,0)$ and blue -- $(x,p,\eta,\xi) = (0,1,0,1)$. The trajectories corresponding to green and blue dots represent hyper-dimensional tori, while that corresponding to red is chaotic. For the dynamics to be ergodic, the trajectory must sample the phase-space as per the following distribution:
\begin{equation}
f_{\text{ex}} \left(x,p,\eta,\xi \right) \propto e^{-\left[\dfrac{x^2}{2} + \dfrac{p^2}{2}+ \dfrac{\eta^2}{2} + \dfrac{\xi^2}{2} \right] },
\label{eq:PB_canonical_distribution}
\end{equation}
However, it is evident from the three trajectories, that the entire phase-space is not sampled as per the canonical distribution (\ref{eq:PB_canonical_distribution}), and hence, the dynamics is non-ergodic.

\begin{figure}[htp]
\includegraphics[scale = 0.45]{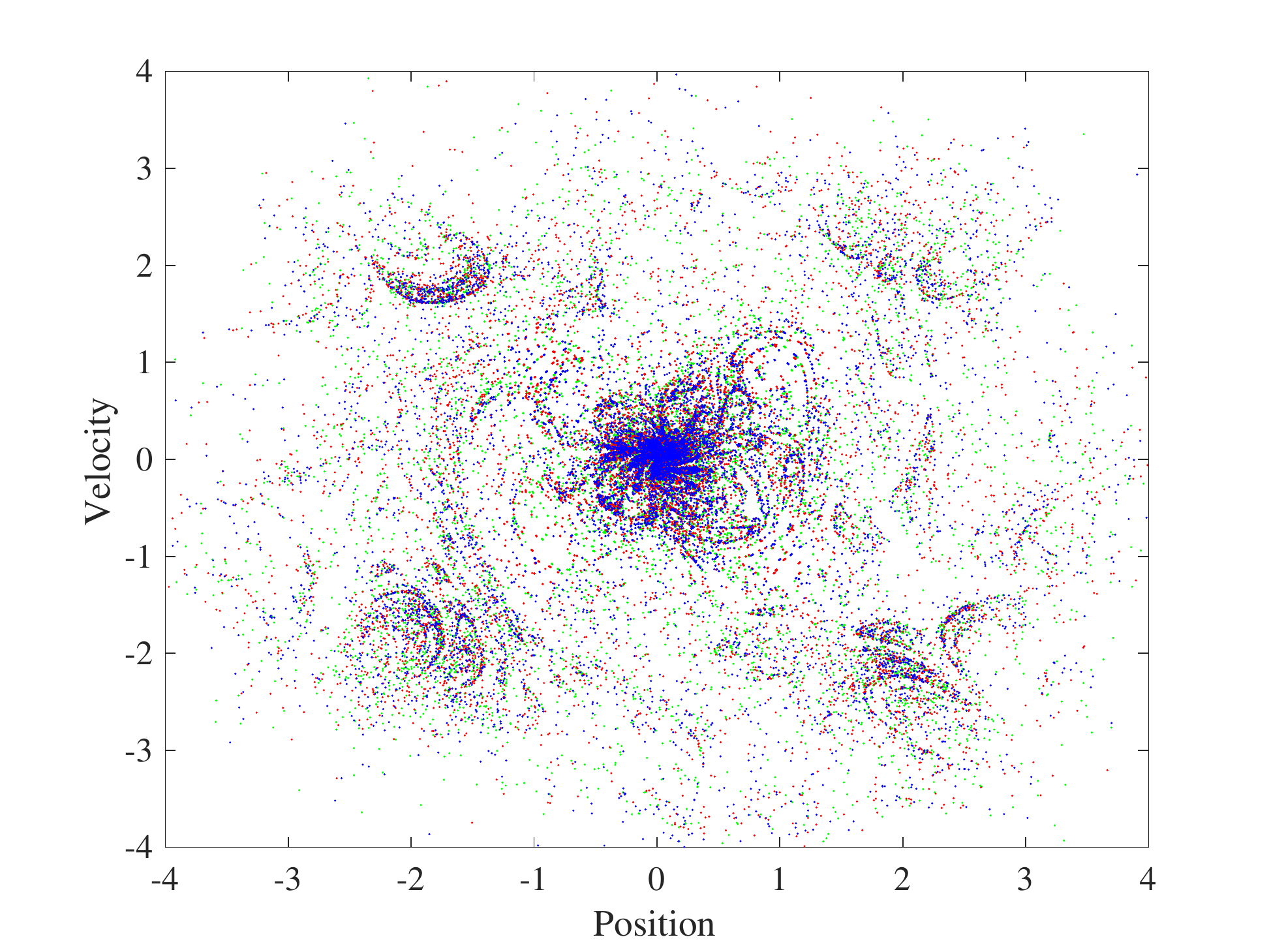}
\caption[CZTH plots]{\label{fig:PB_NEQ} Plot of position and velocity of the Patra-Bhattacharya thermostatted single harmonic oscillator at the double Poincar\'e section $|\eta|=|\xi|<0.001$ for three different initial conditions: green -- $(x,p,\eta,\xi) = (1,0,0,0)$, red -- $(x,p,\eta,\xi) = (4,1,0,0)$ and blue -- $(x,p,\eta,\xi) = (0,1,0,1)$. The multifractal nature of the dynamics is evident suggesting a robust heat flow in the Patra-Bhattacharya thermostatted oscillator.}
\end{figure}

We now subject the oscillator to a position-dependent temperature field: $T(x) = 1.0 + 0.2 \tanh (x)$. For the same initial conditions as that of the equilibrium case, the double Poincar\'e section plot is shown in figure (\ref{fig:PB_NEQ}). The multifractal nature of the dynamics is evident, suggesting a robust heat flow.

The Patra-Bhattacharya thermostat is a particular case of the generalized Baur-Bulgac-Kusnezov thermostat, which we discuss next.

\subsection{Bauer-Bulgac-Kusnezov Thermostat}
Back in the early 1990s, before the concept of Rugh's temperature arose, Bauer, Bulgac and Kusnezov \cite{bbk_pra,bbk_annals} argued that the coupling form, initially chosen in the Guessing Method, for the Nos\'e-Hoover thermostat could be generalized in a manner that the configurational degrees of freedom get modified simultaneously. They sought a better coupling than the Nos\'e-Hoover method for improving its ergodic characteristics. As it turns out, their approach can be used for controlling Rugh's temperature as well.  

Bauer and coworkers improved the Guessing Method for any arbitrary coupling between the configurational and momentum degrees of freedom and the reservoirs -- denoted by $h_1 \left( \upsilon  \right)$ and $h_2 \left( \chi  \right)$. The generalized form of coupling is:
\begin{equation}
\begin{array}{ccl}
 {{{\dot{x}}}_{i}} & = & \dfrac{p_i}{m}-{{h}_{1}} \left( \upsilon  \right)C\left( {{x}_{i}},{{p}_{i}} \right) \\ 
 {{{\dot{p}}}_{i}} & = & - \dfrac{\partial \Phi}{\partial x_i } -{{h}_{2}}\left( \chi  \right)D\left( {{x}_{i}},{{p}_{i}} \right) \\ 
\end{array}
\label{eq:two_bbk_one}
\end{equation}
Here, $C(x_i,p_i)$ and $D(x_i,p_i)$ are arbitrary phase functions. The temporal evolution of $\upsilon$ and $\chi$, the thermostat variables, can be obtained as:
\begin{equation}
\begin{array}{rcl}
\dot{\chi } & = & \dfrac{1}{{{Q}_{\chi }}}\sum\limits_{i=1}^{3N}\left[ {\dfrac{{{p}_{i}}}{m}D\left( {{x}_{i}},{{p}_{i}} \right)-{{k}_{B}}T_0\dfrac{\partial D\left( {{x}_{i}},{{p}_{i}} \right)}{\partial {{p}_{i}}}} \right] \\ 
\dot{\upsilon } & = & \dfrac{1}{{{Q}_{\upsilon }}}\sum\limits_{i=1}^{3N}\left[ {\dfrac{\partial \Phi }{\partial {{x}_{i}}}C\left( {{x}_{i}},{{p}_{i}} \right)-{{k}_{B}}T_0\dfrac{\partial C\left( {{x}_{i}},{{p}_{i}} \right)}{\partial {{x}_{i}}}} \right],
\end{array}
\label{eq:two_bbk_two}
\end{equation}
where, $Q_\upsilon$ and $Q_\chi$ are the masses associated with the reservoir variables $\upsilon$ and $\chi$. A family of thermostat schemes may be obtained by appropriately selecting the coupling functions $h_1$ and $h_2$ along with $C(\ldots)$ and $D(\ldots)$. For example, the Nos\'e-Hoover thermostat is obtained if: $h_1(\upsilon) = \dot{\upsilon} = 0$, $h_2(\chi) = \eta$ and $D(x_i,p_i) = p_i$. The Patra-Bhattacharya thermostat is obtained if $h_1(\upsilon) = \xi,h_2(\chi) = \eta, C(x_i,p_i) = \partial \Phi/\partial x_i$ and $D(x_i,p_i) = p_i$. 

\begin{figure}[htp]
\includegraphics[scale = 0.45]{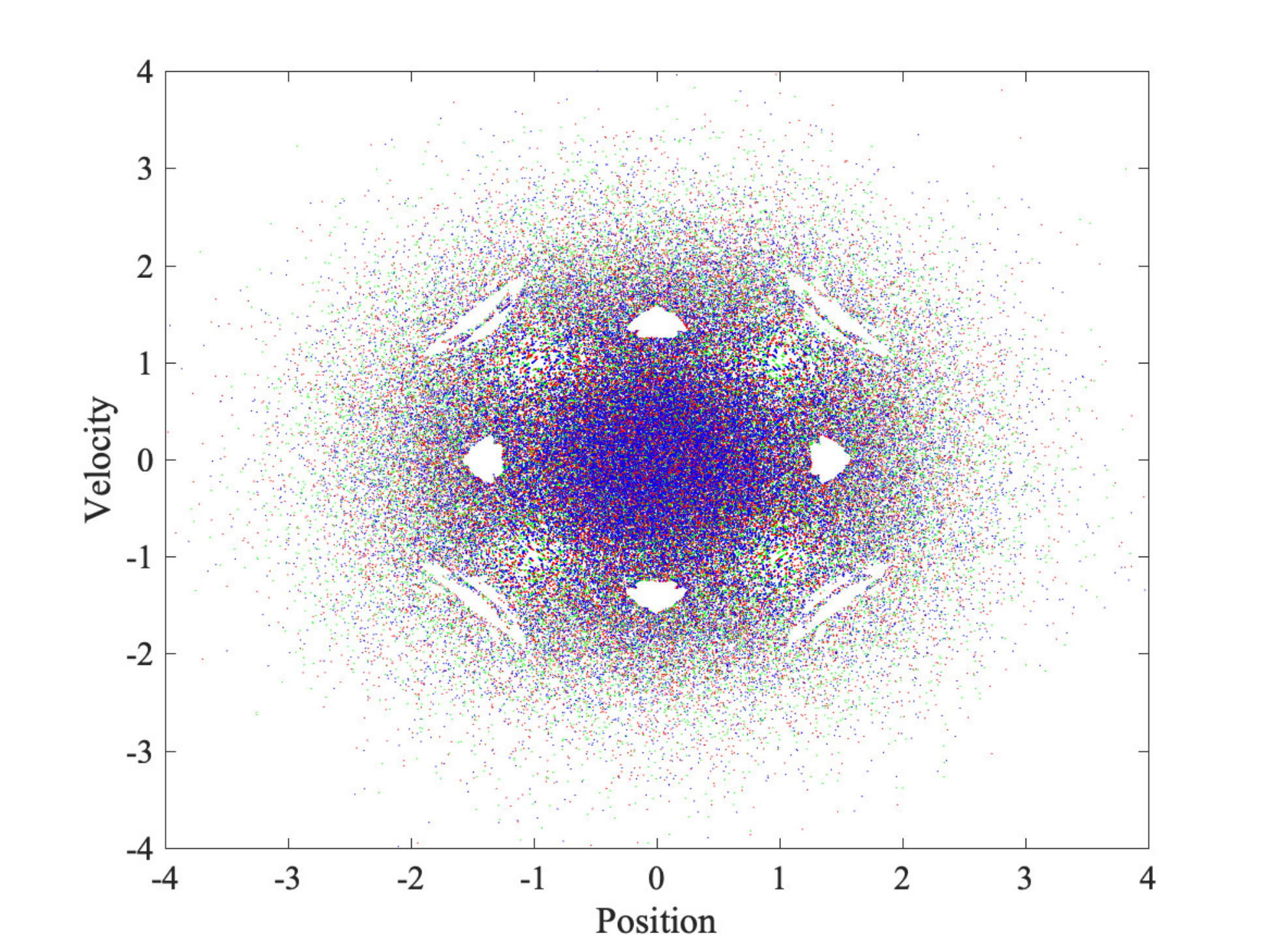}
\caption[CZTH plots]{\label{fig:BBK_EQ} Plot of position and velocity for the Bauer-Bulgac-Kusnezov thermostatted single harmonic oscillator at the double Poincar\'e section $|\upsilon|=|\chi|<0.001$ for three different initial conditions: green -- $(x,p,\chi,\upsilon) = (1,0,0,0)$, red -- $(x,p,\chi,\upsilon) = (4,1,0,0)$ and blue -- $(x,p,\chi,\upsilon) = (0,1,0,1)$. Although all trajectories are chaotic, the entire phase-space is not filled by a single trajectory, and consequently, the dynamics is non-ergodic. Note, however, that the ergodic characteristics have shown a marked improvement over the Patra-Bhattacharya thermostat for the same initial conditions.}
\end{figure}

The benefit of this thermostat is that the equations of motion have improved ergodic characteristics than either the Nos\'e-Hoover dynamics or the Patra-Bhattacharya dynamics. For example, when a single harmonic oscillator, of unit mass and stiffness constant, is coupled to this thermostat using cubic coupling, the equations of motion become:
\begin{equation}
\begin{array}{rcl}
\dot{x}=p-\upsilon^3 x, & \dot{p}=-x-{{\chi }^{3}}p, \\
\dot{\chi }=\dfrac{1}{{{Q}_{\chi }}}\left[ p^2 - k_BT_0 \right], & \dot{\upsilon }=\dfrac{1}{{{Q}_{\upsilon }}}\left[ x^2 - k_BT_0 \right].
\end{array}
\label{eq:two_bbk_three}
\end{equation}
When these equations are solved using the $4^{th}$ order Runge-Kutta method with $Q_\upsilon = Q_\chi = k_BT_0 = 1$, the ergodic characteristics show marked improvement over the Patra-Bhattacharya thermostat, as shown in figure (\ref{fig:BBK_EQ}). For the same initial conditions chosen for the Patra-Bhattacharya thermostat, the Bauer-Bulgac-Kusnezov thermostatted oscillator has the ability to sample more allowable phase-space as the cubic nature of coupling aids phase-space mixing.
Note that the extended phase-space density for this specific coupling type is given by:
\begin{equation}
    f_\text{ex} \propto \exp \left[-\left(\dfrac{p^2}{2} + \dfrac{x^2}{2} + \dfrac{\upsilon^4}{4} + \dfrac{\chi^4}{4}\right)\right]
\end{equation}

The Rugh's temperature may be controlled with the following choice of the variables $C$ and $D$:
\begin{equation}
C(x_i,p_i) = \dfrac{\partial \Phi}{\partial x_i} \text{ and } D(x_i,p_i) = p_i
\end{equation}
The Bauer-Bulgac-Kusnezov thermostat permits coupling of an arbitrarily higher order, but that comes at a price -- the equations of motion become stiff. They require very small time steps for an accurate solution. Overall, this thermostat has only been used for pedagogical purposes, despite it being time-reversible and showing improved ergodic characteristics. To the best of our knowledge, no generalized symplectic algorithm exists for it. Further, no Hamiltonian function is known to exist as well. However, being a part of the Nos\'e-Hoover family, we think that it satisfies the different laws of thermodynamics, and is able to allow heat flow from a hotter thermostat to a colder thermostat spontaneously.

\section{Configurational Temperature and Its Control}
The generalized temperature-curvature relationship, shown in equation (\ref{eq:two_eleven}), opened up the possibility of defining the temperature of a microscopic system solely in terms of the configurational variables of the particles. If $B$ in equation (\ref{eq:two_eleven}) is chosen such that it is a scalar functional of only the particles' coordinates and not momenta, the temperature of the system depends only on the microscopic configuration of the particles. Hence, this temperature is called the configurational temperature. Perhaps, the most common way of defining configurational temperature, $T_C$, is obtained by choosing $B = \Phi(\mathbf{x})$ \cite{butler_98,jepps_00,lue_configT}:
\begin{equation}
\dfrac{1}{k_BT_C} = \dfrac{ \langle \nabla^2 \Phi \rangle}{\langle (\nabla \Phi)^2 \rangle},
\label{eq:two_thirteen}
\end{equation}
Remarkably, the same expression appears, almost half a century ago, in Landau and Lifshitz' textbook on statistical physics \cite{landau2013course}, where it is shown that $T_C$ can be obtained through a single integration by parts of the expression:
\begin{equation}
\langle \nabla_{x}^2H \rangle = \dfrac{\langle \left(\nabla_{x}H \right)^2 \rangle }{k_BT_C}.
\end{equation}
Here, the averaging is performed with respect to the canonical distribution.

Other common ways of defining configurational temperature draw inspiration from higher order measures of kinetic temperature. If we take $ B = \Phi^{2}$ in the generalized temperature-curvature relationship we obtain the second order configurational temperature:
\begin{equation}
\dfrac{1}{k_BT_{C,2}} = \dfrac{ \langle \Phi \nabla^2 \Phi + \left \|(\nabla \Phi)^2 \right \|\rangle}{\langle \Phi \left \|(\nabla \Phi)^2 \right \|\rangle},
\label{eq:SO_Tcon}
\end{equation}
Likewise, one can obtain even higher order measures of configurational temperature. 

Configurational temperature is especially suitable for non-equilibrium problems. Take the case of a microscopic system undergoing shock loading. The equipartition theorem breaks down here since the motion of the particles is much faster in the direction of shockwave propagation than in the other two directions \cite{zhao_KT,he_KT}. The kinetic temperature, as a result, is no longer a scalar quantity but behaves like a tensor, which is difficult to reconcile with the thermodynamic definition of temperature. Using configurational temperature, on the other hand, bypasses this problem. Since the equipartition theorem is not necessary for defining configurational temperature, $T_C$ can be used for temperature computation in problems of this kind. It must be noted that in strongly non-equilibrium problems, $T_k \neq T_R \neq T_C$, and it is not known which of the three gives a closer approximation to ``reality''. Using $T_C$ over $T_k$ is advantageous in another class of non-equilibrium problems -- shear flow. Here, the streaming velocity must be known beforehand for calculating the peculiar kinetic energy, which serves as an input to $T_k$. The inability of correctly determining the streaming velocity leads to several problems like stabilization of string phases \cite{erpenbeck1984shear,evans1986shear,evans1992conditions}, creation of anti-symmetric stress components \cite{travis1995thermostats}, etc. Using $T_C$ is also advantageous in thermostatting biological molecules, which are usually geometrically constrained, and more often than not, comprise several non-translational degrees of freedom, like planar rotation, bond rotation, etc. Thermostatting the three translation degrees of freedom, as is done for $T_k$, may not be sufficient \cite{braga_05}.

However, calculating $T_C$ is full of difficulties in systems where the particles are non-interacting or interact negligibly, such as in perfect gases. In these cases, both the numerator, $\langle \left( \nabla \Phi \right)^2 \rangle$) and the denominator, $\langle \nabla^2 \Phi \rangle$, of $T_C \to 0$, making $T_C$ indeterminate. The Rugh's temperature, $T_R$, as well the kinetic temperature, $T_k$, in the limit of perfect gases, are finite, though. Note that due to the requirement of numerically computing the diagonal elements of the Hessian matrix, determining $T_C$ is a computational nightmare for systems where inter-particle potential comprises multi-body interactions or is non-analytical, similar to $T_R$.

To the best of our knowledge, three separate configurational temperature control algorithms have been developed. Of the three, we discuss two algorithms and omit the thermostat by Delhommelle and Evans \cite{delhommelle2001configurational} which involves spatial gradient of $T_C$ making the equations stiff in some situations. The remaining two approaches are discussed next. 

\subsection{Braga-Travis Thermostat}
Configurational temperature control is necessary for situations where the kinetic temperature control results in spurious results such as, non-equilibrium molecular dynamics simulations of certain heat-driven processes \cite{erpenbeck_84,evans_86} and flowing systems with spatial and possibly time-varying streaming velocity. In non-equilibrium problems of shockwave propagation, where the kinetic temperature in one direction is significantly higher than the rest two \cite{zhao_KT,he_KT}, thermostatting the entire simulation domain at a fixed kinetic temperature may not be correct. In these situations, controlling the configurational temperature is superior over controlling the kinetic temperature. While in equilibrium, controlling any one kind of temperature necessary implies an automatic control of all other measures of temperature, in non-equilibrium, there is no theory which suggests that the different measures of temperature should agree with each other. 

For resolving these issues, the first configurational thermostat was developed on the lines of the Gaussian isokinetic thermostat \cite{config_thermostat_01,config_thermostat_02}. Gauss' principle of least constraint was used with the holonomic constraint (without the average) shown in equation (\ref{eq:two_thirteen}). However, the utility of this thermostat is found to be low since the final equations of motion contain terms involving the third-order derivative of $\Phi(\mathbf{x})$, which is computationally very difficult to obtain. The equations of motion are stiff as well. In a bid to alleviate these problems, a separate configurational thermostat was created for thermostatting the \textit{slow configurational} variables selectively \cite{config_thermostat_03} through the Smoluchowski equation. Under the assumption that the momentum variables relax much faster than the configurational variables, the momentum evolution equations were completely dropped. 

Braga and Travis \cite{braga_05,bt_thermostat_review,bt_thermostat_pressure} adopted a different approach for thermostatting the configurational degrees of freedom. Their approach is similar to the Nos\'e-Hoover thermostat, and can be derived using the Guessing Method of Hoover. Consider the following two equations:
\begin{equation}
\begin{array}{ccl}
{{\dot x}_i} & = & \dfrac{{{p_i}}}{{{m_i}}} - \xi \dfrac{{\partial \Phi(\mathbf{x})}}{{\partial {x_i}}}\\
{{\dot p}_i} & = &  - \dfrac{{\partial \Phi(\mathbf{x})}}{{\partial {x_i}}} \\
\dot{\xi} & = & ?
\end{array}
\label{eq:two_bt_one}
\end{equation}
Here, the variable $\xi$ represents the configurational heat reservoir variable, whose evolution at this stage is unknown. $\dot{\xi}$ is obtained such that a canonical distribution is sampled by the dynamics. Since $\xi$ itself must be canonically distributed, the extended phase-space distribution function of the microscopic system and the reservoir is given by:
\begin{equation}
f_{\text{ex}} \left(\mathbf{x,p},\xi \right) \propto e^{-\beta \left[H + \dfrac{Q_\xi \xi^2}{2} \right] }.
\label{eq:BT_extended_distribution}
\end{equation}
Further, the evolution equations must satisfy the steady-state Liouville's equation (\ref{eq:two_thirtysix}) with $\mathbf{\Gamma} = (\mathbf{x,p},\xi)$, which upon solving, gives the evolution of $\dot{\xi}$. The final equations of motion become:
\begin{equation}
\begin{array}{ccl}
\dot{x}_i & = & \dfrac{p_i}{m_i} - \xi \dfrac{\partial \Phi}{\partial x_i}, \\
\dot{p}_i & = & -\dfrac{\partial \Phi}{\partial x_i}  \\
\dot{\xi} & = & \dfrac{1}{Q_\xi} \sum\limits_{i=1}^{3N} \left[ \left(\dfrac{\partial \Phi}{\partial x_i} \right)^2 - k_BT_0 \dfrac{\partial^2 \Phi}{\partial x_i^2} \right], \\
\end{array}
\label{eq:BT_EoM}
\end{equation}
In equilibrium, where $\langle \xi \rangle = 0$, $\langle \dot{\xi} \rangle$ must also equal zero, from which it is trivial to show that the desired temperature, $k_BT_0$, equals the configurational temperature $T_C$. 

The Braga-Travis equations of motion are time-reversible, easy to implement for microscopic systems with pair-wise interaction potential and satisfy the Zeroth Law and the Second Law of thermodynamics. These equations, however, do not have any Hamiltonian associated with them and are non-ergodic. The Braga-Travis thermostat has seen many improvements over the years, and now, several variants exist, such as, the thermostat which includes bond-length constraints \cite{bt_thermostat_pressure} suited for long-chain molecules, a stochastic analog \cite{samoletov_notes} with improved ergodic properties, and the Braga-Travis chain thermostat which is the configurational counterpart of the Nos\'e-Hoover chain thermostat \cite{beckedahl2016configurational}.

\subsubsection{Solving the Equations of Motion}
Alternatively, one can use Trotter's expansion to come up with a different symplectic integrator: 
\begin{enumerate}
    \item Initialize the system with suitable initial conditions of positions and momenta, and $\xi = 0$.
    \item Run in a loop until the desired time, $t$, is reached with an incremental time step of $\Delta t$:
    \begin{enumerate}
        \item Compute $G = \dfrac{1}{Q_\xi} \sum\limits_{i=1}^{3N} \left[ \left(\dfrac{\partial \Phi}{\partial x_i} \right)^2 - k_BT_0 \dfrac{\partial^2 \Phi}{\partial x_i^2} \right]$
        \item Propagate $\xi$ by $\Delta t /2$: $\xi \to \xi + G \Delta t/2$.
        \item Propagate all momenta by $\Delta t/2$: $p_i \to p_i + F_i \Delta t/2$ where $ F_i = -\dfrac{\partial \Phi}{\partial x_i}$.
        \item Propagate all positions by $ \Delta t$:
        \begin{equation}
             x_i \to x_i e^{-k \Delta t \xi}  + p_i \Delta t e^{-\frac{k\xi \Delta t }{2}} \left[ \dfrac{\sinh\left(\Delta t \dfrac{k \xi}{2}\right)}{\Delta t \dfrac{k \xi}{2}} \right].
             \nonumber
        \end{equation}
             This propagation assumes that in the vicinity of the position $x_i$, the conservative force field may be approximated using the second order terms of the Taylor's series expansion, so that the potential energy of any particle may be written as $ \Phi = (1/2)kx_{i}^{2}$; $k$ is a constant. 
        \item Propagate all momenta by $\Delta t/2$: $p_i \to p_i + F_i \Delta t/2$ where $ F_i = -\dfrac{\partial \Phi}{\partial x_i}$.
	    \item Compute $G = \dfrac{1}{Q_\xi} \sum\limits_{i=1}^{3N} \left[ \left(\dfrac{\partial \Phi}{\partial x_i} \right)^2 - k_BT_0 \dfrac{\partial^2 \Phi}{\partial x_i^2} \right]$
        \item Propagate $\xi$ by $\Delta t /2$: $\xi \to \xi + G \Delta t/2$.
    \end{enumerate}
\end{enumerate}

\subsubsection{Phase space characteristics using Single Harmonic Oscillator}
\begin{figure}
\includegraphics[scale=0.45]{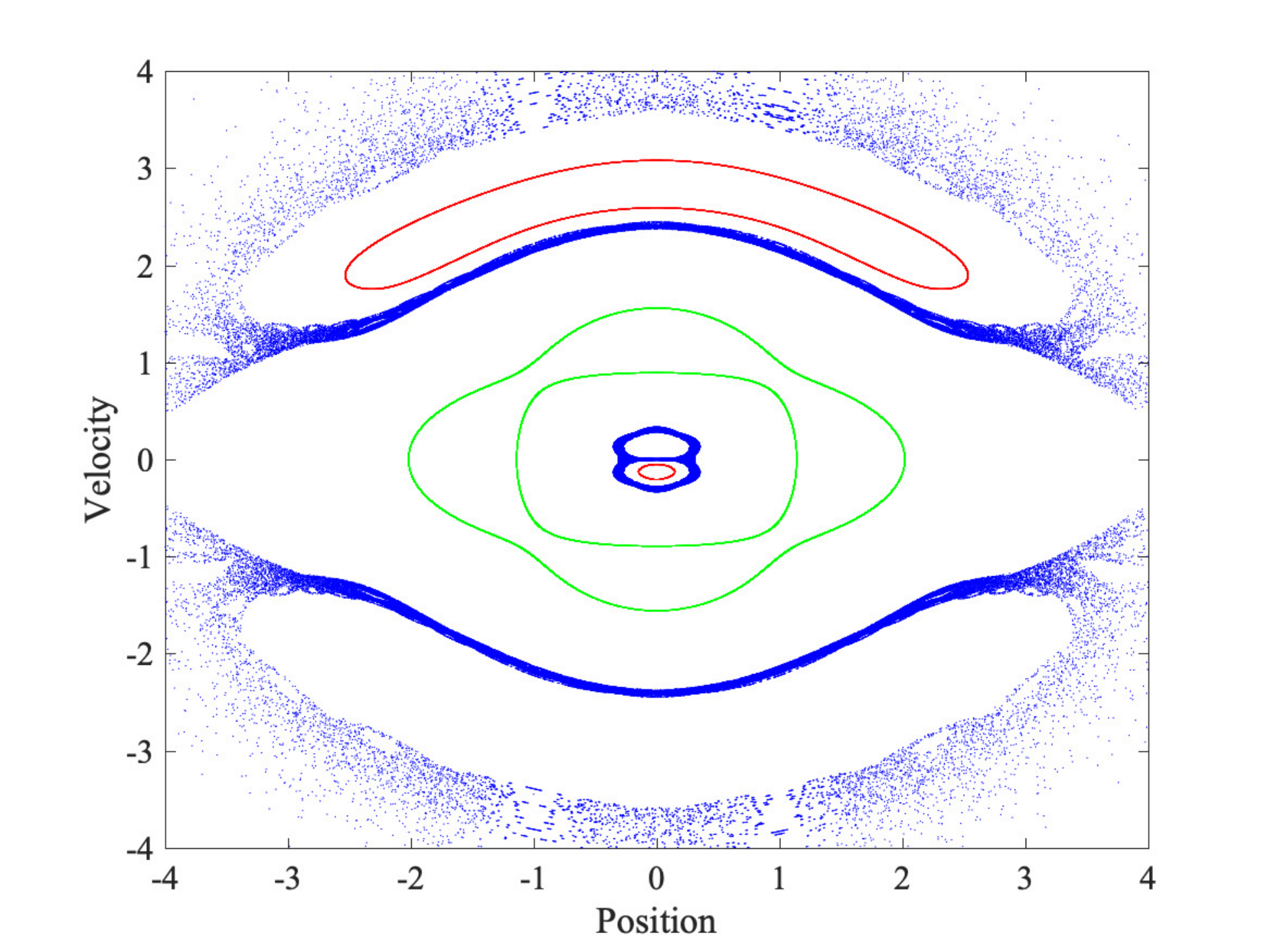}
\caption[Phase space plots for BT dynamics]{\label{fig:BT_EQ_phase_plot} Poincar\'e section plots for the Braga-Travis dynamics at $|\xi| < 0.001$ cross section for three different initial conditions: (a) green = $(x,p,\eta) = (1,1,0)$, (b) red = $(x,p,\eta) = (2,2,1)$, and (c) blue = $(x,p,\eta) = (3,3,3)$. The different initial conditions result in different nature of trajectories, with none being phase-space filling. These figures are reminiscent of the Nos\'e-Hoover dynamics with $x$ and $p$ interchanged. Like the Nos\'e-Hoover thermostat, the lack of ergodicity, and consequently the inability of Braga-Travis thermostat to thermalize the single harmonic oscillator is self evident.}
\end{figure}

When a single harmonic oscillator of unit mass and stiffness, is coupled with the Braga-Travis thermostat kept at $k_BT_0 = 1$, the equations of motion become:
\begin{equation}
\begin{array}{ccl}
\dot{x} & = & p - \xi x, \\
\dot{p} & = & -x \\
\dot{\xi} & = & \dfrac{1}{Q_\xi} \left[ x^2 - 1\right]
\end{array}
\label{eq:BT_SHO}
\end{equation}
These equations of motion bear stark resemblance with the Nos\'e-Hoover thermostatted single harmonic oscillator (having $x$ and $p$ interchanged). So, similar to the Nos\'e-Hoover thermostat, the Braga-Travis thermostat displays non-ergodicity. The Poincar\'e section plot at $|\xi|<0.001$ with $Q_\xi = 1$ is shown for three initial conditions in the figure (\ref{fig:BT_EQ_phase_plot}). The Poincar\'e section plots are similar to that of the Nos\'e-Hoover thermostat, with $x$ and $p$ interchanged.
For the single harmonic oscillator, if Braga-Travis thermostat were ergodic, then the dynamics would have sampled the phase-space according to the following distribution:
\begin{equation}
f_{\text{ex}} \left(x,p,\xi \right) \propto \exp \left[ - \left( \dfrac{p^2}{2} + \dfrac{x^2}{2} + \dfrac{\xi^2}{2} \right)\right].
\label{eq:BT_extended_distribution_SHO}
\end{equation}
From the Poincar\'e section plots, it is quite clear that the phase-space is not sampled by the Braga-Travis thermostat as per the equation (\ref{eq:BT_extended_distribution_SHO}). Nevertheless, Braga-Travis thermostat is gaining popularity for simulating non-equilibrium situations described before, and if the problem being studied is a large system $N \gg 1$, then the issue of large Poincar\'e recurrence time overtakes the issue of non-ergodicity of the dynamics. 

Due to the similarity of the phase-space plots with the Nos\'e-Hoover thermostatted oscillator, we do not subject the oscillator coupled with the Braga-Travis thermostat to the non-equilibrium position dependent temperature field.

\subsection{$C_{1,2}$ Thermostat}
The $C_{1,2}$ thermostat was proposed to improve the ergodic characteristics of the Braga-Travis thermostat \cite{patra2015ergodic}. It is an extension of the Braga-Travis thermostat for simultaneously controlling the first two orders of configurational temperature by means of two thermostat variables. The algorithm is similar to the Hoover-Holian thermostat, where two thermostat variables are employed for controlling the first two moments of kinetic energy. Controlling the second-order configurational temperature, $T_{C,2}$, shown in equation (\ref{eq:SO_Tcon}), along with the first order configurational temperature, $T_C$, shown in equation (\ref{eq:two_thirteen}) removes the errors associated with the incorrect sampling of the phase-space especially in very small-scale systems.

This thermostat can also be developed using Hoover's Guessing Method. Consider two independent heat reservoirs, one controlling $T_C$ and the other controlling $T_{C,2}$. Let the effect of these heat reservoirs be represented through the variables $\xi$ and $\eta$, respectively. We assume that the thermostats are coupled to the system in such a manner that the temporal evolution of positions and momenta are governed by:
\begin{equation}
\begin{array}{ccl}
{{\dot x}_i} & = & \dfrac{{{p_i}}}{{{m_i}}} - \xi \dfrac{{\partial \Phi(\mathbf{x})}}{{\partial {x_i}}} - 2\eta \Phi(\mathbf{x}) \dfrac{\partial \Phi(\mathbf{x})}{\partial x_i}, \\
{{\dot p}_i} & = &  - \dfrac{{\partial \Phi(\mathbf{x})}}{{\partial {x_i}}}.
\end{array}
\label{eq:c12_eom1}
\end{equation}
The extended phase-space dynamics satisfies the following extended distribution:
\begin{equation}
f_{\text{ex}} \left(\mathbf{x,p},\eta,\xi \right) \propto e^{-\beta \left[H + \dfrac{Q_\eta \eta^2}{2} + \dfrac{Q_\xi \xi^2}{2} \right] },
\label{eq:C12_extended_distribution}
\end{equation}
where, $H = \sum p_i^2/2m + \Phi(\mathbf{x})$. The steady state Liouville's equation (\ref{eq:two_thirtysix}) is solved, where $\mathbf{\Gamma} = (\mathbf{x,p},\eta,\xi)$, in conjunction with the extended phase-space distribution to obtain the evolution of $\dot{\eta}$ and $\dot{\xi}$. The final equations of motion become:
\begin{equation}
\begin{array}{ccl}
\dot{x}_i & = & \dfrac{p_i}{m_i} - \xi \dfrac{\partial \Phi}{\partial x_i} - 2 \eta \Phi \dfrac{\partial \Phi}{\partial x_i}, \\
\dot{p}_i & = & -\dfrac{\partial \Phi}{\partial x_i}, \\
\dot{\xi} & = & \dfrac{1}{Q_\xi} \sum\limits_{i=1}^{3N} \left[ \left(\dfrac{\partial \Phi}{\partial x_i} \right)^2 - k_BT_0 \dfrac{\partial^2 \Phi}{\partial x_i^2} \right], \\
\dot{\eta} & = & \dfrac{1}{Q_\eta} \sum\limits_{i=1}^{3N} \left[ \Phi \left(\dfrac{\partial \Phi}{\partial x_i} \right)^2 - k_BT_0 \left( \Phi \dfrac{\partial^2 \Phi}{\partial x_i^2} + \left(\dfrac{\partial \Phi}{\partial x_i} \right)^2 \right) \right].
\end{array}
\label{eq:C12_EOM}
\end{equation}
The pseudo energy, in this case is given by:
\begin{equation}
\begin{array}{rcl}
E_{C_{12}} & = & \Phi(\mathbf{x}) + \sum\limits_{i=1}^{3N}\dfrac{p_i^2}{2m} + Q_\xi \dfrac{\xi^2}{2} + Q_\eta \dfrac{\eta^2}{2} \\
& & + \sum\limits_{i=1}^{3N} \left[ \eta \left(\Phi + k_BT_0 \right) \left(\dfrac{\partial \Phi}{\partial x_i} \right)^2 + \xi k_BT_0 \dfrac{\partial^2 \Phi}{\partial x_i^2} \right. \\
& & \left. + \eta k_B T_0 \left(\Phi + \dfrac{\partial^2 \Phi}{\partial x_i^2} \right) \right]
\end{array}
\label{eq:C12_Constant_Motion}
\end{equation}
Although, the $C_{12}$ thermostat improves upon the ergodic properties vis-\'a-vis the Braga-Travis thermostat, it suffers from the same problem as that of the latter when it comes to numerical implementation -- the $C_{12}$ thermostat requires the computation of the diagonal elements of the Hessian matrix, which is often not directly available analytically and obtaining it numerically is a computationally exhaustive task. Like the Braga-Travis thermostat, the $C_{12}$ equations of motion cannot be obtained from  Hamiltonian basis. The $C_{12}$ thermostat can be extended for controlling the first three orders of configurational temperature, however, the resulting equations of motion are stiff.

\subsubsection{Solving the Equations of Motion}
No symplectic algorithm has been proposed so far for this thermostat, and as a result, the equations are motion are usually integrated using either Predictor-Corrector algorithm (for multi-particle system) or the Runge-Kutta method (for very small-scale systems). We highlight a possible implementation using the Gear's Predictor-Corrector algorithm:
\begin{enumerate}
    \item Use Taylor's series for predicting the positions, velocities and accelerations of the particles:
    \begin{equation}
        \begin{array}{rcl}
             x_p(t+\Delta t) & = & x(t) + v(t) \Delta t + \dfrac{1}{2} a(t) \Delta t^2 + \dfrac{1}{6} b(t) \Delta t^3 \\
             v_p(t+\Delta t) & = & v(t) + a(t) \Delta t + \dfrac{1}{2} b(t) \Delta t^2 \\
             a_p(t+\Delta t) & = & a(t) + b(t) \Delta t 
        \end{array}
        \nonumber
    \end{equation}
    \item Compute the updated accelerations, $a_n$, based on the predicted positions, $x_p$. 
    \item Compute the updated velocities, $v_n$, through: $v_n = v_p - \xi \dfrac{\partial \Phi}{\partial x_p} - 2 \eta \Phi \dfrac{\partial \Phi}{\partial x_i}$.
    \item The error in acceleration is: $\Delta a = a_n - a_p$. Based on this error, the following corrections are made:
    \begin{equation}
        \begin{array}{rcl}
             x(t+\Delta t) & = & x_p(t+ \Delta t) + c_0 \Delta a \\
             v(t+\Delta t) & = & v_n(t+ \Delta t) + c_1 \Delta a \\
             a(t+\Delta t) & = & a_n(t + \Delta t) \\
             b(t+\Delta t) & = & b(t) + c_2 \Delta a
        \end{array}
    \nonumber
    \end{equation}
    The constants $(c_0,c_1,c_2) = (1/6,5/6,1/3)$.
    \item Update $\eta$ and $\xi$ using the new positions, velocities and accelerations.
\end{enumerate}

\subsubsection{Phase-Space Characteristics using a Single Harmonic Oscillator}
A single harmonic oscillator of unit mass and stiffness is coupled with the $C_{12}$ thermostat to assess the ergodic properties. The relevant equations of motion are:
\begin{equation}
\begin{array}{rcl}
\dot{x} & = & p - \eta x - \xi x^3\\
\dot{p} & = & -x \\
\dot{\eta} & = & \dfrac{1}{Q_\eta} \left( x^2 - k_BT_0 \right) \\
\dot{\xi} & = & \dfrac{1}{Q_\xi} \left( x^4 - 3 k_BT_0 x^2 \right),
\end{array}
\label{eq:C12_SHO}
\end{equation}
Assuming, $Q_\eta = Q_\xi = k_B T_0 = 1$, the dynamics bears extreme resemblance with the Hoover-Holian thermostat. The dynamics is, therefore, expected to have similar ergodic properties like the Hoover-Holian thermostat as well. 

\section{Virial Temperature and Its Control}
Long before the generalized temperature-curvature relationship came into the picture, the virial theorem provided the only means for relating the kinetic energy of the system with the potential energy. The virial theorem has been proved both in the framework of the classical thermodynamics and statistical mechanics \cite{landau2013course}. The validity of the virial theorem does not necessitate the ergodicity of the underlying dynamics, so long as the connection of kinetic energy with temperature is not invoked. Mathematically, the virial theorem can be expressed as:
\begin{equation}
\left\langle \sum{{{x}_{i}}\frac{\partial \Phi }{\partial {{x}_{i}}}} \right\rangle_t =\left\langle \sum{\frac{p_{i}^{2}}{m}} \right\rangle_t
\label{eq:two_vt_one}
\end{equation}
where, $\langle \ldots \rangle_t$ denotes time average. Note that the expression (\ref{eq:two_vt_one}) remains valid in equilibrium as well as in non-equilibrium steady states. If one now brings equipartition theorem into the picture to relate the kinetic energy of the system with the temperature, which necessary involves ergodicity, the usual expression of the virial theorem may be rewritten as:
\begin{equation}
\left\langle \sum{{{x}_{i}}\frac{\partial \Phi }{\partial {{x}_{i}}}} \right\rangle =3N{{k}_{B}}T_0
\label{eq:two_vt_two}
\end{equation}
Thus, it becomes evident that one can control the kinetic temperature indirectly through the Clausius’ Virial theorem. Similar expression for temperature may be obtained from the generalized temperature-curvature relationship, (\ref{eq:two_eleven}), with $B = \sum{x_{i}^{2}/2}$:
\begin{equation}
\begin{array}{rcl}
{{k}_{B}}T_0 & = & \dfrac{\left\langle \sum{{{x}_{i}}\dfrac{\partial \Phi }{\partial {{x}_{i}}}} \right\rangle }{\left\langle 3N \right\rangle } \\
\Rightarrow \left\langle \sum{{{x}_{i}}\dfrac{\partial \Phi }{\partial {{x}_{i}}}} \right\rangle & = & 3N{{k}_{B}}T_0.
\end{array}
\label{eq:two_vt_three}
\end{equation}
From both approaches, it is evident that the mathematical expression of temperature is devoid of any momentum terms. In other words, the virial temperature may be treated as a special case of the configurational temperature. 

First attempts of controlling the Virial temperature began in the early 1990s \cite{virial_thermostat,virial_thermostat_2}. One may use Hoover's Guessing method to obtain the following equations of motion:
\begin{equation}
{{\dot{x}}_{i}}=\frac{{{p}_{i}}}{m}-\eta {{x}_{i}},{{\dot{p}}_{i}}=-\frac{\partial \Phi }{\partial {{x}_{i}}},\dot{\eta }=\frac{1}{{{Q}_{\eta }}}\left[ \sum{{{x}_{i}}\frac{\partial \Phi }{\partial {{x}_{i}}}}-3N{{k}_{B}}T_0 \right].
\label{eq:two_vt_four}
\end{equation}
These equations of motion satisfy the generalized BBK equations, (\ref{eq:two_bbk_one}) and (\ref{eq:two_bbk_two}). 

It is interesting to note that these equations of motion have a Hamiltonian basis \cite{samoletov2007thermostats}. Consider the Hamiltonian:
\begin{equation}
    H = e^{-\mu}\sum \dfrac{p_i^2}{2m} + e^{-\mu}\Phi(e^\mu \mathbf{\bar{x}}) + e^\mu \dfrac{\bar{\eta}^2}{2Q_\eta} - 3e^{-\mu}Nk_BT_0\mu
    \label{eq:VT_Hamiltonian}
\end{equation}
The canonical variables $\bar{\mathbf{x}}$ and $\bar{\eta}$ are connected with the real variables $\mathbf{x}$ and $\eta$, respectively, through the transformations: $\mathbf{\bar{x}} = e^\mu \mathbf{x}$ and $\bar{\eta} = e^{-\mu} Q_\eta \eta$. The Hamiltonian generates equations (\ref{eq:two_vt_four}) provided the constant of motion: $I = \sum p_i^2 / 2m + \Phi(\mathbf{x}) + 0.5 Q_\eta \eta^2 + 3 Nk_B T_0 \mu = 0$, where $\mu = \int \eta dt$.

A strong form of equation (\ref{eq:two_vt_two}), where, the virial of forces $x_i \frac{\partial \Phi}{\partial x_i}$ is equal to $3Nk_BT_0$ at every time step, may be used to develop an isovirial thermostat \cite{config_thermostat_03}. The isovirial thermostat is analogous to the Gaussian Isokinetic thermostat but for the ``slow'' configurational variables. The steps for developing this thermostat are similar to that of the Gaussian Isokinetic thermostat with minor differences. Unlike the Gaussian isokinetic thermostat, the strong form of the Virial equation represents a holonomic constraint, and so, the differential constraint equation in the acceleration space is obtained by differentiating the Virial equation twice. Upon obtaining the differential constraint equation, the rest of the steps are the same as that of GIK thermostat, and have been left as an exercise for the readers. The resulting dynamics, however, does not sample from a canonical distribution. Virial thermostat has not received widespread attention from researchers, and to the best of our knowledge, its applications have been limited to very small-scale systems.

For a single harmonic oscillator, with $m = 1, k = 1$ and $k_BT_0 = 1$, the equations of motion for the Virial thermostat becomes:
\begin{equation}
\dot{x}=p-\eta x,\dot{p}=-x,\dot{\eta }=\frac{1}{{{Q}_{\eta }}}\left[ \sum{{{x}^{2}}}-{{k}_{B}}T_0 \right].
\label{eq:two_vt_five}
\end{equation}
Notice that these equations are identical with the equations of motion of the Braga-Travis thermostat. Thus, the ergodic and the non-equilibrium transport characteristics of this thermostat are the same as that of the Braga-Travis thermostat for small-scale systems. However, for larger systems, it seems that the Virial thermostat is better than the Braga-Travis thermostat, since the former's implementation does not require the computation of the Hessian matrix.

\section{Summary \& Challenges Ahead -- Open Questions}
In this review, we have presented different deterministic temperature control algorithms typically used in molecular dynamics simulations. While most of the algorithms are offshoots of the extended system method of Nos\'e and Hoover, each algorithm has different advantages and disadvantages. Apart from the philosophy behind their development, we highlight  -- (i) the situations where they are useful, (ii) the constant of motion and Hamiltonian, if any, from which the equations of motion can be obtained, (iii) the symplectic algorithms, if any, to solve the equations of motion numerically, (iv) the ergodic characteristics in equilibrium using a single harmonic oscillator, and (v) the ability to ensure non-equilibrium thermal transport in a single harmonic oscillator. When it comes to equilibrium and small-scale systems, some of the thermostats behave similarly, for example, Nos\'e-Hoover, Braga-Travis, Virial, Bauer-Bulgac-Kusnezov and Patra-Bhattacharya thermostat are all non-ergodic. On the other hand, Martyna-Klein-Tuckerman, Hoover-Holian, and $C_{1,2}$ thermostats are ergodic. In fact, for a single harmonic oscillator, some of the thermostats have similar equations of motion. 

We now discuss some of the outstanding challenges and open questions associated with thermostatted dynamics:
\begin{itemize}
    \item Almost all deterministic thermostats discussed in this review, satisfy the extended phase-space Liouville's equation. In most of the cases, the Liouville's measure is a product of the usual canonical distribution with independent normally distributed thermostatted variables. For example, in the Nos\'e-Hoover thermostat, Liouville's measure is the product of the canonically distributed configurational and momentum variables with the normally distributed thermostat variable. Consider a situation where the system comprises of particles interacting harmonically with each other. For such a case, the extended phase-space distribution is the sum of independent normal variables. A philosophical issue exists with the extended phase-space distribution being the sum of independent normal distributions -- the momentum evolution equation of any particle in this case is: $\dot{p}_i = -\frac{\partial \Phi}{x_i} - \eta p_i$. As $\Phi$ is a harmonic function, $\partial \Phi/\partial x_i$ is linear in $x$. For the LHS, since $p_i$ is a normal random variate, so is $\dot{p}_i$. However, in the RHS, while $\partial \Phi/\partial x_i$ is a normal random variate, $\eta p_i$ is a product of two normal random variables, which does not yield a normal random variate. Thus, from a statistical viewpoint, the momentum evolution equation cannot simultaneously result in momentum being normally distributed with $\eta$ being normal distributed as well. This philosophical issue persists for all deterministic thermostats that rely on the extended system method. Analysis of Nos\'e-Hoover thermostat and its variants discussed here from this statistical viewpoint is yet to emerge, and will possibly be able to shed more light on the ergodic characteristics of this family of thermostats.
    
    \item For larger systems, while in equilibrium the different algorithms result in similar dynamical properties since they satisfy the extended phase-space Liouville's equation, the same cannot be said in non-equilibrium. Researchers have reported non-equilibrium problems wherein  thermostat-dependent properties are obtained even in near-zero non-linearity. This begs the question -- which thermostat yields correct properties. The question remains wide open for the research community to investigate, including the reason why the different thermostats show different properties. The role of a thermostat in non-equilibrium situations is to extract the extra heat from the system. So it seems that the rate at which the heat is extracted plays an important role in non-equilibrium as this rate is different for the different thermostats.

    \item Moving on, one of the important question in thermostatted dynamics is related to ergodicity. While the question of ergodicity is less important in larger systems owing to prohibitively large Poincar\'e recurrence time, it forms the theoretical foundation for linking dynamical systems with statistical mechanics. In this regard, analytical proof of ergodicity has been given only for a limited number of cases. To the best of our knowledge, such analytical proof does not exist for any thermostatted dynamics. A particularly interesting case is that of the two-variable Martyna-Klein-Tuckerman thermostat, where for large values of $Q$, say $Q_\eta = Q_\xi = 10$, one observes regular trajectories which disappear for smaller values of $Q$ such as $Q_\eta = Q_\xi = 1$. More insights into these can be obtained if the dynamics is analyzed analytically. 
    
    \item Thermostat algorithms have been designed to control the temperature of the system to a specific value. However, researchers have reported that the temperature of a system fluctuates \cite{chui1992temperature,hickman2016temperature}. These temperature fluctuations can guide in developing better thermostat algorithms, where apart from controlling the temperature one may control the fluctuations as well. Further, a comparison of temperature fluctuations for the different thermostat algorithms in equilibrium can help in understanding which of the thermostats is, so as to speak, ``the best'' for controlling temperature. Exploration in this domain seems limited owing to the contradicting views concerning the existence of temperature fluctuations in a canonical ensemble \cite{kittel1973nonexistence}. 
    
    \item How do the usual configurational thermostats differ from the virial thermostat? This question is at the heart of making configurational thermostats more popular. The knowledge of Hessian is must for configurational thermostats, and computing it is extremely time-consuming of the order of $\sim N^3$. Virial thermostat, on the other hand, uses the information of only forces. Thus, if the Virial thermostat performs comparably to the other configurational thermostats in different equilibrium and non-equilibrium problems, one can use Virial thermostats in lieu of the configurational thermostats. A starting point could be to compare the auto-correlation functions of velocity and energy in equilibrium and non-equilibrium scenarios. On a related note, a comparison of auto-correlation functions due to the kinetic and configurational thermostats could also shed light on the rate at which the fast (momentum) and the slow (configurational) degrees of freedom are thermostatted.
    
    \item A separate set of questions arise related to temperature measurement -- (i) can one measure the configurational temperature dynamically in real-time instead of post-processing the snapshots of configurations \cite{han2004configurational,zhao2014measuring}, (ii) is it possible to uniquely measure kinetic and configurational temperatures of a system in equilibrium, and (iii) what do we measure in experiments using thermometers in away-from-equilibrium scenario. While researchers have been working towards answering these questions, satisficatory answers are yet to emerge. 
    
    \item Several thermostats have a Hamiltonian associated with them -- Nos\'e, Nos\'e-Hoover, Gaussian isokinetic and Virial thermostats -- in the sense that the equations of motion can be derived from the usual Hamilton's equations upon suitable substitution. It is not necessary for the canonical variables to be the same as that of the real variables. The existence of Hamiltonian makes it easy to relate thermostatted dynamics with the rich physics of dynamical systems. However, several thermostats, especially the multi-variables one, do not have any Hamiltonian. An interesting line of exploration is to create suitable Hamiltonians for these thermostats.
    
    \item Gaussian isokinetic thermostat can be derived from both Gauss' principle of least constraint (the usual method) as well as the Gauss' principle of least action (through the Hamiltonian route). This suggests that the Gauss' principle of least constraint is related to Gauss' principle of least action. While the relationship between them for holonomic constraints is straight forward to show, such is not the case for non-holonomic constraints. Establishing the relationship between the two principles will pave way for an improved understanding of thermostatted dynamical systems.
    
    \item Lyapunov exponents are closely linked with time-reversibility -- in an irreversible dynamical system, the sum of the Lyapunov exponents is less than zero. For a thermostatted system, apart from irreversibility, Lyapunov exponents also relate (in almost all cases) the heat-transfer entropy with the phase-space compressibility. However, such a link has only been (dis)proved for pedagogical systems. Exploration with realistic systems is yet to be observed.

\end{itemize}

Interesting progress along these and newer lines can be expected in the near future.

\begin{acknowledgments}
PKP acknowledges the support for this research provided in part by the Department of Science and Technology, Govt. of India, under the scheme DST-ECR.
\end{acknowledgments}

\bibliography{apssamp}

\providecommand{\noopsort}[1]{}\providecommand{\singleletter}[1]{#1}%
\begin{thebibliography}{122}%
\makeatletter
\providecommand \@ifxundefined [1]{%
 \@ifx{#1\undefined}
}%
\providecommand \@ifnum [1]{%
 \ifnum #1\expandafter \@firstoftwo
 \else \expandafter \@secondoftwo
 \fi
}%
\providecommand \@ifx [1]{%
 \ifx #1\expandafter \@firstoftwo
 \else \expandafter \@secondoftwo
 \fi
}%
\providecommand \natexlab [1]{#1}%
\providecommand \enquote  [1]{``#1''}%
\providecommand \bibnamefont  [1]{#1}%
\providecommand \bibfnamefont [1]{#1}%
\providecommand \citenamefont [1]{#1}%
\providecommand \href@noop [0]{\@secondoftwo}%
\providecommand \href [0]{\begingroup \@sanitize@url \@href}%
\providecommand \@href[1]{\@@startlink{#1}\@@href}%
\providecommand \@@href[1]{\endgroup#1\@@endlink}%
\providecommand \@sanitize@url [0]{\catcode `\\12\catcode `\$12\catcode
  `\&12\catcode `\#12\catcode `\^12\catcode `\_12\catcode `\%12\relax}%
\providecommand \@@startlink[1]{}%
\providecommand \@@endlink[0]{}%
\providecommand \url  [0]{\begingroup\@sanitize@url \@url }%
\providecommand \@url [1]{\endgroup\@href {#1}{\urlprefix }}%
\providecommand \urlprefix  [0]{URL }%
\providecommand \Eprint [0]{\href }%
\providecommand \doibase [0]{https://doi.org/}%
\providecommand \selectlanguage [0]{\@gobble}%
\providecommand \bibinfo  [0]{\@secondoftwo}%
\providecommand \bibfield  [0]{\@secondoftwo}%
\providecommand \translation [1]{[#1]}%
\providecommand \BibitemOpen [0]{}%
\providecommand \bibitemStop [0]{}%
\providecommand \bibitemNoStop [0]{.\EOS\space}%
\providecommand \EOS [0]{\spacefactor3000\relax}%
\providecommand \BibitemShut  [1]{\csname bibitem#1\endcsname}%
\let\auto@bib@innerbib\@empty
\bibitem [{\citenamefont {Goldstein}\ \emph {et~al.}(2001)\citenamefont
  {Goldstein}, \citenamefont {Jr.},\ and\ \citenamefont
  {Safko}}]{goldstein_01}%
  \BibitemOpen
  \bibfield  {author} {\bibinfo {author} {\bibfnamefont {H.}~\bibnamefont
  {Goldstein}}, \bibinfo {author} {\bibfnamefont {C.~P.~P.}\ \bibnamefont
  {Jr.}},\ and\ \bibinfo {author} {\bibfnamefont {J.~L.}\ \bibnamefont
  {Safko}},\ }\href@noop {} {\emph {\bibinfo {title} {Classical Mechanics}}},\
  \bibinfo {edition} {3rd}\ ed.\ (\bibinfo  {publisher} {Addison-Wesley},\
  \bibinfo {year} {2001})\BibitemShut {NoStop}%
\bibitem [{\citenamefont {Martys}\ and\ \citenamefont
  {Mountain}(1999)}]{velocity_verlet_symplectic}%
  \BibitemOpen
  \bibfield  {author} {\bibinfo {author} {\bibfnamefont {N.~S.}\ \bibnamefont
  {Martys}}\ and\ \bibinfo {author} {\bibfnamefont {R.~D.}\ \bibnamefont
  {Mountain}},\ }\bibfield  {title} {\bibinfo {title} {Velocity verlet
  algorithm for dissipative-particle-dynamics-based models of suspensions},\
  }\href@noop {} {\bibfield  {journal} {\bibinfo  {journal} {Phys. Rev. E}\
  }\textbf {\bibinfo {volume} {59}},\ \bibinfo {pages} {3733} (\bibinfo {year}
  {1999})}\BibitemShut {NoStop}%
\bibitem [{\citenamefont {Sprott}\ \emph
  {et~al.}(2014{\natexlab{a}})\citenamefont {Sprott}, \citenamefont {Hoover},\
  and\ \citenamefont {Hoover}}]{sprott_hoover_14}%
  \BibitemOpen
  \bibfield  {author} {\bibinfo {author} {\bibfnamefont {J.~C.}\ \bibnamefont
  {Sprott}}, \bibinfo {author} {\bibfnamefont {W.~G.}\ \bibnamefont {Hoover}},\
  and\ \bibinfo {author} {\bibfnamefont {C.~G.}\ \bibnamefont {Hoover}},\
  }\bibfield  {title} {\bibinfo {title} {Heat conduction, and the lack thereof,
  in time-reversible dynamical systems: Generalized nos\'e-hoover oscillators
  with a temperature gradient},\ }\href@noop {} {\bibfield  {journal} {\bibinfo
   {journal} {Physical Review E}\ }\textbf {\bibinfo {volume} {89}},\ \bibinfo
  {pages} {042914} (\bibinfo {year} {2014}{\natexlab{a}})}\BibitemShut
  {NoStop}%
\bibitem [{\citenamefont {S.}(1993)}]{Domokos_ergodicity}%
  \BibitemOpen
  \bibfield  {author} {\bibinfo {author} {\bibfnamefont {D.}~\bibnamefont
  {S.}},\ }\bibfield  {title} {\bibinfo {title} {Ergodicity of classical
  billiard balls},\ }\href@noop {} {\bibfield  {journal} {\bibinfo  {journal}
  {Physica A: Statistical Mechanics and its Applications}\ }\textbf {\bibinfo
  {volume} {194}},\ \bibinfo {pages} {86 } (\bibinfo {year}
  {1993})}\BibitemShut {NoStop}%
\bibitem [{\citenamefont {Sinai}(1979)}]{sinai_79}%
  \BibitemOpen
  \bibfield  {author} {\bibinfo {author} {\bibfnamefont {Y.~G.}\ \bibnamefont
  {Sinai}},\ }\bibfield  {title} {\bibinfo {title} {Ergodic properties of the
  lorentz gas},\ }\href@noop {} {\bibfield  {journal} {\bibinfo  {journal}
  {Functional Analysis and Its Applications}\ }\textbf {\bibinfo {volume}
  {13}},\ \bibinfo {pages} {192} (\bibinfo {year} {1979})}\BibitemShut
  {NoStop}%
\bibitem [{\citenamefont {Legoll}\ \emph
  {et~al.}(2007{\natexlab{a}})\citenamefont {Legoll}, \citenamefont {Luskin},\
  and\ \citenamefont {Moeckel}}]{legoll}%
  \BibitemOpen
  \bibfield  {author} {\bibinfo {author} {\bibfnamefont {F.}~\bibnamefont
  {Legoll}}, \bibinfo {author} {\bibfnamefont {M.}~\bibnamefont {Luskin}},\
  and\ \bibinfo {author} {\bibfnamefont {R.}~\bibnamefont {Moeckel}},\
  }\bibfield  {title} {\bibinfo {title} {Non-ergodicity of the nos\'e-hoover
  thermostatted harmonic oscillator},\ }\href@noop {} {\bibfield  {journal}
  {\bibinfo  {journal} {Archive for Rational Mechanics and Analysis}\ }\textbf
  {\bibinfo {volume} {184}},\ \bibinfo {pages} {449} (\bibinfo {year}
  {2007}{\natexlab{a}})}\BibitemShut {NoStop}%
\bibitem [{\citenamefont {Bulgac}\ and\ \citenamefont
  {Kusnezov}(1990)}]{bbk_pra}%
  \BibitemOpen
  \bibfield  {author} {\bibinfo {author} {\bibfnamefont {A.}~\bibnamefont
  {Bulgac}}\ and\ \bibinfo {author} {\bibfnamefont {D.}~\bibnamefont
  {Kusnezov}},\ }\bibfield  {title} {\bibinfo {title} {Canonical ensemble
  averages from pseudomicrocanonical dynamics},\ }\href@noop {} {\bibfield
  {journal} {\bibinfo  {journal} {Physical Review A}\ }\textbf {\bibinfo
  {volume} {42}},\ \bibinfo {pages} {5045} (\bibinfo {year}
  {1990})}\BibitemShut {NoStop}%
\bibitem [{\citenamefont {Posch}\ \emph {et~al.}(1986)\citenamefont {Posch},
  \citenamefont {Hoover},\ and\ \citenamefont {Vesely}}]{hoover_86}%
  \BibitemOpen
  \bibfield  {author} {\bibinfo {author} {\bibfnamefont {H.~A.}\ \bibnamefont
  {Posch}}, \bibinfo {author} {\bibfnamefont {W.~G.}\ \bibnamefont {Hoover}},\
  and\ \bibinfo {author} {\bibfnamefont {F.~J.}\ \bibnamefont {Vesely}},\
  }\bibfield  {title} {\bibinfo {title} {Canonical dynamics of the nos\'e
  oscillator: Stability, order, and chaos},\ }\href@noop {} {\bibfield
  {journal} {\bibinfo  {journal} {Physical Review A}\ }\textbf {\bibinfo
  {volume} {33}},\ \bibinfo {pages} {4253} (\bibinfo {year}
  {1986})}\BibitemShut {NoStop}%
\bibitem [{\citenamefont {Watanabe}\ and\ \citenamefont
  {Kobayashi}(2007{\natexlab{a}})}]{watanabe_07a}%
  \BibitemOpen
  \bibfield  {author} {\bibinfo {author} {\bibfnamefont {H.}~\bibnamefont
  {Watanabe}}\ and\ \bibinfo {author} {\bibfnamefont {H.}~\bibnamefont
  {Kobayashi}},\ }\bibfield  {title} {\bibinfo {title} {Ergodicity of a
  thermostat family of the nos\'{e}-hoover type},\ }\href@noop {} {\bibfield
  {journal} {\bibinfo  {journal} {Physical Review E}\ }\textbf {\bibinfo
  {volume} {75}},\ \bibinfo {pages} {040102} (\bibinfo {year}
  {2007}{\natexlab{a}})}\BibitemShut {NoStop}%
\bibitem [{\citenamefont {Zeh}(2001)}]{zeh2001dieter}%
  \BibitemOpen
  \bibfield  {author} {\bibinfo {author} {\bibfnamefont {H.}~\bibnamefont
  {Zeh}},\ }\href@noop {} {\bibinfo {title} {The physical basis of the
  direction of time}} (\bibinfo {year} {2001})\BibitemShut {NoStop}%
\bibitem [{\citenamefont {Patra}\ and\ \citenamefont
  {Bhattacharya}(2014)}]{patra2014nonergodicity}%
  \BibitemOpen
  \bibfield  {author} {\bibinfo {author} {\bibfnamefont {P.~K.}\ \bibnamefont
  {Patra}}\ and\ \bibinfo {author} {\bibfnamefont {B.}~\bibnamefont
  {Bhattacharya}},\ }\bibfield  {title} {\bibinfo {title} {Nonergodicity of the
  nose-hoover chain thermostat in computationally achievable time},\
  }\href@noop {} {\bibfield  {journal} {\bibinfo  {journal} {Physical Review
  E}\ }\textbf {\bibinfo {volume} {90}},\ \bibinfo {pages} {043304} (\bibinfo
  {year} {2014})}\BibitemShut {NoStop}%
\bibitem [{\citenamefont {Patra}\ \emph {et~al.}(2015)\citenamefont {Patra},
  \citenamefont {Sprott}, \citenamefont {Hoover},\ and\ \citenamefont
  {Hoover}}]{patra2015deterministic}%
  \BibitemOpen
  \bibfield  {author} {\bibinfo {author} {\bibfnamefont {P.~K.}\ \bibnamefont
  {Patra}}, \bibinfo {author} {\bibfnamefont {J.~C.}\ \bibnamefont {Sprott}},
  \bibinfo {author} {\bibfnamefont {W.~G.}\ \bibnamefont {Hoover}},\ and\
  \bibinfo {author} {\bibfnamefont {C.~G.}\ \bibnamefont {Hoover}},\ }\bibfield
   {title} {\bibinfo {title} {Deterministic time-reversible thermostats: chaos,
  ergodicity, and the zeroth law of thermodynamics},\ }\href@noop {} {\bibfield
   {journal} {\bibinfo  {journal} {Molecular Physics}\ }\textbf {\bibinfo
  {volume} {113}},\ \bibinfo {pages} {2863} (\bibinfo {year}
  {2015})}\BibitemShut {NoStop}%
\bibitem [{\citenamefont {Posch}\ and\ \citenamefont
  {Hoover}(1997{\natexlab{a}})}]{posch_hoover}%
  \BibitemOpen
  \bibfield  {author} {\bibinfo {author} {\bibfnamefont {H.~A.}\ \bibnamefont
  {Posch}}\ and\ \bibinfo {author} {\bibfnamefont {W.~G.}\ \bibnamefont
  {Hoover}},\ }\bibfield  {title} {\bibinfo {title} {Time-reversible
  dissipative attractors in three and four phase-space dimensions},\
  }\href@noop {} {\bibfield  {journal} {\bibinfo  {journal} {Physical Review
  E}\ }\textbf {\bibinfo {volume} {55}},\ \bibinfo {pages} {6803} (\bibinfo
  {year} {1997}{\natexlab{a}})}\BibitemShut {NoStop}%
\bibitem [{\citenamefont {Bond}\ \emph
  {et~al.}(1999{\natexlab{a}})\citenamefont {Bond}, \citenamefont
  {Leimkuhler},\ and\ \citenamefont {Laird}}]{bond1999nose}%
  \BibitemOpen
  \bibfield  {author} {\bibinfo {author} {\bibfnamefont {S.~D.}\ \bibnamefont
  {Bond}}, \bibinfo {author} {\bibfnamefont {B.~J.}\ \bibnamefont
  {Leimkuhler}},\ and\ \bibinfo {author} {\bibfnamefont {B.~B.}\ \bibnamefont
  {Laird}},\ }\bibfield  {title} {\bibinfo {title} {The nos{\'e}--poincar{\'e}
  method for constant temperature molecular dynamics},\ }\href@noop {}
  {\bibfield  {journal} {\bibinfo  {journal} {Journal of Computational
  Physics}\ }\textbf {\bibinfo {volume} {151}},\ \bibinfo {pages} {114}
  (\bibinfo {year} {1999}{\natexlab{a}})}\BibitemShut {NoStop}%
\bibitem [{\citenamefont {Sweet}(2004)}]{sweet2004hamilton}%
  \BibitemOpen
  \bibfield  {author} {\bibinfo {author} {\bibfnamefont {C.~R.}\ \bibnamefont
  {Sweet}},\ }\emph {\bibinfo {title} {Hamilton thermostatting techniques for
  molecular dynamics simulation}},\ \href@noop {} {Ph.D. thesis},\ \bibinfo
  {school} {University of Leicester (United Kingdom)} (\bibinfo {year}
  {2004})\BibitemShut {NoStop}%
\bibitem [{\citenamefont {Leimkuhler}\ and\ \citenamefont
  {Sweet}(2004)}]{leimkuhler2004canonical}%
  \BibitemOpen
  \bibfield  {author} {\bibinfo {author} {\bibfnamefont {B.~J.}\ \bibnamefont
  {Leimkuhler}}\ and\ \bibinfo {author} {\bibfnamefont {C.~R.}\ \bibnamefont
  {Sweet}},\ }\bibfield  {title} {\bibinfo {title} {The canonical ensemble via
  symplectic integrators using nos{\'e} and nos{\'e}--poincar{\'e} chains},\
  }\href@noop {} {\bibfield  {journal} {\bibinfo  {journal} {The Journal of
  chemical physics}\ }\textbf {\bibinfo {volume} {121}},\ \bibinfo {pages}
  {108} (\bibinfo {year} {2004})}\BibitemShut {NoStop}%
\bibitem [{\citenamefont {Haile}(1992)}]{haile_book}%
  \BibitemOpen
  \bibfield  {author} {\bibinfo {author} {\bibfnamefont {J.}~\bibnamefont
  {Haile}},\ }\href@noop {} {\emph {\bibinfo {title} {Molecular dynamics
  simulation}}},\ Vol.~\bibinfo {volume} {27}\ (\bibinfo  {publisher} {Wiley,
  New York},\ \bibinfo {year} {1992})\BibitemShut {NoStop}%
\bibitem [{\citenamefont {Powles}(1963)}]{powles_63}%
  \BibitemOpen
  \bibfield  {author} {\bibinfo {author} {\bibfnamefont {J.~G.}\ \bibnamefont
  {Powles}},\ }\bibfield  {title} {\bibinfo {title} {Negative absolute
  temperatures and rotating temperatures},\ }\href@noop {} {\bibfield
  {journal} {\bibinfo  {journal} {Contemporary Physics}\ }\textbf {\bibinfo
  {volume} {4}},\ \bibinfo {pages} {338} (\bibinfo {year} {1963})}\BibitemShut
  {NoStop}%
\bibitem [{\citenamefont {Jaynes}(1957{\natexlab{a}})}]{jaynes_57}%
  \BibitemOpen
  \bibfield  {author} {\bibinfo {author} {\bibfnamefont {E.~T.}\ \bibnamefont
  {Jaynes}},\ }\bibfield  {title} {\bibinfo {title} {Information theory and
  statistical mechanics},\ }\href {https://doi.org/10.1103/PhysRev.106.620}
  {\bibfield  {journal} {\bibinfo  {journal} {Phys. Rev.}\ }\textbf {\bibinfo
  {volume} {106}},\ \bibinfo {pages} {620} (\bibinfo {year}
  {1957}{\natexlab{a}})}\BibitemShut {NoStop}%
\bibitem [{\citenamefont {Jaynes}(1957{\natexlab{b}})}]{jaynes_57b}%
  \BibitemOpen
  \bibfield  {author} {\bibinfo {author} {\bibfnamefont {E.~T.}\ \bibnamefont
  {Jaynes}},\ }\bibfield  {title} {\bibinfo {title} {Information theory and
  statistical mechanics. ii},\ }\href {https://doi.org/10.1103/PhysRev.108.171}
  {\bibfield  {journal} {\bibinfo  {journal} {Phys. Rev.}\ }\textbf {\bibinfo
  {volume} {108}},\ \bibinfo {pages} {171} (\bibinfo {year}
  {1957}{\natexlab{b}})}\BibitemShut {NoStop}%
\bibitem [{\citenamefont {Shannon}(1948)}]{shannon_48}%
  \BibitemOpen
  \bibfield  {author} {\bibinfo {author} {\bibfnamefont {C.}~\bibnamefont
  {Shannon}},\ }\bibfield  {title} {\bibinfo {title} {A mathematical theory of
  communication},\ }\href {https://doi.org/10.1002/j.1538-7305.1948.tb01338.x}
  {\bibfield  {journal} {\bibinfo  {journal} {Bell System Technical Journal,
  The}\ }\textbf {\bibinfo {volume} {27}},\ \bibinfo {pages} {379} (\bibinfo
  {year} {1948})}\BibitemShut {NoStop}%
\bibitem [{\citenamefont {Casas-V\'{a}zquez}\ and\ \citenamefont
  {Jou}(2003)}]{casas_review_03}%
  \BibitemOpen
  \bibfield  {author} {\bibinfo {author} {\bibfnamefont {J.}~\bibnamefont
  {Casas-V\'{a}zquez}}\ and\ \bibinfo {author} {\bibfnamefont {D.}~\bibnamefont
  {Jou}},\ }\bibfield  {title} {\bibinfo {title} {Temperature in
  non-equilibrium states: a review of open problems and current proposals},\
  }\href@noop {} {\bibfield  {journal} {\bibinfo  {journal} {Reports on
  Progress in Physics}\ }\textbf {\bibinfo {volume} {66}},\ \bibinfo {pages}
  {1937} (\bibinfo {year} {2003})}\BibitemShut {NoStop}%
\bibitem [{\citenamefont {Huang}(2001)}]{huang_book}%
  \BibitemOpen
  \bibfield  {author} {\bibinfo {author} {\bibfnamefont {K.}~\bibnamefont
  {Huang}},\ }\href@noop {} {\emph {\bibinfo {title} {Introduction to
  statistical physics}}}\ (\bibinfo  {publisher} {CRC Press},\ \bibinfo {year}
  {2001})\BibitemShut {NoStop}%
\bibitem [{\citenamefont {Allen}\ \emph {et~al.}(2008)\citenamefont {Allen},
  \citenamefont {Tildesley},\ and\ \citenamefont {Banavar}}]{allen_tildesley}%
  \BibitemOpen
  \bibfield  {author} {\bibinfo {author} {\bibfnamefont {M.~P.}\ \bibnamefont
  {Allen}}, \bibinfo {author} {\bibfnamefont {D.~J.}\ \bibnamefont
  {Tildesley}},\ and\ \bibinfo {author} {\bibfnamefont {J.~R.}\ \bibnamefont
  {Banavar}},\ }\bibfield  {title} {\bibinfo {title} {Computer simulation of
  liquids},\ }\href@noop {} {\bibfield  {journal} {\bibinfo  {journal} {Physics
  Today}\ }\textbf {\bibinfo {volume} {42}},\ \bibinfo {pages} {105} (\bibinfo
  {year} {2008})}\BibitemShut {NoStop}%
\bibitem [{\citenamefont {Woodcock}(1971)}]{woodcock_71}%
  \BibitemOpen
  \bibfield  {author} {\bibinfo {author} {\bibfnamefont {L.~V.}\ \bibnamefont
  {Woodcock}},\ }\bibfield  {title} {\bibinfo {title} {Isothermal molecular
  dynamics calculations for liquid salts},\ }\href@noop {} {\bibfield
  {journal} {\bibinfo  {journal} {Chemical Physics Letters}\ }\textbf {\bibinfo
  {volume} {10}},\ \bibinfo {pages} {257} (\bibinfo {year} {1971})}\BibitemShut
  {NoStop}%
\bibitem [{\citenamefont {Martin-L{\"o}f}(1979)}]{martin1979equivalence}%
  \BibitemOpen
  \bibfield  {author} {\bibinfo {author} {\bibfnamefont {A.}~\bibnamefont
  {Martin-L{\"o}f}},\ }\bibfield  {title} {\bibinfo {title} {The equivalence of
  ensembles and the gibbs phase rule for classical lattice systems},\
  }\href@noop {} {\bibfield  {journal} {\bibinfo  {journal} {Journal of
  Statistical Physics}\ }\textbf {\bibinfo {volume} {20}},\ \bibinfo {pages}
  {557} (\bibinfo {year} {1979})}\BibitemShut {NoStop}%
\bibitem [{\citenamefont {H{\"u}nenberger}(2005)}]{hunenberger}%
  \BibitemOpen
  \bibfield  {author} {\bibinfo {author} {\bibfnamefont {P.~H.}\ \bibnamefont
  {H{\"u}nenberger}},\ }\bibfield  {title} {\bibinfo {title} {Thermostat
  algorithms for molecular dynamics simulations},\ }in\ \href
  {https://doi.org/10.1007/b99427} {\emph {\bibinfo {booktitle} {Advanced
  Computer Simulation}}},\ \bibinfo {series} {Advances in Polymer Science},
  Vol.\ \bibinfo {volume} {173},\ \bibinfo {editor} {edited by\ \bibinfo
  {editor} {\bibfnamefont {C.}~\bibnamefont {Dr.~Holm}}\ and\ \bibinfo {editor}
  {\bibfnamefont {K.}~\bibnamefont {Prof. Dr.~Kremer}}}\ (\bibinfo  {publisher}
  {Springer Berlin Heidelberg},\ \bibinfo {year} {2005})\ pp.\ \bibinfo {pages}
  {105--149}\BibitemShut {NoStop}%
\bibitem [{\citenamefont {Bussi}\ \emph {et~al.}(2007)\citenamefont {Bussi},
  \citenamefont {Donadio},\ and\ \citenamefont {Parrinello}}]{bussi_07}%
  \BibitemOpen
  \bibfield  {author} {\bibinfo {author} {\bibfnamefont {G.}~\bibnamefont
  {Bussi}}, \bibinfo {author} {\bibfnamefont {D.}~\bibnamefont {Donadio}},\
  and\ \bibinfo {author} {\bibfnamefont {M.}~\bibnamefont {Parrinello}},\
  }\bibfield  {title} {\bibinfo {title} {Canonical sampling through velocity
  rescaling},\ }\href@noop {} {\bibfield  {journal} {\bibinfo  {journal} {The
  Journal of Chemical Physics}\ }\textbf {\bibinfo {volume} {126}},\ \bibinfo
  {eid} {014101} (\bibinfo {year} {2007})}\BibitemShut {NoStop}%
\bibitem [{\citenamefont {Hoover}\ \emph {et~al.}(1982)\citenamefont {Hoover},
  \citenamefont {Ladd},\ and\ \citenamefont {Moran}}]{hoover_gik}%
  \BibitemOpen
  \bibfield  {author} {\bibinfo {author} {\bibfnamefont {W.~G.}\ \bibnamefont
  {Hoover}}, \bibinfo {author} {\bibfnamefont {A.~J.~C.}\ \bibnamefont
  {Ladd}},\ and\ \bibinfo {author} {\bibfnamefont {B.}~\bibnamefont {Moran}},\
  }\bibfield  {title} {\bibinfo {title} {High-strain-rate plastic flow studied
  via nonequilibrium molecular dynamics},\ }\href@noop {} {\bibfield  {journal}
  {\bibinfo  {journal} {Physical Review Letters}\ }\textbf {\bibinfo {volume}
  {48}},\ \bibinfo {pages} {1818} (\bibinfo {year} {1982})},\ \bibinfo {note}
  {pRL}\BibitemShut {NoStop}%
\bibitem [{\citenamefont {Evans}(1983)}]{evans_gik}%
  \BibitemOpen
  \bibfield  {author} {\bibinfo {author} {\bibfnamefont {D.~J.}\ \bibnamefont
  {Evans}},\ }\bibfield  {title} {\bibinfo {title} {Computer ``experiment'' for
  nonlinear thermodynamics of couette flow},\ }\href@noop {} {\bibfield
  {journal} {\bibinfo  {journal} {The Journal of Chemical Physics}\ }\textbf
  {\bibinfo {volume} {78}},\ \bibinfo {pages} {3297} (\bibinfo {year}
  {1983})}\BibitemShut {NoStop}%
\bibitem [{\citenamefont {Evans}\ and\ \citenamefont
  {Morriss}(1983{\natexlab{a}})}]{evans_morriss_gik}%
  \BibitemOpen
  \bibfield  {author} {\bibinfo {author} {\bibfnamefont {D.~J.}\ \bibnamefont
  {Evans}}\ and\ \bibinfo {author} {\bibfnamefont {G.~P.}\ \bibnamefont
  {Morriss}},\ }\bibfield  {title} {\bibinfo {title} {The isothermal/isobaric
  molecular dynamics ensemble},\ }\href@noop {} {\bibfield  {journal} {\bibinfo
   {journal} {Physics Letters A}\ }\textbf {\bibinfo {volume} {98}},\ \bibinfo
  {pages} {433} (\bibinfo {year} {1983}{\natexlab{a}})}\BibitemShut {NoStop}%
\bibitem [{\citenamefont {Evans}\ and\ \citenamefont
  {Morriss}(1983{\natexlab{b}})}]{evans_morriss_gik2}%
  \BibitemOpen
  \bibfield  {author} {\bibinfo {author} {\bibfnamefont {D.~J.}\ \bibnamefont
  {Evans}}\ and\ \bibinfo {author} {\bibfnamefont {G.~P.}\ \bibnamefont
  {Morriss}},\ }\bibfield  {title} {\bibinfo {title} {Isothermal-isobaric
  molecular dynamics},\ }\href@noop {} {\bibfield  {journal} {\bibinfo
  {journal} {Chemical Physics}\ }\textbf {\bibinfo {volume} {77}},\ \bibinfo
  {pages} {63} (\bibinfo {year} {1983}{\natexlab{b}})}\BibitemShut {NoStop}%
\bibitem [{\citenamefont {Evans}\ and\ \citenamefont
  {Morriss}(2008{\natexlab{a}})}]{evans_book}%
  \BibitemOpen
  \bibfield  {author} {\bibinfo {author} {\bibfnamefont {D.~J.}\ \bibnamefont
  {Evans}}\ and\ \bibinfo {author} {\bibfnamefont {G.}~\bibnamefont
  {Morriss}},\ }\href@noop {} {\emph {\bibinfo {title} {Statistical Mechanics
  of Nonequilibrium Liquids}}},\ \bibinfo {edition} {2nd}\ ed.\ (\bibinfo
  {publisher} {Cambridge University Press},\ \bibinfo {year}
  {2008})\BibitemShut {NoStop}%
\bibitem [{\citenamefont {Evans}\ and\ \citenamefont
  {Morriss}(2008{\natexlab{b}})}]{evans2008statistical}%
  \BibitemOpen
  \bibfield  {author} {\bibinfo {author} {\bibfnamefont {D.~J.}\ \bibnamefont
  {Evans}}\ and\ \bibinfo {author} {\bibfnamefont {G.}~\bibnamefont
  {Morriss}},\ }\href@noop {} {\emph {\bibinfo {title} {Statistical mechanics
  of nonequilibrium liquids}}}\ (\bibinfo  {publisher} {Cambridge University
  Press},\ \bibinfo {year} {2008})\BibitemShut {NoStop}%
\bibitem [{\citenamefont {Morriss}\ and\ \citenamefont
  {Dettmann}(1998)}]{morriss1998thermostats}%
  \BibitemOpen
  \bibfield  {author} {\bibinfo {author} {\bibfnamefont {G.~P.}\ \bibnamefont
  {Morriss}}\ and\ \bibinfo {author} {\bibfnamefont {C.~P.}\ \bibnamefont
  {Dettmann}},\ }\bibfield  {title} {\bibinfo {title} {Thermostats: analysis
  and application},\ }\href@noop {} {\bibfield  {journal} {\bibinfo  {journal}
  {Chaos: An Interdisciplinary Journal of Nonlinear Science}\ }\textbf
  {\bibinfo {volume} {8}},\ \bibinfo {pages} {321} (\bibinfo {year}
  {1998})}\BibitemShut {NoStop}%
\bibitem [{\citenamefont {Dettmann}\ and\ \citenamefont
  {Morriss}(1996)}]{dettmann_gik}%
  \BibitemOpen
  \bibfield  {author} {\bibinfo {author} {\bibfnamefont {C.~P.}\ \bibnamefont
  {Dettmann}}\ and\ \bibinfo {author} {\bibfnamefont {G.~P.}\ \bibnamefont
  {Morriss}},\ }\bibfield  {title} {\bibinfo {title} {Hamiltonian formulation
  of the gaussian isokinetic thermostat},\ }\href@noop {} {\bibfield  {journal}
  {\bibinfo  {journal} {Physical Review E}\ }\textbf {\bibinfo {volume} {54}},\
  \bibinfo {pages} {2495} (\bibinfo {year} {1996})}\BibitemShut {NoStop}%
\bibitem [{\citenamefont {Bright}\ \emph {et~al.}(2005)\citenamefont {Bright},
  \citenamefont {Evans},\ and\ \citenamefont {Searles}}]{bright2005new}%
  \BibitemOpen
  \bibfield  {author} {\bibinfo {author} {\bibfnamefont {J.~N.}\ \bibnamefont
  {Bright}}, \bibinfo {author} {\bibfnamefont {D.~J.}\ \bibnamefont {Evans}},\
  and\ \bibinfo {author} {\bibfnamefont {D.~J.}\ \bibnamefont {Searles}},\
  }\bibfield  {title} {\bibinfo {title} {New observations regarding
  deterministic, time-reversible thermostats and gauss’s principle of least
  constraint},\ }\href@noop {} {\bibfield  {journal} {\bibinfo  {journal} {The
  Journal of chemical physics}\ }\textbf {\bibinfo {volume} {122}},\ \bibinfo
  {pages} {194106} (\bibinfo {year} {2005})}\BibitemShut {NoStop}%
\bibitem [{\citenamefont {Evans}\ and\ \citenamefont
  {Morriss}(1986{\natexlab{a}})}]{evans1986shear}%
  \BibitemOpen
  \bibfield  {author} {\bibinfo {author} {\bibfnamefont {D.~J.}\ \bibnamefont
  {Evans}}\ and\ \bibinfo {author} {\bibfnamefont {G.~P.}\ \bibnamefont
  {Morriss}},\ }\bibfield  {title} {\bibinfo {title} {Shear thickening and
  turbulence in simple fluids},\ }\href@noop {} {\bibfield  {journal} {\bibinfo
   {journal} {Physical review letters}\ }\textbf {\bibinfo {volume} {56}},\
  \bibinfo {pages} {2172} (\bibinfo {year} {1986}{\natexlab{a}})}\BibitemShut
  {NoStop}%
\bibitem [{\citenamefont {Morriss}(1988)}]{morriss1988lyapunov}%
  \BibitemOpen
  \bibfield  {author} {\bibinfo {author} {\bibfnamefont {G.~P.}\ \bibnamefont
  {Morriss}},\ }\bibfield  {title} {\bibinfo {title} {Lyapunov dimension of
  two-body planar couette flow},\ }\href@noop {} {\bibfield  {journal}
  {\bibinfo  {journal} {Physical Review A}\ }\textbf {\bibinfo {volume} {37}},\
  \bibinfo {pages} {2118} (\bibinfo {year} {1988})}\BibitemShut {NoStop}%
\bibitem [{\citenamefont {Evans}\ and\ \citenamefont
  {Sarman}(1993)}]{evans1993equivalence}%
  \BibitemOpen
  \bibfield  {author} {\bibinfo {author} {\bibfnamefont {D.~J.}\ \bibnamefont
  {Evans}}\ and\ \bibinfo {author} {\bibfnamefont {S.}~\bibnamefont {Sarman}},\
  }\bibfield  {title} {\bibinfo {title} {Equivalence of thermostatted nonlinear
  responses},\ }\href@noop {} {\bibfield  {journal} {\bibinfo  {journal}
  {Physical Review E}\ }\textbf {\bibinfo {volume} {48}},\ \bibinfo {pages}
  {65} (\bibinfo {year} {1993})}\BibitemShut {NoStop}%
\bibitem [{\citenamefont {Morriss}\ and\ \citenamefont
  {Evans}(1987)}]{morriss1987application}%
  \BibitemOpen
  \bibfield  {author} {\bibinfo {author} {\bibfnamefont {G.~P.}\ \bibnamefont
  {Morriss}}\ and\ \bibinfo {author} {\bibfnamefont {D.~J.}\ \bibnamefont
  {Evans}},\ }\bibfield  {title} {\bibinfo {title} {Application of transient
  correlation functions to shear flow far from equilibrium},\ }\href@noop {}
  {\bibfield  {journal} {\bibinfo  {journal} {Physical Review A}\ }\textbf
  {\bibinfo {volume} {35}},\ \bibinfo {pages} {792} (\bibinfo {year}
  {1987})}\BibitemShut {NoStop}%
\bibitem [{\citenamefont {Delhommelle}\ and\ \citenamefont
  {Evans}(2001{\natexlab{a}})}]{delhommelle2001configurational}%
  \BibitemOpen
  \bibfield  {author} {\bibinfo {author} {\bibfnamefont {J.}~\bibnamefont
  {Delhommelle}}\ and\ \bibinfo {author} {\bibfnamefont {D.~J.}\ \bibnamefont
  {Evans}},\ }\bibfield  {title} {\bibinfo {title} {Configurational temperature
  thermostat for fluids undergoing shear flow: application to liquid
  chlorine},\ }\href@noop {} {\bibfield  {journal} {\bibinfo  {journal}
  {Molecular Physics}\ }\textbf {\bibinfo {volume} {99}},\ \bibinfo {pages}
  {1825} (\bibinfo {year} {2001}{\natexlab{a}})}\BibitemShut {NoStop}%
\bibitem [{\citenamefont {Zhang}\ \emph {et~al.}(1999)\citenamefont {Zhang},
  \citenamefont {Searles}, \citenamefont {Evans}, \citenamefont {den
  Toom~Hansen},\ and\ \citenamefont {Isbister}}]{zhang1999kinetic}%
  \BibitemOpen
  \bibfield  {author} {\bibinfo {author} {\bibfnamefont {F.}~\bibnamefont
  {Zhang}}, \bibinfo {author} {\bibfnamefont {D.~J.}\ \bibnamefont {Searles}},
  \bibinfo {author} {\bibfnamefont {D.~J.}\ \bibnamefont {Evans}}, \bibinfo
  {author} {\bibfnamefont {J.~S.}\ \bibnamefont {den Toom~Hansen}},\ and\
  \bibinfo {author} {\bibfnamefont {D.~J.}\ \bibnamefont {Isbister}},\
  }\bibfield  {title} {\bibinfo {title} {Kinetic energy conserving integrators
  for gaussian thermostatted sllod},\ }\href@noop {} {\bibfield  {journal}
  {\bibinfo  {journal} {The Journal of chemical physics}\ }\textbf {\bibinfo
  {volume} {111}},\ \bibinfo {pages} {18} (\bibinfo {year} {1999})}\BibitemShut
  {NoStop}%
\bibitem [{\citenamefont {Nos\'{e}}(1984{\natexlab{a}})}]{nose}%
  \BibitemOpen
  \bibfield  {author} {\bibinfo {author} {\bibfnamefont {S.}~\bibnamefont
  {Nos\'{e}}},\ }\bibfield  {title} {\bibinfo {title} {A unified formulation of
  the constant temperature molecular dynamics methods},\ }\href@noop {}
  {\bibfield  {journal} {\bibinfo  {journal} {The Journal of Chemical Physics}\
  }\textbf {\bibinfo {volume} {81}},\ \bibinfo {pages} {511} (\bibinfo {year}
  {1984}{\natexlab{a}})}\BibitemShut {NoStop}%
\bibitem [{\citenamefont {Nos\'{e}}(1984{\natexlab{b}})}]{nose_thermostat}%
  \BibitemOpen
  \bibfield  {author} {\bibinfo {author} {\bibfnamefont {S.}~\bibnamefont
  {Nos\'{e}}},\ }\bibfield  {title} {\bibinfo {title} {A molecular dynamics
  method for simulations in the canonical ensemble},\ }\href@noop {} {\bibfield
   {journal} {\bibinfo  {journal} {Molecular Physics}\ }\textbf {\bibinfo
  {volume} {52}},\ \bibinfo {pages} {255} (\bibinfo {year}
  {1984}{\natexlab{b}})}\BibitemShut {NoStop}%
\bibitem [{\citenamefont {Shuichi}(1991)}]{nose_review}%
  \BibitemOpen
  \bibfield  {author} {\bibinfo {author} {\bibfnamefont {N.}~\bibnamefont
  {Shuichi}},\ }\bibfield  {title} {\bibinfo {title} {Constant temperature
  molecular dynamics methods},\ }\href {https://doi.org/10.1143/PTPS.103.1}
  {\bibfield  {journal} {\bibinfo  {journal} {Progress of Theoretical Physics
  Supplement}\ }\textbf {\bibinfo {volume} {103}},\ \bibinfo {pages} {1}
  (\bibinfo {year} {1991})}\BibitemShut {NoStop}%
\bibitem [{\citenamefont {Hoover}(1985)}]{nose_hoover}%
  \BibitemOpen
  \bibfield  {author} {\bibinfo {author} {\bibfnamefont {W.~G.}\ \bibnamefont
  {Hoover}},\ }\bibfield  {title} {\bibinfo {title} {Canonical dynamics:
  Equilibrium phase-space distributions},\ }\href@noop {} {\bibfield  {journal}
  {\bibinfo  {journal} {Physical Review A}\ }\textbf {\bibinfo {volume} {31}},\
  \bibinfo {pages} {1695} (\bibinfo {year} {1985})},\ \bibinfo {note}
  {pRA}\BibitemShut {NoStop}%
\bibitem [{\citenamefont {Hoover}(1991)}]{hoover_computational_stat_mech}%
  \BibitemOpen
  \bibfield  {author} {\bibinfo {author} {\bibfnamefont {W.~G.}\ \bibnamefont
  {Hoover}},\ }\href@noop {} {\emph {\bibinfo {title} {Computational
  Statistical Mechanics}}}\ (\bibinfo  {publisher} {Elsevier},\ \bibinfo
  {address} {Amsterdam},\ \bibinfo {year} {1991})\BibitemShut {NoStop}%
\bibitem [{\citenamefont {Dettmann}\ and\ \citenamefont
  {Morriss}(1997)}]{dettmann1997hamiltonian}%
  \BibitemOpen
  \bibfield  {author} {\bibinfo {author} {\bibfnamefont {C.}~\bibnamefont
  {Dettmann}}\ and\ \bibinfo {author} {\bibfnamefont {G.}~\bibnamefont
  {Morriss}},\ }\bibfield  {title} {\bibinfo {title} {Hamiltonian reformulation
  and pairing of lyapunov exponents for nos{\'e}-hoover dynamics},\ }\href@noop
  {} {\bibfield  {journal} {\bibinfo  {journal} {Physical Review E}\ }\textbf
  {\bibinfo {volume} {55}},\ \bibinfo {pages} {3693} (\bibinfo {year}
  {1997})}\BibitemShut {NoStop}%
\bibitem [{\citenamefont {Bond}\ \emph
  {et~al.}(1999{\natexlab{b}})\citenamefont {Bond}, \citenamefont
  {Leimkuhler},\ and\ \citenamefont {Laird}}]{nose_poincare}%
  \BibitemOpen
  \bibfield  {author} {\bibinfo {author} {\bibfnamefont {S.~D.}\ \bibnamefont
  {Bond}}, \bibinfo {author} {\bibfnamefont {B.~J.}\ \bibnamefont
  {Leimkuhler}},\ and\ \bibinfo {author} {\bibfnamefont {B.~B.}\ \bibnamefont
  {Laird}},\ }\bibfield  {title} {\bibinfo {title} {The nos\'{e}-poincar\'{e}
  method for constant temperature molecular dynamics},\ }\href@noop {}
  {\bibfield  {journal} {\bibinfo  {journal} {Journal of Computational
  Physics}\ }\textbf {\bibinfo {volume} {151}},\ \bibinfo {pages} {114}
  (\bibinfo {year} {1999}{\natexlab{b}})}\BibitemShut {NoStop}%
\bibitem [{\citenamefont {Posch}\ and\ \citenamefont
  {Hoover}(1997{\natexlab{b}})}]{posch1997time}%
  \BibitemOpen
  \bibfield  {author} {\bibinfo {author} {\bibfnamefont {H.}~\bibnamefont
  {Posch}}\ and\ \bibinfo {author} {\bibfnamefont {W.~G.}\ \bibnamefont
  {Hoover}},\ }\bibfield  {title} {\bibinfo {title} {Time-reversible
  dissipative attractors in three and four phase-space dimensions},\
  }\href@noop {} {\bibfield  {journal} {\bibinfo  {journal} {Physical Review
  E}\ }\textbf {\bibinfo {volume} {55}},\ \bibinfo {pages} {6803} (\bibinfo
  {year} {1997}{\natexlab{b}})}\BibitemShut {NoStop}%
\bibitem [{\citenamefont {Searles}\ and\ \citenamefont
  {Evans}(2001)}]{searles2001fluctuation}%
  \BibitemOpen
  \bibfield  {author} {\bibinfo {author} {\bibfnamefont {D.~J.}\ \bibnamefont
  {Searles}}\ and\ \bibinfo {author} {\bibfnamefont {D.~J.}\ \bibnamefont
  {Evans}},\ }\bibfield  {title} {\bibinfo {title} {Fluctuation theorem for
  heat flow},\ }\href@noop {} {\bibfield  {journal} {\bibinfo  {journal}
  {International journal of thermophysics}\ }\textbf {\bibinfo {volume} {22}},\
  \bibinfo {pages} {123} (\bibinfo {year} {2001})}\BibitemShut {NoStop}%
\bibitem [{\citenamefont {Watanabe}\ and\ \citenamefont
  {Kobayashi}(2007{\natexlab{b}})}]{watanabe2007ergodicity}%
  \BibitemOpen
  \bibfield  {author} {\bibinfo {author} {\bibfnamefont {H.}~\bibnamefont
  {Watanabe}}\ and\ \bibinfo {author} {\bibfnamefont {H.}~\bibnamefont
  {Kobayashi}},\ }\bibfield  {title} {\bibinfo {title} {Ergodicity of a
  thermostat family of the nos{\'e}-hoover type},\ }\href@noop {} {\bibfield
  {journal} {\bibinfo  {journal} {Physical Review E}\ }\textbf {\bibinfo
  {volume} {75}},\ \bibinfo {pages} {040102} (\bibinfo {year}
  {2007}{\natexlab{b}})}\BibitemShut {NoStop}%
\bibitem [{\citenamefont {Jellinek}\ and\ \citenamefont
  {Berry}(1989)}]{jellinek_berry}%
  \BibitemOpen
  \bibfield  {author} {\bibinfo {author} {\bibfnamefont {J.}~\bibnamefont
  {Jellinek}}\ and\ \bibinfo {author} {\bibfnamefont {R.~S.}\ \bibnamefont
  {Berry}},\ }\bibfield  {title} {\bibinfo {title} {Generalization of nos\'e's
  isothermal molecular dynamics: Necessary and sufficient conditions of
  dynamical simulations of statistical ensembles},\ }\href@noop {} {\bibfield
  {journal} {\bibinfo  {journal} {Physical Review A}\ }\textbf {\bibinfo
  {volume} {40}},\ \bibinfo {pages} {2816} (\bibinfo {year}
  {1989})}\BibitemShut {NoStop}%
\bibitem [{\citenamefont {Bra{\'n}ka}\ and\ \citenamefont
  {Wojciechowski}(2000)}]{branka2000generalization}%
  \BibitemOpen
  \bibfield  {author} {\bibinfo {author} {\bibfnamefont {A.}~\bibnamefont
  {Bra{\'n}ka}}\ and\ \bibinfo {author} {\bibfnamefont {K.}~\bibnamefont
  {Wojciechowski}},\ }\bibfield  {title} {\bibinfo {title} {Generalization of
  nos{\'e} and nos{\'e}-hoover isothermal dynamics},\ }\href@noop {} {\bibfield
   {journal} {\bibinfo  {journal} {Physical Review E}\ }\textbf {\bibinfo
  {volume} {62}},\ \bibinfo {pages} {3281} (\bibinfo {year}
  {2000})}\BibitemShut {NoStop}%
\bibitem [{\citenamefont {Bravetti}\ and\ \citenamefont
  {Tapias}(2016)}]{bravetti2016thermostat}%
  \BibitemOpen
  \bibfield  {author} {\bibinfo {author} {\bibfnamefont {A.}~\bibnamefont
  {Bravetti}}\ and\ \bibinfo {author} {\bibfnamefont {D.}~\bibnamefont
  {Tapias}},\ }\bibfield  {title} {\bibinfo {title} {Thermostat algorithm for
  generating target ensembles},\ }\href@noop {} {\bibfield  {journal} {\bibinfo
   {journal} {Physical Review E}\ }\textbf {\bibinfo {volume} {93}},\ \bibinfo
  {pages} {022139} (\bibinfo {year} {2016})}\BibitemShut {NoStop}%
\bibitem [{\citenamefont {Sadus}(2002)}]{sadus2002molecular}%
  \BibitemOpen
  \bibfield  {author} {\bibinfo {author} {\bibfnamefont {R.~J.}\ \bibnamefont
  {Sadus}},\ }\href@noop {} {\emph {\bibinfo {title} {Molecular simulation of
  fluids}}}\ (\bibinfo  {publisher} {Elsevier},\ \bibinfo {year}
  {2002})\BibitemShut {NoStop}%
\bibitem [{\citenamefont {Patra}\ \emph {et~al.}(2016)\citenamefont {Patra},
  \citenamefont {Hoover}, \citenamefont {Hoover},\ and\ \citenamefont
  {Sprott}}]{patra2016equivalence}%
  \BibitemOpen
  \bibfield  {author} {\bibinfo {author} {\bibfnamefont {P.~K.}\ \bibnamefont
  {Patra}}, \bibinfo {author} {\bibfnamefont {W.~G.}\ \bibnamefont {Hoover}},
  \bibinfo {author} {\bibfnamefont {C.~G.}\ \bibnamefont {Hoover}},\ and\
  \bibinfo {author} {\bibfnamefont {J.~C.}\ \bibnamefont {Sprott}},\ }\bibfield
   {title} {\bibinfo {title} {The equivalence of dissipation from gibbs’
  entropy production with phase-volume loss in ergodic heat-conducting
  oscillators},\ }\href@noop {} {\bibfield  {journal} {\bibinfo  {journal}
  {International Journal of Bifurcation and Chaos}\ }\textbf {\bibinfo {volume}
  {26}},\ \bibinfo {pages} {1650089} (\bibinfo {year} {2016})}\BibitemShut
  {NoStop}%
\bibitem [{\citenamefont {Watanabe}\ and\ \citenamefont
  {Kobayashi}(2007{\natexlab{c}})}]{watanabe_07b}%
  \BibitemOpen
  \bibfield  {author} {\bibinfo {author} {\bibfnamefont {H.}~\bibnamefont
  {Watanabe}}\ and\ \bibinfo {author} {\bibfnamefont {H.}~\bibnamefont
  {Kobayashi}},\ }\bibfield  {title} {\bibinfo {title} {Ergodicity of the
  nos\'{e}-hoover method},\ }\href@noop {} {\bibfield  {journal} {\bibinfo
  {journal} {Molecular Simulation}\ }\textbf {\bibinfo {volume} {33}},\
  \bibinfo {pages} {77} (\bibinfo {year} {2007}{\natexlab{c}})}\BibitemShut
  {NoStop}%
\bibitem [{\citenamefont {Sprott}\ \emph
  {et~al.}(2014{\natexlab{b}})\citenamefont {Sprott}, \citenamefont {Hoover},\
  and\ \citenamefont {Hoover}}]{sprott2014heat}%
  \BibitemOpen
  \bibfield  {author} {\bibinfo {author} {\bibfnamefont {J.~C.}\ \bibnamefont
  {Sprott}}, \bibinfo {author} {\bibfnamefont {W.~G.}\ \bibnamefont {Hoover}},\
  and\ \bibinfo {author} {\bibfnamefont {C.~G.}\ \bibnamefont {Hoover}},\
  }\bibfield  {title} {\bibinfo {title} {Heat conduction, and the lack thereof,
  in time-reversible dynamical systems: Generalized nos{\'e}-hoover oscillators
  with a temperature gradient},\ }\href@noop {} {\bibfield  {journal} {\bibinfo
   {journal} {Physical Review E}\ }\textbf {\bibinfo {volume} {89}},\ \bibinfo
  {pages} {042914} (\bibinfo {year} {2014}{\natexlab{b}})}\BibitemShut
  {NoStop}%
\bibitem [{\citenamefont {Yu}\ \emph {et~al.}(2016)\citenamefont {Yu},
  \citenamefont {Zhang}, \citenamefont {Zhang}, \citenamefont {Liu},
  \citenamefont {Tang},\ and\ \citenamefont {Li}}]{yu2016simulation}%
  \BibitemOpen
  \bibfield  {author} {\bibinfo {author} {\bibfnamefont {J.}~\bibnamefont
  {Yu}}, \bibinfo {author} {\bibfnamefont {S.}~\bibnamefont {Zhang}}, \bibinfo
  {author} {\bibfnamefont {Q.}~\bibnamefont {Zhang}}, \bibinfo {author}
  {\bibfnamefont {R.}~\bibnamefont {Liu}}, \bibinfo {author} {\bibfnamefont
  {M.}~\bibnamefont {Tang}},\ and\ \bibinfo {author} {\bibfnamefont
  {X.}~\bibnamefont {Li}},\ }\bibfield  {title} {\bibinfo {title} {Simulation
  study and experiment verification of the creep mechanism of a nickel-based
  single crystal superalloy obtained from microstructural evolution},\
  }\href@noop {} {\bibfield  {journal} {\bibinfo  {journal} {RSC advances}\
  }\textbf {\bibinfo {volume} {6}},\ \bibinfo {pages} {107748} (\bibinfo {year}
  {2016})}\BibitemShut {NoStop}%
\bibitem [{\citenamefont {Hoover}\ and\ \citenamefont
  {Kum}(1997)}]{hoover_kum}%
  \BibitemOpen
  \bibfield  {author} {\bibinfo {author} {\bibfnamefont {W.~G.}\ \bibnamefont
  {Hoover}}\ and\ \bibinfo {author} {\bibfnamefont {O.}~\bibnamefont {Kum}},\
  }\bibfield  {title} {\bibinfo {title} {Ergodicity, mixing, and time
  reversibility for atomistic nonequilibrium steady states},\ }\href@noop {}
  {\bibfield  {journal} {\bibinfo  {journal} {Phys. Rev. E}\ }\textbf {\bibinfo
  {volume} {56}},\ \bibinfo {pages} {5517} (\bibinfo {year}
  {1997})}\BibitemShut {NoStop}%
\bibitem [{\citenamefont {Martyna}\ \emph {et~al.}(1992)\citenamefont
  {Martyna}, \citenamefont {Klein},\ and\ \citenamefont {Tuckerman}}]{mkt}%
  \BibitemOpen
  \bibfield  {author} {\bibinfo {author} {\bibfnamefont {G.~J.}\ \bibnamefont
  {Martyna}}, \bibinfo {author} {\bibfnamefont {M.~L.}\ \bibnamefont {Klein}},\
  and\ \bibinfo {author} {\bibfnamefont {M.}~\bibnamefont {Tuckerman}},\
  }\bibfield  {title} {\bibinfo {title} {Nos[e-acute]--hoover chains: The
  canonical ensemble via continuous dynamics},\ }\href@noop {} {\bibfield
  {journal} {\bibinfo  {journal} {The Journal of Chemical Physics}\ }\textbf
  {\bibinfo {volume} {97}},\ \bibinfo {pages} {2635} (\bibinfo {year}
  {1992})}\BibitemShut {NoStop}%
\bibitem [{\citenamefont {Hoover}\ and\ \citenamefont {Holian}(1996)}]{hh}%
  \BibitemOpen
  \bibfield  {author} {\bibinfo {author} {\bibfnamefont {W.~G.}\ \bibnamefont
  {Hoover}}\ and\ \bibinfo {author} {\bibfnamefont {B.~L.}\ \bibnamefont
  {Holian}},\ }\bibfield  {title} {\bibinfo {title} {Kinetic moments method for
  the canonical ensemble distribution},\ }\href@noop {} {\bibfield  {journal}
  {\bibinfo  {journal} {Physics Letters A}\ }\textbf {\bibinfo {volume}
  {211}},\ \bibinfo {pages} {253} (\bibinfo {year} {1996})}\BibitemShut
  {NoStop}%
\bibitem [{\citenamefont {Tobias}\ \emph {et~al.}(1993)\citenamefont {Tobias},
  \citenamefont {Martyna},\ and\ \citenamefont {Klein}}]{tobias_93}%
  \BibitemOpen
  \bibfield  {author} {\bibinfo {author} {\bibfnamefont {D.~J.}\ \bibnamefont
  {Tobias}}, \bibinfo {author} {\bibfnamefont {G.~J.}\ \bibnamefont
  {Martyna}},\ and\ \bibinfo {author} {\bibfnamefont {M.~L.}\ \bibnamefont
  {Klein}},\ }\bibfield  {title} {\bibinfo {title} {Molecular dynamics
  simulations of a protein in the canonical ensemble},\ }\href@noop {}
  {\bibfield  {journal} {\bibinfo  {journal} {The Journal of Physical
  Chemistry}\ }\textbf {\bibinfo {volume} {97}},\ \bibinfo {pages} {12959}
  (\bibinfo {year} {1993})}\BibitemShut {NoStop}%
\bibitem [{\citenamefont {Martyna}\ \emph {et~al.}(1996)\citenamefont
  {Martyna}, \citenamefont {Tuckerman}, \citenamefont {Tobias},\ and\
  \citenamefont {Klein}}]{martyna1996explicit}%
  \BibitemOpen
  \bibfield  {author} {\bibinfo {author} {\bibfnamefont {G.~J.}\ \bibnamefont
  {Martyna}}, \bibinfo {author} {\bibfnamefont {M.~E.}\ \bibnamefont
  {Tuckerman}}, \bibinfo {author} {\bibfnamefont {D.~J.}\ \bibnamefont
  {Tobias}},\ and\ \bibinfo {author} {\bibfnamefont {M.~L.}\ \bibnamefont
  {Klein}},\ }\bibfield  {title} {\bibinfo {title} {Explicit reversible
  integrators for extended systems dynamics},\ }\href@noop {} {\bibfield
  {journal} {\bibinfo  {journal} {Molecular Physics}\ }\textbf {\bibinfo
  {volume} {87}},\ \bibinfo {pages} {1117} (\bibinfo {year}
  {1996})}\BibitemShut {NoStop}%
\bibitem [{\citenamefont {Legoll}\ \emph
  {et~al.}(2007{\natexlab{b}})\citenamefont {Legoll}, \citenamefont {Luskin},\
  and\ \citenamefont {Moeckel}}]{legoll2007non}%
  \BibitemOpen
  \bibfield  {author} {\bibinfo {author} {\bibfnamefont {F.}~\bibnamefont
  {Legoll}}, \bibinfo {author} {\bibfnamefont {M.}~\bibnamefont {Luskin}},\
  and\ \bibinfo {author} {\bibfnamefont {R.}~\bibnamefont {Moeckel}},\
  }\bibfield  {title} {\bibinfo {title} {Non-ergodicity of the nos{\'e}--hoover
  thermostatted harmonic oscillator},\ }\href@noop {} {\bibfield  {journal}
  {\bibinfo  {journal} {Archive for rational mechanics and analysis}\ }\textbf
  {\bibinfo {volume} {184}},\ \bibinfo {pages} {449} (\bibinfo {year}
  {2007}{\natexlab{b}})}\BibitemShut {NoStop}%
\bibitem [{\citenamefont {Legoll}\ \emph {et~al.}(2009)\citenamefont {Legoll},
  \citenamefont {Luskin},\ and\ \citenamefont {Moeckel}}]{legoll2009non}%
  \BibitemOpen
  \bibfield  {author} {\bibinfo {author} {\bibfnamefont {F.}~\bibnamefont
  {Legoll}}, \bibinfo {author} {\bibfnamefont {M.}~\bibnamefont {Luskin}},\
  and\ \bibinfo {author} {\bibfnamefont {R.}~\bibnamefont {Moeckel}},\
  }\bibfield  {title} {\bibinfo {title} {Non-ergodicity of nos{\'e}--hoover
  dynamics},\ }\href@noop {} {\bibfield  {journal} {\bibinfo  {journal}
  {Nonlinearity}\ }\textbf {\bibinfo {volume} {22}},\ \bibinfo {pages} {1673}
  (\bibinfo {year} {2009})}\BibitemShut {NoStop}%
\bibitem [{\citenamefont {Tuckerman}\ and\ \citenamefont
  {Martyna}(2000)}]{tuckerman_00}%
  \BibitemOpen
  \bibfield  {author} {\bibinfo {author} {\bibfnamefont {M.~E.}\ \bibnamefont
  {Tuckerman}}\ and\ \bibinfo {author} {\bibfnamefont {G.~J.}\ \bibnamefont
  {Martyna}},\ }\bibfield  {title} {\bibinfo {title} {Understanding modern
  molecular dynamics:? techniques and applications},\ }\href@noop {} {\bibfield
   {journal} {\bibinfo  {journal} {The Journal of Physical Chemistry B}\
  }\textbf {\bibinfo {volume} {104}},\ \bibinfo {pages} {159} (\bibinfo {year}
  {2000})}\BibitemShut {NoStop}%
\bibitem [{\citenamefont {Hoover}\ and\ \citenamefont
  {Posch}(1994)}]{hoover1994second}%
  \BibitemOpen
  \bibfield  {author} {\bibinfo {author} {\bibfnamefont {W.}~\bibnamefont
  {Hoover}}\ and\ \bibinfo {author} {\bibfnamefont {H.}~\bibnamefont {Posch}},\
  }\bibfield  {title} {\bibinfo {title} {Second-law irreversibility and
  phase-space dimensionality loss from time-reversible nonequilibrium
  steady-state lyapunov spectra},\ }\href@noop {} {\bibfield  {journal}
  {\bibinfo  {journal} {Physical Review E}\ }\textbf {\bibinfo {volume} {49}},\
  \bibinfo {pages} {1913} (\bibinfo {year} {1994})}\BibitemShut {NoStop}%
\bibitem [{\citenamefont {Hoover}\ and\ \citenamefont
  {Posch}(1998)}]{hoover1998multifractals}%
  \BibitemOpen
  \bibfield  {author} {\bibinfo {author} {\bibfnamefont {W.~G.}\ \bibnamefont
  {Hoover}}\ and\ \bibinfo {author} {\bibfnamefont {H.~A.}\ \bibnamefont
  {Posch}},\ }\bibfield  {title} {\bibinfo {title} {Multifractals from
  stochastic many-body molecular dynamics},\ }\href@noop {} {\bibfield
  {journal} {\bibinfo  {journal} {Physics Letters A}\ }\textbf {\bibinfo
  {volume} {246}},\ \bibinfo {pages} {247} (\bibinfo {year}
  {1998})}\BibitemShut {NoStop}%
\bibitem [{\citenamefont {Hoover}(1998)}]{hoover1998liouville}%
  \BibitemOpen
  \bibfield  {author} {\bibinfo {author} {\bibfnamefont {W.~G.}\ \bibnamefont
  {Hoover}},\ }\bibfield  {title} {\bibinfo {title} {Liouville’s theorems,
  gibbs’ entropy, and multifractal distributions for nonequilibrium steady
  states},\ }\href@noop {} {\bibfield  {journal} {\bibinfo  {journal} {The
  Journal of chemical physics}\ }\textbf {\bibinfo {volume} {109}},\ \bibinfo
  {pages} {4164} (\bibinfo {year} {1998})}\BibitemShut {NoStop}%
\bibitem [{\citenamefont {Hoover}\ \emph {et~al.}(2007)\citenamefont {Hoover},
  \citenamefont {Hoover}, \citenamefont {Posch},\ and\ \citenamefont
  {Codelli}}]{hoover2007second}%
  \BibitemOpen
  \bibfield  {author} {\bibinfo {author} {\bibfnamefont {W.~G.}\ \bibnamefont
  {Hoover}}, \bibinfo {author} {\bibfnamefont {C.}~\bibnamefont {Hoover}},
  \bibinfo {author} {\bibfnamefont {H.}~\bibnamefont {Posch}},\ and\ \bibinfo
  {author} {\bibfnamefont {J.}~\bibnamefont {Codelli}},\ }\bibfield  {title}
  {\bibinfo {title} {The second law of thermodynamics and multifractal
  distribution functions: Bin counting, pair correlations, and the
  kaplan--yorke conjecture},\ }\href@noop {} {\bibfield  {journal} {\bibinfo
  {journal} {Communications in Nonlinear Science and Numerical Simulation}\
  }\textbf {\bibinfo {volume} {12}},\ \bibinfo {pages} {214} (\bibinfo {year}
  {2007})}\BibitemShut {NoStop}%
\bibitem [{\citenamefont {Plimpton}(1993)}]{plimpton1993fast}%
  \BibitemOpen
  \bibfield  {author} {\bibinfo {author} {\bibfnamefont {S.}~\bibnamefont
  {Plimpton}},\ }\href@noop {} {\emph {\bibinfo {title} {Fast parallel
  algorithms for short-range molecular dynamics}}},\ \bibinfo {type} {Tech.
  Rep.}\ (\bibinfo  {institution} {Sandia National Labs., Albuquerque, NM
  (United States)},\ \bibinfo {year} {1993})\BibitemShut {NoStop}%
\bibitem [{\citenamefont {Berendsen}\ \emph {et~al.}(1995)\citenamefont
  {Berendsen}, \citenamefont {van~der Spoel},\ and\ \citenamefont {van
  Drunen}}]{berendsen1995gromacs}%
  \BibitemOpen
  \bibfield  {author} {\bibinfo {author} {\bibfnamefont {H.~J.}\ \bibnamefont
  {Berendsen}}, \bibinfo {author} {\bibfnamefont {D.}~\bibnamefont {van~der
  Spoel}},\ and\ \bibinfo {author} {\bibfnamefont {R.}~\bibnamefont {van
  Drunen}},\ }\bibfield  {title} {\bibinfo {title} {Gromacs: a message-passing
  parallel molecular dynamics implementation},\ }\href@noop {} {\bibfield
  {journal} {\bibinfo  {journal} {Computer physics communications}\ }\textbf
  {\bibinfo {volume} {91}},\ \bibinfo {pages} {43} (\bibinfo {year}
  {1995})}\BibitemShut {NoStop}%
\bibitem [{\citenamefont {Liu}\ and\ \citenamefont
  {Tuckerman}(2000)}]{liu2000generalized}%
  \BibitemOpen
  \bibfield  {author} {\bibinfo {author} {\bibfnamefont {Y.}~\bibnamefont
  {Liu}}\ and\ \bibinfo {author} {\bibfnamefont {M.~E.}\ \bibnamefont
  {Tuckerman}},\ }\bibfield  {title} {\bibinfo {title} {Generalized gaussian
  moment thermostatting: A new continuous dynamical approach to the canonical
  ensemble},\ }\href@noop {} {\bibfield  {journal} {\bibinfo  {journal} {The
  Journal of Chemical Physics}\ }\textbf {\bibinfo {volume} {112}},\ \bibinfo
  {pages} {1685} (\bibinfo {year} {2000})}\BibitemShut {NoStop}%
\bibitem [{\citenamefont {Campisi}\ \emph {et~al.}(2012)\citenamefont
  {Campisi}, \citenamefont {Zhan}, \citenamefont {Talkner},\ and\ \citenamefont
  {H{\"a}nggi}}]{CZTH_thermostat}%
  \BibitemOpen
  \bibfield  {author} {\bibinfo {author} {\bibfnamefont {M.}~\bibnamefont
  {Campisi}}, \bibinfo {author} {\bibfnamefont {F.}~\bibnamefont {Zhan}},
  \bibinfo {author} {\bibfnamefont {P.}~\bibnamefont {Talkner}},\ and\ \bibinfo
  {author} {\bibfnamefont {P.}~\bibnamefont {H{\"a}nggi}},\ }\bibfield  {title}
  {\bibinfo {title} {Logarithmic oscillators: ideal hamiltonian thermostats},\
  }\href@noop {} {\bibfield  {journal} {\bibinfo  {journal} {Physical review
  letters}\ }\textbf {\bibinfo {volume} {108}},\ \bibinfo {pages} {250601}
  (\bibinfo {year} {2012})}\BibitemShut {NoStop}%
\bibitem [{\citenamefont {Campisi}\ and\ \citenamefont
  {H{\"a}nggi}(2013)}]{CZTH_thermostat2}%
  \BibitemOpen
  \bibfield  {author} {\bibinfo {author} {\bibfnamefont {M.}~\bibnamefont
  {Campisi}}\ and\ \bibinfo {author} {\bibfnamefont {P.}~\bibnamefont
  {H{\"a}nggi}},\ }\bibfield  {title} {\bibinfo {title} {Thermostated
  hamiltonian dynamics with log oscillators},\ }\href@noop {} {\bibfield
  {journal} {\bibinfo  {journal} {The Journal of Physical Chemistry B}\
  }\textbf {\bibinfo {volume} {117}},\ \bibinfo {pages} {12829} (\bibinfo
  {year} {2013})}\BibitemShut {NoStop}%
\bibitem [{\citenamefont {Patra}\ and\ \citenamefont
  {Bhattacharya}(2018)}]{patra2018zeroth}%
  \BibitemOpen
  \bibfield  {author} {\bibinfo {author} {\bibfnamefont {P.~K.}\ \bibnamefont
  {Patra}}\ and\ \bibinfo {author} {\bibfnamefont {B.}~\bibnamefont
  {Bhattacharya}},\ }\bibfield  {title} {\bibinfo {title} {Zeroth law
  investigation on the logarithmic thermostat},\ }\href@noop {} {\bibfield
  {journal} {\bibinfo  {journal} {Scientific reports}\ }\textbf {\bibinfo
  {volume} {8}},\ \bibinfo {pages} {1} (\bibinfo {year} {2018})}\BibitemShut
  {NoStop}%
\bibitem [{\citenamefont {Hoover}\ and\ \citenamefont
  {Hoover}(2013)}]{hoover_czth}%
  \BibitemOpen
  \bibfield  {author} {\bibinfo {author} {\bibfnamefont {W.~G.}\ \bibnamefont
  {Hoover}}\ and\ \bibinfo {author} {\bibfnamefont {C.~G.}\ \bibnamefont
  {Hoover}},\ }\bibfield  {title} {\bibinfo {title} {Hamiltonian thermostats
  fail to promote heat flow},\ }\href@noop {} {\bibfield  {journal} {\bibinfo
  {journal} {Communications in Nonlinear Science and Numerical Simulation}\
  }\textbf {\bibinfo {volume} {18}},\ \bibinfo {pages} {3365} (\bibinfo {year}
  {2013})}\BibitemShut {NoStop}%
\bibitem [{\citenamefont {Sponseller}\ and\ \citenamefont
  {Blaisten-Barojas}(2014)}]{sponseller_14}%
  \BibitemOpen
  \bibfield  {author} {\bibinfo {author} {\bibfnamefont {D.}~\bibnamefont
  {Sponseller}}\ and\ \bibinfo {author} {\bibfnamefont {E.}~\bibnamefont
  {Blaisten-Barojas}},\ }\bibfield  {title} {\bibinfo {title} {Failure of
  logarithmic oscillators to serve as a thermostat for small atomic clusters},\
  }\href@noop {} {\bibfield  {journal} {\bibinfo  {journal} {Physical Review
  E}\ }\textbf {\bibinfo {volume} {89}},\ \bibinfo {pages} {021301} (\bibinfo
  {year} {2014})}\BibitemShut {NoStop}%
\bibitem [{\citenamefont {Chen}\ \emph {et~al.}(2017)\citenamefont {Chen},
  \citenamefont {He},\ and\ \citenamefont {Zhao}}]{chen2017violation}%
  \BibitemOpen
  \bibfield  {author} {\bibinfo {author} {\bibfnamefont {K.}~\bibnamefont
  {Chen}}, \bibinfo {author} {\bibfnamefont {D.}~\bibnamefont {He}},\ and\
  \bibinfo {author} {\bibfnamefont {H.}~\bibnamefont {Zhao}},\ }\bibfield
  {title} {\bibinfo {title} {Violation of the virial theorem and generalized
  equipartition theorem for logarithmic oscillators serving as a thermostat},\
  }\href@noop {} {\bibfield  {journal} {\bibinfo  {journal} {Scientific
  reports}\ }\textbf {\bibinfo {volume} {7}},\ \bibinfo {pages} {1} (\bibinfo
  {year} {2017})}\BibitemShut {NoStop}%
\bibitem [{\citenamefont {Rugh}(1997)}]{Rugh}%
  \BibitemOpen
  \bibfield  {author} {\bibinfo {author} {\bibfnamefont {H.~H.}\ \bibnamefont
  {Rugh}},\ }\bibfield  {title} {\bibinfo {title} {Dynamical approach to
  temperature},\ }\href@noop {} {\bibfield  {journal} {\bibinfo  {journal}
  {Physical Review Letters}\ }\textbf {\bibinfo {volume} {78}},\ \bibinfo
  {pages} {772} (\bibinfo {year} {1997})}\BibitemShut {NoStop}%
\bibitem [{\citenamefont {Rugh}(1998)}]{rugh_98}%
  \BibitemOpen
  \bibfield  {author} {\bibinfo {author} {\bibfnamefont {H.~H.}\ \bibnamefont
  {Rugh}},\ }\bibfield  {title} {\bibinfo {title} {A geometric, dynamical
  approach to thermodynamics},\ }\href@noop {} {\bibfield  {journal} {\bibinfo
  {journal} {Journal of Physics A: Mathematical and General}\ }\textbf
  {\bibinfo {volume} {31}},\ \bibinfo {pages} {7761} (\bibinfo {year}
  {1998})}\BibitemShut {NoStop}%
\bibitem [{\citenamefont {Jepps}\ \emph {et~al.}(2000)\citenamefont {Jepps},
  \citenamefont {Ayton},\ and\ \citenamefont {Evans}}]{jepps_00}%
  \BibitemOpen
  \bibfield  {author} {\bibinfo {author} {\bibfnamefont {O.~G.}\ \bibnamefont
  {Jepps}}, \bibinfo {author} {\bibfnamefont {G.}~\bibnamefont {Ayton}},\ and\
  \bibinfo {author} {\bibfnamefont {D.~J.}\ \bibnamefont {Evans}},\ }\bibfield
  {title} {\bibinfo {title} {Microscopic expressions for the thermodynamic
  temperature},\ }\href@noop {} {\bibfield  {journal} {\bibinfo  {journal}
  {Physical Review E}\ }\textbf {\bibinfo {volume} {62}},\ \bibinfo {pages}
  {4757} (\bibinfo {year} {2000})}\BibitemShut {NoStop}%
\bibitem [{\citenamefont {Morriss}\ and\ \citenamefont
  {Rondoni}(1999)}]{morriss_99}%
  \BibitemOpen
  \bibfield  {author} {\bibinfo {author} {\bibfnamefont {G.~P.}\ \bibnamefont
  {Morriss}}\ and\ \bibinfo {author} {\bibfnamefont {L.}~\bibnamefont
  {Rondoni}},\ }\bibfield  {title} {\bibinfo {title} {Definition of temperature
  in equilibrium and nonequilibrium systems},\ }\href
  {https://doi.org/10.1103/PhysRevE.59.R5} {\bibfield  {journal} {\bibinfo
  {journal} {Phys. Rev. E}\ }\textbf {\bibinfo {volume} {59}},\ \bibinfo
  {pages} {R5} (\bibinfo {year} {1999})}\BibitemShut {NoStop}%
\bibitem [{\citenamefont {Hoover}\ and\ \citenamefont
  {Hoover}(2008)}]{hoover2008nonequilibrium}%
  \BibitemOpen
  \bibfield  {author} {\bibinfo {author} {\bibfnamefont {W.~G.}\ \bibnamefont
  {Hoover}}\ and\ \bibinfo {author} {\bibfnamefont {C.~G.}\ \bibnamefont
  {Hoover}},\ }\bibfield  {title} {\bibinfo {title} {Nonequilibrium temperature
  and thermometry in heat-conducting $\phi$ 4 models},\ }\href@noop {}
  {\bibfield  {journal} {\bibinfo  {journal} {Physical Review E}\ }\textbf
  {\bibinfo {volume} {77}},\ \bibinfo {pages} {041104} (\bibinfo {year}
  {2008})}\BibitemShut {NoStop}%
\bibitem [{\citenamefont {Hoover}\ and\ \citenamefont
  {Hoover}(2007)}]{hoover2007hamiltonian}%
  \BibitemOpen
  \bibfield  {author} {\bibinfo {author} {\bibfnamefont {W.~G.}\ \bibnamefont
  {Hoover}}\ and\ \bibinfo {author} {\bibfnamefont {C.~G.}\ \bibnamefont
  {Hoover}},\ }\bibfield  {title} {\bibinfo {title} {Hamiltonian dynamics of
  thermostated systems: Two-temperature heat-conducting $\phi$ 4 chains},\
  }\href@noop {} {\bibfield  {journal} {\bibinfo  {journal} {The Journal of
  chemical physics}\ }\textbf {\bibinfo {volume} {126}},\ \bibinfo {pages}
  {164113} (\bibinfo {year} {2007})}\BibitemShut {NoStop}%
\bibitem [{\citenamefont {Baranyai}\ \emph {et~al.}(1992)\citenamefont
  {Baranyai}, \citenamefont {Evans},\ and\ \citenamefont
  {Daivis}}]{baranyai1992isothermal}%
  \BibitemOpen
  \bibfield  {author} {\bibinfo {author} {\bibfnamefont {A.}~\bibnamefont
  {Baranyai}}, \bibinfo {author} {\bibfnamefont {D.~J.}\ \bibnamefont
  {Evans}},\ and\ \bibinfo {author} {\bibfnamefont {P.~J.}\ \bibnamefont
  {Daivis}},\ }\bibfield  {title} {\bibinfo {title} {Isothermal shear-induced
  heat flow},\ }\href@noop {} {\bibfield  {journal} {\bibinfo  {journal}
  {Physical Review A}\ }\textbf {\bibinfo {volume} {46}},\ \bibinfo {pages}
  {7593} (\bibinfo {year} {1992})}\BibitemShut {NoStop}%
\bibitem [{\citenamefont {Ayton}\ \emph {et~al.}(1999)\citenamefont {Ayton},
  \citenamefont {Jepps},\ and\ \citenamefont {EVANS}}]{ayton1999validity}%
  \BibitemOpen
  \bibfield  {author} {\bibinfo {author} {\bibfnamefont {G.}~\bibnamefont
  {Ayton}}, \bibinfo {author} {\bibfnamefont {O.~G.}\ \bibnamefont {Jepps}},\
  and\ \bibinfo {author} {\bibfnamefont {D.~J.}\ \bibnamefont {EVANS}},\
  }\bibfield  {title} {\bibinfo {title} {On the validity of fourier's law in
  systems with spatially varying strain rates},\ }\href@noop {} {\bibfield
  {journal} {\bibinfo  {journal} {Molecular Physics}\ }\textbf {\bibinfo
  {volume} {96}},\ \bibinfo {pages} {915} (\bibinfo {year} {1999})}\BibitemShut
  {NoStop}%
\bibitem [{\citenamefont {Daivis}\ \emph {et~al.}(2012)\citenamefont {Daivis},
  \citenamefont {Dalton},\ and\ \citenamefont {Morishita}}]{daivis2012effect}%
  \BibitemOpen
  \bibfield  {author} {\bibinfo {author} {\bibfnamefont {P.~J.}\ \bibnamefont
  {Daivis}}, \bibinfo {author} {\bibfnamefont {B.~A.}\ \bibnamefont {Dalton}},\
  and\ \bibinfo {author} {\bibfnamefont {T.}~\bibnamefont {Morishita}},\
  }\bibfield  {title} {\bibinfo {title} {Effect of kinetic and configurational
  thermostats on calculations of the first normal stress coefficient in
  nonequilibrium molecular dynamics simulations},\ }\href@noop {} {\bibfield
  {journal} {\bibinfo  {journal} {Physical Review E}\ }\textbf {\bibinfo
  {volume} {86}},\ \bibinfo {pages} {056707} (\bibinfo {year}
  {2012})}\BibitemShut {NoStop}%
\bibitem [{\citenamefont {Basconi}\ and\ \citenamefont
  {Shirts}(2013)}]{basconi2013effects}%
  \BibitemOpen
  \bibfield  {author} {\bibinfo {author} {\bibfnamefont {J.~E.}\ \bibnamefont
  {Basconi}}\ and\ \bibinfo {author} {\bibfnamefont {M.~R.}\ \bibnamefont
  {Shirts}},\ }\bibfield  {title} {\bibinfo {title} {Effects of temperature
  control algorithms on transport properties and kinetics in molecular dynamics
  simulations},\ }\href@noop {} {\bibfield  {journal} {\bibinfo  {journal}
  {Journal of chemical theory and computation}\ }\textbf {\bibinfo {volume}
  {9}},\ \bibinfo {pages} {2887} (\bibinfo {year} {2013})}\BibitemShut
  {NoStop}%
\bibitem [{\citenamefont {Cho}\ and\ \citenamefont
  {Joannopoulos}(1992)}]{cho_joannopoulos}%
  \BibitemOpen
  \bibfield  {author} {\bibinfo {author} {\bibfnamefont {K.}~\bibnamefont
  {Cho}}\ and\ \bibinfo {author} {\bibfnamefont {J.~D.}\ \bibnamefont
  {Joannopoulos}},\ }\bibfield  {title} {\bibinfo {title} {Ergodicity and
  dynamical properties of constant-temperature molecular dynamics},\
  }\href@noop {} {\bibfield  {journal} {\bibinfo  {journal} {Physical Review
  A}\ }\textbf {\bibinfo {volume} {45}},\ \bibinfo {pages} {7089} (\bibinfo
  {year} {1992})}\BibitemShut {NoStop}%
\bibitem [{\citenamefont {Patra}\ and\ \citenamefont
  {Bhattacharya}(2016)}]{patra2016heat}%
  \BibitemOpen
  \bibfield  {author} {\bibinfo {author} {\bibfnamefont {P.~K.}\ \bibnamefont
  {Patra}}\ and\ \bibinfo {author} {\bibfnamefont {B.}~\bibnamefont
  {Bhattacharya}},\ }\bibfield  {title} {\bibinfo {title} {Heat pump without
  particle transport or external work on the medium achieved by differential
  thermostatting of the phase space},\ }\href@noop {} {\bibfield  {journal}
  {\bibinfo  {journal} {Physical Review E}\ }\textbf {\bibinfo {volume} {93}},\
  \bibinfo {pages} {033308} (\bibinfo {year} {2016})}\BibitemShut {NoStop}%
\bibitem [{\citenamefont {Kusnezov}\ \emph {et~al.}(1990)\citenamefont
  {Kusnezov}, \citenamefont {Bulgac},\ and\ \citenamefont
  {Bauer}}]{bbk_annals}%
  \BibitemOpen
  \bibfield  {author} {\bibinfo {author} {\bibfnamefont {D.}~\bibnamefont
  {Kusnezov}}, \bibinfo {author} {\bibfnamefont {A.}~\bibnamefont {Bulgac}},\
  and\ \bibinfo {author} {\bibfnamefont {W.}~\bibnamefont {Bauer}},\ }\bibfield
   {title} {\bibinfo {title} {Canonical ensembles from chaos},\ }\href@noop {}
  {\bibfield  {journal} {\bibinfo  {journal} {Annals of Physics}\ }\textbf
  {\bibinfo {volume} {204}},\ \bibinfo {pages} {155} (\bibinfo {year}
  {1990})}\BibitemShut {NoStop}%
\bibitem [{\citenamefont {Butler}\ \emph {et~al.}(1998)\citenamefont {Butler},
  \citenamefont {Ayton}, \citenamefont {Jepps},\ and\ \citenamefont
  {Evans}}]{butler_98}%
  \BibitemOpen
  \bibfield  {author} {\bibinfo {author} {\bibfnamefont {B.~D.}\ \bibnamefont
  {Butler}}, \bibinfo {author} {\bibfnamefont {G.}~\bibnamefont {Ayton}},
  \bibinfo {author} {\bibfnamefont {O.~G.}\ \bibnamefont {Jepps}},\ and\
  \bibinfo {author} {\bibfnamefont {D.~J.}\ \bibnamefont {Evans}},\ }\bibfield
  {title} {\bibinfo {title} {Configurational temperature: Verification of monte
  carlo simulations},\ }\href@noop {} {\bibfield  {journal} {\bibinfo
  {journal} {The Journal of Chemical Physics}\ }\textbf {\bibinfo {volume}
  {109}},\ \bibinfo {pages} {6519} (\bibinfo {year} {1998})}\BibitemShut
  {NoStop}%
\bibitem [{\citenamefont {Lue}\ and\ \citenamefont
  {Evans}(2000)}]{lue_configT}%
  \BibitemOpen
  \bibfield  {author} {\bibinfo {author} {\bibfnamefont {L.}~\bibnamefont
  {Lue}}\ and\ \bibinfo {author} {\bibfnamefont {D.~J.}\ \bibnamefont
  {Evans}},\ }\bibfield  {title} {\bibinfo {title} {Configurational temperature
  for systems with constraints},\ }\href
  {https://doi.org/10.1103/PhysRevE.62.4764} {\bibfield  {journal} {\bibinfo
  {journal} {Phys. Rev. E}\ }\textbf {\bibinfo {volume} {62}},\ \bibinfo
  {pages} {4764} (\bibinfo {year} {2000})}\BibitemShut {NoStop}%
\bibitem [{\citenamefont {Landau}\ and\ \citenamefont
  {Lifshitz}(2013)}]{landau2013course}%
  \BibitemOpen
  \bibfield  {author} {\bibinfo {author} {\bibfnamefont {L.~D.}\ \bibnamefont
  {Landau}}\ and\ \bibinfo {author} {\bibfnamefont {E.~M.}\ \bibnamefont
  {Lifshitz}},\ }\href@noop {} {\emph {\bibinfo {title} {Course of theoretical
  physics}}}\ (\bibinfo  {publisher} {Elsevier},\ \bibinfo {year}
  {2013})\BibitemShut {NoStop}%
\bibitem [{\citenamefont {Zhao}\ \emph {et~al.}(2013)\citenamefont {Zhao},
  \citenamefont {An}, \citenamefont {Li}, \citenamefont {Wu}, \citenamefont
  {Goddard},\ and\ \citenamefont {Luo}}]{zhao_KT}%
  \BibitemOpen
  \bibfield  {author} {\bibinfo {author} {\bibfnamefont {F.~P.}\ \bibnamefont
  {Zhao}}, \bibinfo {author} {\bibfnamefont {Q.}~\bibnamefont {An}}, \bibinfo
  {author} {\bibfnamefont {B.}~\bibnamefont {Li}}, \bibinfo {author}
  {\bibfnamefont {H.~A.}\ \bibnamefont {Wu}}, \bibinfo {author} {\bibfnamefont
  {W.~A.}\ \bibnamefont {Goddard}},\ and\ \bibinfo {author} {\bibfnamefont
  {S.~N.}\ \bibnamefont {Luo}},\ }\bibfield  {title} {\bibinfo {title} {Shock
  response of a model structured nanofoam of cu},\ }\href@noop {} {\bibfield
  {journal} {\bibinfo  {journal} {Journal of Applied Physics}\ }\textbf
  {\bibinfo {volume} {113}},\ \bibinfo {eid} {063516} (\bibinfo {year}
  {2013})}\BibitemShut {NoStop}%
\bibitem [{\citenamefont {He}\ \emph {et~al.}(2012)\citenamefont {He},
  \citenamefont {Duan}, \citenamefont {Shao}, \citenamefont {Wang},\ and\
  \citenamefont {Qin}}]{he_KT}%
  \BibitemOpen
  \bibfield  {author} {\bibinfo {author} {\bibfnamefont {A.~M.}\ \bibnamefont
  {He}}, \bibinfo {author} {\bibfnamefont {S.}~\bibnamefont {Duan}}, \bibinfo
  {author} {\bibfnamefont {J.~L.}\ \bibnamefont {Shao}}, \bibinfo {author}
  {\bibfnamefont {P.}~\bibnamefont {Wang}},\ and\ \bibinfo {author}
  {\bibfnamefont {C.}~\bibnamefont {Qin}},\ }\bibfield  {title} {\bibinfo
  {title} {Shock melting of single crystal copper with a nanovoid: Molecular
  dynamics simulations},\ }\href@noop {} {\bibfield  {journal} {\bibinfo
  {journal} {Journal of Applied Physics}\ }\textbf {\bibinfo {volume} {112}},\
  \bibinfo {eid} {074116} (\bibinfo {year} {2012})}\BibitemShut {NoStop}%
\bibitem [{\citenamefont {Erpenbeck}(1984{\natexlab{a}})}]{erpenbeck1984shear}%
  \BibitemOpen
  \bibfield  {author} {\bibinfo {author} {\bibfnamefont {J.~J.}\ \bibnamefont
  {Erpenbeck}},\ }\bibfield  {title} {\bibinfo {title} {Shear viscosity of the
  hard-sphere fluid via nonequilibrium molecular dynamics},\ }\href@noop {}
  {\bibfield  {journal} {\bibinfo  {journal} {Physical review letters}\
  }\textbf {\bibinfo {volume} {52}},\ \bibinfo {pages} {1333} (\bibinfo {year}
  {1984}{\natexlab{a}})}\BibitemShut {NoStop}%
\bibitem [{\citenamefont {Evans}\ \emph {et~al.}(1992)\citenamefont {Evans},
  \citenamefont {Cui}, \citenamefont {Hanley},\ and\ \citenamefont
  {Straty}}]{evans1992conditions}%
  \BibitemOpen
  \bibfield  {author} {\bibinfo {author} {\bibfnamefont {D.~J.}\ \bibnamefont
  {Evans}}, \bibinfo {author} {\bibfnamefont {S.}~\bibnamefont {Cui}}, \bibinfo
  {author} {\bibfnamefont {H.}~\bibnamefont {Hanley}},\ and\ \bibinfo {author}
  {\bibfnamefont {G.}~\bibnamefont {Straty}},\ }\bibfield  {title} {\bibinfo
  {title} {Conditions for the existence of a reentrant solid phase in a sheared
  atomic fluid},\ }\href@noop {} {\bibfield  {journal} {\bibinfo  {journal}
  {Physical Review A}\ }\textbf {\bibinfo {volume} {46}},\ \bibinfo {pages}
  {6731} (\bibinfo {year} {1992})}\BibitemShut {NoStop}%
\bibitem [{\citenamefont {Travis}\ \emph {et~al.}(1995)\citenamefont {Travis},
  \citenamefont {Daivis},\ and\ \citenamefont {Evans}}]{travis1995thermostats}%
  \BibitemOpen
  \bibfield  {author} {\bibinfo {author} {\bibfnamefont {K.~P.}\ \bibnamefont
  {Travis}}, \bibinfo {author} {\bibfnamefont {P.~J.}\ \bibnamefont {Daivis}},\
  and\ \bibinfo {author} {\bibfnamefont {D.~J.}\ \bibnamefont {Evans}},\
  }\bibfield  {title} {\bibinfo {title} {Thermostats for molecular fluids
  undergoing shear flow: Application to liquid chlorine},\ }\href@noop {}
  {\bibfield  {journal} {\bibinfo  {journal} {The Journal of chemical physics}\
  }\textbf {\bibinfo {volume} {103}},\ \bibinfo {pages} {10638} (\bibinfo
  {year} {1995})}\BibitemShut {NoStop}%
\bibitem [{\citenamefont {Braga}\ and\ \citenamefont
  {Travis}(2005)}]{braga_05}%
  \BibitemOpen
  \bibfield  {author} {\bibinfo {author} {\bibfnamefont {C.}~\bibnamefont
  {Braga}}\ and\ \bibinfo {author} {\bibfnamefont {K.~P.}\ \bibnamefont
  {Travis}},\ }\bibfield  {title} {\bibinfo {title} {A configurational
  temperature nos[e-acute]-hoover thermostat},\ }\href@noop {} {\bibfield
  {journal} {\bibinfo  {journal} {The Journal of Chemical Physics}\ }\textbf
  {\bibinfo {volume} {123}},\ \bibinfo {pages} {134101} (\bibinfo {year}
  {2005})}\BibitemShut {NoStop}%
\bibitem [{\citenamefont {Erpenbeck}(1984{\natexlab{b}})}]{erpenbeck_84}%
  \BibitemOpen
  \bibfield  {author} {\bibinfo {author} {\bibfnamefont {J.~J.}\ \bibnamefont
  {Erpenbeck}},\ }\bibfield  {title} {\bibinfo {title} {Shear viscosity of the
  hard-sphere fluid via nonequilibrium molecular dynamics},\ }\href@noop {}
  {\bibfield  {journal} {\bibinfo  {journal} {Physical Review Letters}\
  }\textbf {\bibinfo {volume} {52}},\ \bibinfo {pages} {1333} (\bibinfo {year}
  {1984}{\natexlab{b}})}\BibitemShut {NoStop}%
\bibitem [{\citenamefont {Evans}\ and\ \citenamefont
  {Morriss}(1986{\natexlab{b}})}]{evans_86}%
  \BibitemOpen
  \bibfield  {author} {\bibinfo {author} {\bibfnamefont {D.~J.}\ \bibnamefont
  {Evans}}\ and\ \bibinfo {author} {\bibfnamefont {G.~P.}\ \bibnamefont
  {Morriss}},\ }\bibfield  {title} {\bibinfo {title} {Shear thickening and
  turbulence in simple fluids},\ }\href@noop {} {\bibfield  {journal} {\bibinfo
   {journal} {Physical Review Letters}\ }\textbf {\bibinfo {volume} {56}},\
  \bibinfo {pages} {2172} (\bibinfo {year} {1986}{\natexlab{b}})}\BibitemShut
  {NoStop}%
\bibitem [{\citenamefont {Delhommelle}\ and\ \citenamefont
  {Evans}(2001{\natexlab{b}})}]{config_thermostat_01}%
  \BibitemOpen
  \bibfield  {author} {\bibinfo {author} {\bibfnamefont {J.}~\bibnamefont
  {Delhommelle}}\ and\ \bibinfo {author} {\bibfnamefont {D.~J.}\ \bibnamefont
  {Evans}},\ }\bibfield  {title} {\bibinfo {title} {Configurational temperature
  thermostat for fluids undergoing shear flow: application to liquid
  chlorine},\ }\href@noop {} {\bibfield  {journal} {\bibinfo  {journal}
  {Molecular Physics}\ }\textbf {\bibinfo {volume} {99}},\ \bibinfo {pages}
  {1825} (\bibinfo {year} {2001}{\natexlab{b}})}\BibitemShut {NoStop}%
\bibitem [{\citenamefont {Lue}\ \emph {et~al.}(2002)\citenamefont {Lue},
  \citenamefont {Jepps}, \citenamefont {Delhommelle},\ and\ \citenamefont
  {Evans}}]{config_thermostat_02}%
  \BibitemOpen
  \bibfield  {author} {\bibinfo {author} {\bibfnamefont {L.}~\bibnamefont
  {Lue}}, \bibinfo {author} {\bibfnamefont {O.~G.}\ \bibnamefont {Jepps}},
  \bibinfo {author} {\bibfnamefont {J.}~\bibnamefont {Delhommelle}},\ and\
  \bibinfo {author} {\bibfnamefont {D.~J.}\ \bibnamefont {Evans}},\ }\bibfield
  {title} {\bibinfo {title} {Configurational thermostats for molecular
  systems},\ }\href@noop {} {\bibfield  {journal} {\bibinfo  {journal}
  {Molecular Physics}\ }\textbf {\bibinfo {volume} {100}},\ \bibinfo {pages}
  {2387} (\bibinfo {year} {2002})}\BibitemShut {NoStop}%
\bibitem [{\citenamefont {Samoletov}\ \emph
  {et~al.}(2007{\natexlab{a}})\citenamefont {Samoletov}, \citenamefont
  {Dettmann},\ and\ \citenamefont {Chaplain}}]{config_thermostat_03}%
  \BibitemOpen
  \bibfield  {author} {\bibinfo {author} {\bibfnamefont {A.~A.}\ \bibnamefont
  {Samoletov}}, \bibinfo {author} {\bibfnamefont {C.~P.}\ \bibnamefont
  {Dettmann}},\ and\ \bibinfo {author} {\bibfnamefont {M.~A.~J.}\ \bibnamefont
  {Chaplain}},\ }\bibfield  {title} {\bibinfo {title} {Thermostats for ``slow''
  configurational modes},\ }\href@noop {} {\bibfield  {journal} {\bibinfo
  {journal} {Journal of Statistical Physics}\ }\textbf {\bibinfo {volume}
  {128}},\ \bibinfo {pages} {1321} (\bibinfo {year}
  {2007}{\natexlab{a}})}\BibitemShut {NoStop}%
\bibitem [{\citenamefont {Travis}\ and\ \citenamefont
  {Braga}(2006)}]{bt_thermostat_review}%
  \BibitemOpen
  \bibfield  {author} {\bibinfo {author} {\bibfnamefont {K.~P.}\ \bibnamefont
  {Travis}}\ and\ \bibinfo {author} {\bibfnamefont {C.}~\bibnamefont {Braga}},\
  }\bibfield  {title} {\bibinfo {title} {Configurational temperature and
  pressure molecular dynamics: review of current methodology and applications
  to the shear flow of a simple fluid},\ }\href@noop {} {\bibfield  {journal}
  {\bibinfo  {journal} {Molecular Physics}\ }\textbf {\bibinfo {volume}
  {104}},\ \bibinfo {pages} {3735} (\bibinfo {year} {2006})}\BibitemShut
  {NoStop}%
\bibitem [{\citenamefont {Travis}\ and\ \citenamefont
  {Braga}(2008)}]{bt_thermostat_pressure}%
  \BibitemOpen
  \bibfield  {author} {\bibinfo {author} {\bibfnamefont {K.~P.}\ \bibnamefont
  {Travis}}\ and\ \bibinfo {author} {\bibfnamefont {C.}~\bibnamefont {Braga}},\
  }\bibfield  {title} {\bibinfo {title} {Configurational temperature control
  for atomic and molecular systems},\ }\href@noop {} {\bibfield  {journal}
  {\bibinfo  {journal} {The Journal of Chemical Physics}\ }\textbf {\bibinfo
  {volume} {128}},\ \bibinfo {pages} {014111} (\bibinfo {year}
  {2008})}\BibitemShut {NoStop}%
\bibitem [{\citenamefont {Samoletov}\ \emph {et~al.}(2010)\citenamefont
  {Samoletov}, \citenamefont {Dettmann},\ and\ \citenamefont
  {Chaplain}}]{samoletov_notes}%
  \BibitemOpen
  \bibfield  {author} {\bibinfo {author} {\bibfnamefont {A.~A.}\ \bibnamefont
  {Samoletov}}, \bibinfo {author} {\bibfnamefont {C.~P.}\ \bibnamefont
  {Dettmann}},\ and\ \bibinfo {author} {\bibfnamefont {M.~A.~J.}\ \bibnamefont
  {Chaplain}},\ }\bibfield  {title} {\bibinfo {title} {Notes on configurational
  thermostat schemes},\ }\href@noop {} {\bibfield  {journal} {\bibinfo
  {journal} {The Journal of Chemical Physics}\ }\textbf {\bibinfo {volume}
  {132}},\ \bibinfo {pages} {246101} (\bibinfo {year} {2010})}\BibitemShut
  {NoStop}%
\bibitem [{\citenamefont {Beckedahl}\ \emph {et~al.}(2016)\citenamefont
  {Beckedahl}, \citenamefont {Obaga}, \citenamefont {Uken}, \citenamefont
  {Sergi},\ and\ \citenamefont {Ferrario}}]{beckedahl2016configurational}%
  \BibitemOpen
  \bibfield  {author} {\bibinfo {author} {\bibfnamefont {D.}~\bibnamefont
  {Beckedahl}}, \bibinfo {author} {\bibfnamefont {E.~O.}\ \bibnamefont
  {Obaga}}, \bibinfo {author} {\bibfnamefont {D.~A.}\ \bibnamefont {Uken}},
  \bibinfo {author} {\bibfnamefont {A.}~\bibnamefont {Sergi}},\ and\ \bibinfo
  {author} {\bibfnamefont {M.}~\bibnamefont {Ferrario}},\ }\bibfield  {title}
  {\bibinfo {title} {On the configurational temperature nos{\`e}--hoover
  thermostat},\ }\href@noop {} {\bibfield  {journal} {\bibinfo  {journal}
  {Physica A: Statistical Mechanics and its Applications}\ }\textbf {\bibinfo
  {volume} {461}},\ \bibinfo {pages} {19} (\bibinfo {year} {2016})}\BibitemShut
  {NoStop}%
\bibitem [{\citenamefont {Patra}\ and\ \citenamefont
  {Bhattacharya}(2015)}]{patra2015ergodic}%
  \BibitemOpen
  \bibfield  {author} {\bibinfo {author} {\bibfnamefont {P.~K.}\ \bibnamefont
  {Patra}}\ and\ \bibinfo {author} {\bibfnamefont {B.}~\bibnamefont
  {Bhattacharya}},\ }\bibfield  {title} {\bibinfo {title} {An ergodic
  configurational thermostat using selective control of higher order
  temperatures},\ }\href@noop {} {\bibfield  {journal} {\bibinfo  {journal}
  {The Journal of chemical physics}\ }\textbf {\bibinfo {volume} {142}},\
  \bibinfo {pages} {194103} (\bibinfo {year} {2015})}\BibitemShut {NoStop}%
\bibitem [{\citenamefont {Hamilton}(1990)}]{virial_thermostat}%
  \BibitemOpen
  \bibfield  {author} {\bibinfo {author} {\bibfnamefont {I.~P.}\ \bibnamefont
  {Hamilton}},\ }\bibfield  {title} {\bibinfo {title} {Modified nos\'e-hoover
  equation for a one-dimensional oscillator: Enforcement of the virial
  theorem},\ }\href@noop {} {\bibfield  {journal} {\bibinfo  {journal}
  {Physical Review A}\ }\textbf {\bibinfo {volume} {42}},\ \bibinfo {pages}
  {7467} (\bibinfo {year} {1990})}\BibitemShut {NoStop}%
\bibitem [{\citenamefont {L’Heureux}\ and\ \citenamefont
  {Hamilton}(1993)}]{virial_thermostat_2}%
  \BibitemOpen
  \bibfield  {author} {\bibinfo {author} {\bibfnamefont {I.}~\bibnamefont
  {L’Heureux}}\ and\ \bibinfo {author} {\bibfnamefont {I.}~\bibnamefont
  {Hamilton}},\ }\bibfield  {title} {\bibinfo {title} {Canonically modified
  nos\'e-hoover equation with explicit inclusion of the virial},\ }\href@noop
  {} {\bibfield  {journal} {\bibinfo  {journal} {Physical Review E}\ }\textbf
  {\bibinfo {volume} {47}},\ \bibinfo {pages} {1411} (\bibinfo {year}
  {1993})}\BibitemShut {NoStop}%
\bibitem [{\citenamefont {Samoletov}\ \emph
  {et~al.}(2007{\natexlab{b}})\citenamefont {Samoletov}, \citenamefont
  {Dettmann},\ and\ \citenamefont {Chaplain}}]{samoletov2007thermostats}%
  \BibitemOpen
  \bibfield  {author} {\bibinfo {author} {\bibfnamefont {A.~A.}\ \bibnamefont
  {Samoletov}}, \bibinfo {author} {\bibfnamefont {C.~P.}\ \bibnamefont
  {Dettmann}},\ and\ \bibinfo {author} {\bibfnamefont {M.~A.}\ \bibnamefont
  {Chaplain}},\ }\bibfield  {title} {\bibinfo {title} {Thermostats for ``slow''
  configurational modes},\ }\href@noop {} {\bibfield  {journal} {\bibinfo
  {journal} {Journal of Statistical Physics}\ }\textbf {\bibinfo {volume}
  {128}},\ \bibinfo {pages} {1321} (\bibinfo {year}
  {2007}{\natexlab{b}})}\BibitemShut {NoStop}%
\bibitem [{\citenamefont {Chui}\ \emph {et~al.}(1992)\citenamefont {Chui},
  \citenamefont {Swanson}, \citenamefont {Adriaans}, \citenamefont {Nissen},\
  and\ \citenamefont {Lipa}}]{chui1992temperature}%
  \BibitemOpen
  \bibfield  {author} {\bibinfo {author} {\bibfnamefont {T.}~\bibnamefont
  {Chui}}, \bibinfo {author} {\bibfnamefont {D.}~\bibnamefont {Swanson}},
  \bibinfo {author} {\bibfnamefont {M.}~\bibnamefont {Adriaans}}, \bibinfo
  {author} {\bibfnamefont {J.}~\bibnamefont {Nissen}},\ and\ \bibinfo {author}
  {\bibfnamefont {J.}~\bibnamefont {Lipa}},\ }\bibfield  {title} {\bibinfo
  {title} {Temperature fluctuations in the canonical ensemble},\ }\href@noop {}
  {\bibfield  {journal} {\bibinfo  {journal} {Physical review letters}\
  }\textbf {\bibinfo {volume} {69}},\ \bibinfo {pages} {3005} (\bibinfo {year}
  {1992})}\BibitemShut {NoStop}%
\bibitem [{\citenamefont {Hickman}\ and\ \citenamefont
  {Mishin}(2016)}]{hickman2016temperature}%
  \BibitemOpen
  \bibfield  {author} {\bibinfo {author} {\bibfnamefont {J.}~\bibnamefont
  {Hickman}}\ and\ \bibinfo {author} {\bibfnamefont {Y.}~\bibnamefont
  {Mishin}},\ }\bibfield  {title} {\bibinfo {title} {Temperature fluctuations
  in canonical systems: Insights from molecular dynamics simulations},\
  }\href@noop {} {\bibfield  {journal} {\bibinfo  {journal} {Physical Review
  B}\ }\textbf {\bibinfo {volume} {94}},\ \bibinfo {pages} {184311} (\bibinfo
  {year} {2016})}\BibitemShut {NoStop}%
\bibitem [{\citenamefont {Kittel}(1973)}]{kittel1973nonexistence}%
  \BibitemOpen
  \bibfield  {author} {\bibinfo {author} {\bibfnamefont {C.}~\bibnamefont
  {Kittel}},\ }\bibfield  {title} {\bibinfo {title} {On the nonexistence of
  temperature fluctuations in small systems},\ }\href@noop {} {\bibfield
  {journal} {\bibinfo  {journal} {American Journal of Physics}\ }\textbf
  {\bibinfo {volume} {41}},\ \bibinfo {pages} {1211} (\bibinfo {year}
  {1973})}\BibitemShut {NoStop}%
\bibitem [{\citenamefont {Han}\ and\ \citenamefont
  {Grier}(2004)}]{han2004configurational}%
  \BibitemOpen
  \bibfield  {author} {\bibinfo {author} {\bibfnamefont {Y.}~\bibnamefont
  {Han}}\ and\ \bibinfo {author} {\bibfnamefont {D.~G.}\ \bibnamefont
  {Grier}},\ }\bibfield  {title} {\bibinfo {title} {Configurational temperature
  of charge-stabilized colloidal monolayers},\ }\href@noop {} {\bibfield
  {journal} {\bibinfo  {journal} {Physical review letters}\ }\textbf {\bibinfo
  {volume} {92}},\ \bibinfo {pages} {148301} (\bibinfo {year}
  {2004})}\BibitemShut {NoStop}%
\bibitem [{\citenamefont {Zhao}\ and\ \citenamefont
  {Schr{\"o}ter}(2014)}]{zhao2014measuring}%
  \BibitemOpen
  \bibfield  {author} {\bibinfo {author} {\bibfnamefont {S.-C.}\ \bibnamefont
  {Zhao}}\ and\ \bibinfo {author} {\bibfnamefont {M.}~\bibnamefont
  {Schr{\"o}ter}},\ }\bibfield  {title} {\bibinfo {title} {Measuring the
  configurational temperature of a binary disc packing},\ }\href@noop {}
  {\bibfield  {journal} {\bibinfo  {journal} {Soft matter}\ }\textbf {\bibinfo
  {volume} {10}},\ \bibinfo {pages} {4208} (\bibinfo {year}
  {2014})}\BibitemShut {NoStop}%
\end{thebibliography}%

\end{document}